\documentclass[a4paper, 12pt, twoside, openright]{Thesis} 
\graphicspath{{Figures/}}  

\usepackage[square, numbers, comma, sort&compress]{natbib}  
\usepackage{verbatim}  
\usepackage{vector}  
\usepackage{mathtools}

\usepackage{parskip} 
\usepackage{indentfirst} 

\hypersetup{urlcolor=blue, colorlinks=true}  

\DeclareMathOperator{\arccot}{arccot}

\begin{document}

\def\la{\langle}
\def\ra{\rangle}
\def\beq{\begin{equation}}
\def\eeq{\end{equation}}
\def\beqa{\begin{eqnarray}}
\def\eeqa{\end{eqnarray}}
\def\blankpage{
\newpage
\begin{equation}
\nonumber
\end{equation}
\newpage}
\def\ssst{\scriptscriptstyle}   

\newcommand{\beqas}{\begin{eqnarray*}}
\newcommand{\eeqas}{\end{eqnarray*}}
\newcommand {\fexp} [1] {\exp \left( #1 \right)}
\newcommand {\fabs}[1] {\left| #1 \right|}
\newcommand{\cR}{{\cal{R}}}
\newcommand{\new}[1]{{\it #1}}

\frontmatter	  

\begin{titlepage}
\thispagestyle{empty} 
\hspace{-0.3cm}\includegraphics[scale=0.25]{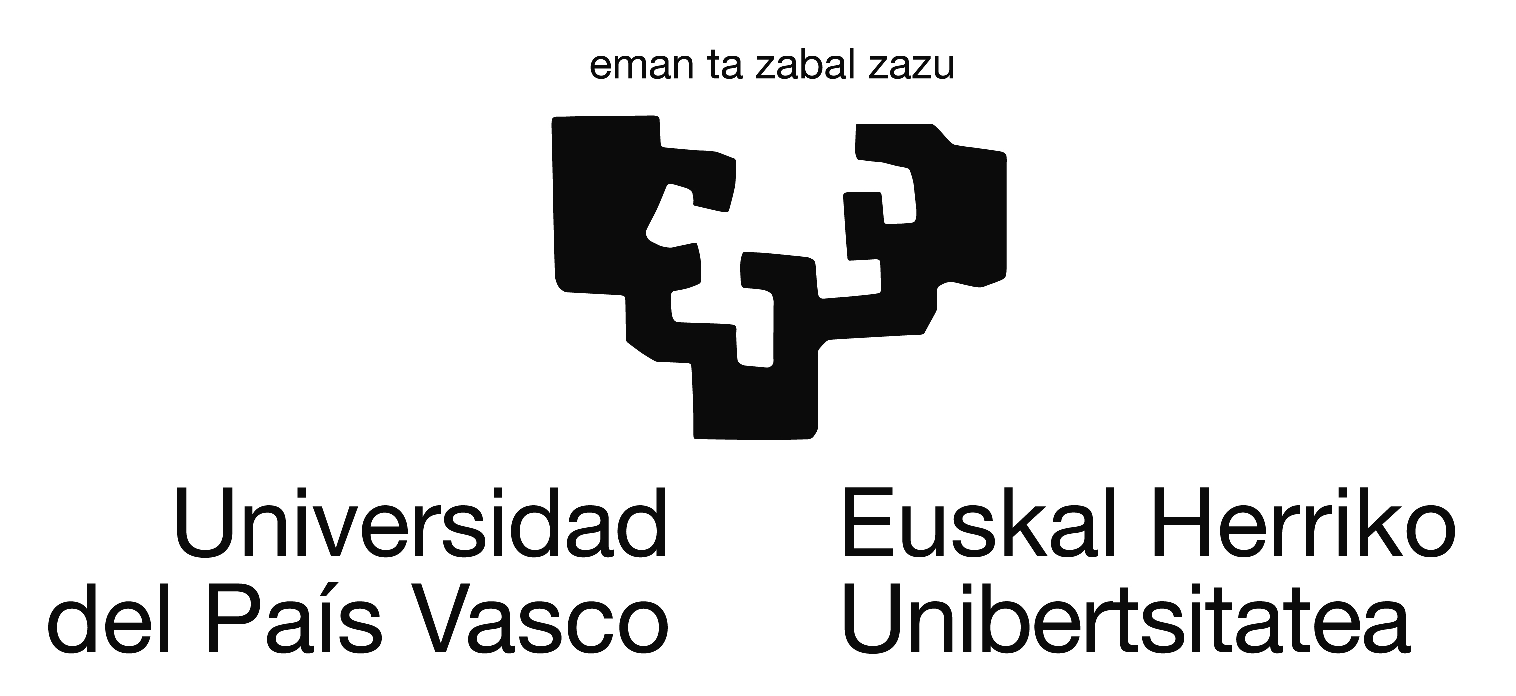}\hspace{0.3cm}
\bigskip
{\centering \large
\par \vspace{1cm}

\hrule\vspace*{0.5cm}

{\LARGE \bf {Shortcuts to adiabaticity in the double well}}

\vspace{0.7cm}\hrule \vspace{2.75cm}
{\LARGE \bf{Sof\'ia Mart\'inez Garaot}}\\
\vspace{1.25cm}
{\it{Supervisor:}} \\
\vspace{0.1cm}
{\large \bf {Prof. Juan Gonzalo Muga Francisco}}\\
\vspace{2.2cm}
\begin{figure}[h]
{\centering {

}\par}
\end{figure}
\vspace{1.0cm}
Departamento de Qu\'{\i}mica-F\'{\i}sica\\
Facultad de Ciencia y Tecnolog\'ia\\
Universidad del Pa\'is Vasco/Euskal Herriko Unibertsitatea\\ (UPV/EHU)\\
\vspace*{1.0cm}
\hspace{5.5cm}Leioa, Febrero 2016} \pagebreak
\end{titlepage}




\setstretch{1.3}  

\fancyhead{}  
\rhead{\thepage}  
\lhead{}  

\pagestyle{fancy}  

\clearpage  

\pagestyle{empty}  
\null\vfill

\vfill\vfill\vfill\vfill\vfill\vfill\null
\clearpage  

\pagestyle{empty}  

\null\vfill
\textit{``El aprendizaje es experiencia, todo lo dem\'as es informaci\'on.''}

\begin{flushright}
Albert Einstein 
\end{flushright}

\vfill\vfill\vfill\vfill\vfill\vfill\null
\clearpage  

%
%
%

\pagestyle{empty}  

\null\vfill

\vfill\vfill\vfill\vfill\vfill\vfill\null
\clearpage  

\setstretch{1.3}  

\acknowledgements{
\addtocontents{toc}{\vspace{1em}}  
Al final de este largo camino me gustar\'ia agradecer a todos aquellos que de alguna forma u otra han estado ayud\'andome y apoy\'andome. Mi primera idea era poner una larga lista de nombres en la que quedasen reflejadas todas y cada una de las personas que me han acompa\~nado durante esta etapa pero finalmente he pensado que todo aquel que me conoce encontrar\'a un hueco donde ubicar su nombre en estas l\'ineas.

En primer lugar tengo que agradecer a mi director por haber confiado en m\'i y haberme guiado durante estos a\~nos. A mis colaboradores por haberme ense\~nado tanto. A mis compa\~neros de despacho porque todos ellos son parte de este trabajo. Al resto de compa\~neros de la universidad por todos los momentos vividos. A todas mis amigas y amigos por estar siempre a mi lado. A mi familia, a mis padres y a Pau. Y por \'ultimo, a ti, por animarme cada d\'ia. 
Gracias a todos.

\it{Esta Tesis ha sido financiada a trav\'es de una beca predoctoral concedida por la Universidad del Pa\'is Vasco/Euskal Herriko Unibertsitatea (UPV/EHU).}

}

\clearpage  

\pagestyle{fancy}  

\lhead{\emph{Contents}}  
\tableofcontents  

\lhead{\emph{List of Figures}}  
\listoffigures  


\setstretch{1.5}  
\clearpage  

\setstretch{1.3}  
\pagestyle{empty}  
\dedicatory{Zuentzat}

\addtocontents{toc}{\vspace{1em}}  
\chapter{List of publications} 
\label{Publications}
\lhead{\emph{List of publications}} 
\hspace{-0.5 cm}{{\large\bf \rm I)}} {\bf The results of this Thesis are based on the following articles}
\section*{{\large\bf Published  Articles}}

\begin{enumerate}

\item E. Torrontegui, S. Mart\'inez-Garaot, A. Ruschhaupt, and J. G. Muga
\\{\it Shortcuts to adiabaticity: Fast-forward approach}\\ \href{http://journals.aps.org/pra/abstract/10.1103/PhysRevA.86.013601}{Phys. Rev. A {\bf 86}, 013601 (2012).}

\item E. Torrontegui, S. Mart\'inez-Garaot, M. Modugno, X. Chen, and J. G. Muga
\\{\it Engineering fast and stable splitting of matter waves}\\ \href{http://journals.aps.org/pra/abstract/10.1103/PhysRevA.87.033630}{Phys. Rev. A {\bf 87}, 033630 (2013).}

\item S. Mart\' inez-Garaot, E. Torrontegui, X. Chen, M. Modugno, D. Gu\'ery-Odelin, S.-Y. Tseng, and J. G. Muga
\\{\it Vibrational Mode Multiplexing of Ultracold Atoms}\\ \href{http://journals.aps.org/prl/abstract/10.1103/PhysRevLett.111.213001}{Phys. Rev. Lett. {\bf 111}, 213001 (2013).}

\item S. Mart\' inez-Garaot, S.-Y. Tseng, and J. G. Muga
\\{\it Compact and high conversion efficiency mode-sorting asymmetric Y junction using shortcuts to adiabaticity}\\ \href{https://www.osapublishing.org/ol/abstract.cfm?uri=ol-39-8-2306}{Opt. Lett. {\bf 39}, 2306 (2014).}

\item S. Mart\' inez-Garaot, E. Torrontegui, and J. G. Muga
\\{\it Shortcuts to adiabaticity in three-level systems using Lie transforms}\\ \href{http://journals.aps.org/pra/abstract/10.1103/PhysRevA.89.053408}{Phys. Rev. A {\bf 89}, 053408 (2014).}

\item S. Mart\' inez-Garaot, A. Ruschhaupt, J. Gillet, Th. Busch, and J. G. Muga
\\{\it Fast quasiadiabatic dynamics}\\ \href{http://journals.aps.org/pra/abstract/10.1103/PhysRevA.92.043406}{Phys. Rev. A {\bf 92}, 043406 (2015).}

\item S. Mart\' inez-Garaot, M. Palmero, D. Gu\'ery-Odelin, and J. G. Muga
\\{\it Fast bias inversion of a double well without residual particle excitation}
\\ \href{http://journals.aps.org/pra/abstract/10.1103/PhysRevA.92.053406}{Phys. Rev. A {\bf 92}, 053406 (2015).}

\vspace{1.25 cm}

\hspace{-0.65  cm}{{\large\bf \rm II)}} {\bf Other articles produced during the Thesis period}
\section*{{\large\bf Published  Articles not included in this Thesis}}

\item S. Ib\' a\~ nez, S. Mart\' inez-Garaot, X. Chen, E. Torrontegui, and J. G. Muga
\\{\it Shortcuts to adiabaticity for non-Hermitian systems}\\ \href{http://journals.aps.org/pra/abstract/10.1103/PhysRevA.84.023415}{Phys. Rev. A {\bf 84}, 023415 (2011).}

\item S. Ib\' a\~ nez, S. Mart\' inez-Garaot, X. Chen, E. Torrontegui, and J. G. Muga
\\{\it Erratum: Shortcuts to adiabaticity for non-Hermitian systems}\\ \href{http://journals.aps.org/pra/abstract/10.1103/PhysRevA.86.019901}{Phys. Rev. A {\bf 86}, 019901 (2012).}

\item E. Torrontegui, S. Mart\' inez-Garaot, and J. G. Muga
\\{\it Hamiltonian engineering via invariants and dynamical algebra}\\ \href{http://journals.aps.org/pra/abstract/10.1103/PhysRevA.89.043408}{Phys. Rev. A {\bf 89}, 043408 (2014).}

\item M. Palmero, S. Mart\' inez-Garaot, J. Alonso, J. P. Home, and J. G. Muga
\\{\it Fast expansions and compressions of trapped-ion chains}\\ \href{http://journals.aps.org/pra/abstract/10.1103/PhysRevA.91.053411}{Phys. Rev. A {\bf 91}, 053411 (2015).}

\item M. Palmero, S. Mart\' inez-Garaot, U. G. Poschinger, A. Ruschhaupt, and J. G. Muga
\\{\it Fast separation of two trapped ions}
\\ \href{http://iopscience.iop.org/article/10.1088/1367-2630/17/9/093031/meta}{New J. Phys. {\bf 17}, 093031 (2015).}

\hspace{-0.65  cm}{{\large\bf \rm III)}} {\bf Review Articles}
\item E. Torrontegui, S. Ib\' a\~ nez, S. Mart\' inez-Garaot, M. Modugno, A. del Campo, D. Gu\' ery-Odelin, A. Ruschhaupt, X. Chen, and J. G. Muga \\
{\it Shortcuts to adiabaticity}\\ \href{http://www.sciencedirect.com/science/article/pii/B9780124080904000025}{Adv. At. Mol. Opt. Phys. {\bf 62}, 117 (2013).}

\end{enumerate}

\mainmatter	  
\pagestyle{fancy}  


\addtocontents{toc}{\vspace{0.6em}}

\addcontentsline{toc}{chapter}{Introduction}
\chapter*{Introduction} 
\label{Introduction}
\lhead{\emph{Introduction}} 

During the last three decades, the research in quantum optics  has experienced a phenomenal boost, largely driven by the rapid progress in microfabrication technologies, precision measurements, coherent radiation sources, and theoretical work. 
Many quantum optical systems are employed to test and illustrate the fundamental notions of quantum theory. They have also practical applications for communications,  quantum information processing, metrology and the development of new quantum-based technologies, whose physical aspects have by now become an integral part of quantum optics. 
Frequently, the manipulated systems are quite simple, such as one or a few ions or neutral atoms in harmonic or double wells. 
Bose-Einstein condensates involve of course many more atoms, but may still be described by
mean-field theories.    
Controlling these systems accurately has become a major goal in contemporary Physics.
Serge Haroche and David J. Wineland won the Nobel Prize in 2012 after developing methods for manipulating individual ions  
in Paul traps or photons in cavities while preserving their quantum-mechanical nature.     
 
This Thesis contributes to this goal by proposing fast operations for one to few ultra cold atoms, or Bose-Einstein condensates, in a double well potential, extending the results as well to optical waveguide systems.     
``Fast'' is to be understood with respect to adiabatic processes. 
The ``adiabatic'' concept may have two different meanings: the thermodynamical one and the quantum one. In thermodynamics, an adiabatic process is the one in which there is no heat transfer between system and environment. In quantum mechanics, as stated by Born and Fock (1928) in the adiabatic theorem: ``a physical system remains in its instantaneous eigenstate when a given perturbation is acting on it slowly enough and if there is a gap between the eigenvalue and the rest of the Hamiltonian's spectrum''.  
In terms of the instantaneous eigenvalues $E_n$ and their corresponding instantaneous eigenvectors $|\phi_n\rangle$, the adiabaticity condition, i.e., the condition that has to be satisfied to follow the adiabatic dynamics, can be written as
\beq
\label{adiabatic}
\hbar \left |\frac{ \langle \phi_n(t)|\partial_t \phi_m(t)\rangle}{ E_n(t)-E_m(t)} \right | \ll 1, \, \, n\neq m. 
\nonumber
\eeq 
In this Thesis, we shall always understand ``adiabatic'' in the quantum-mechanical sense.
Quantum adiabatic processes are in principle useful to drive or prepare states in a robust and controllable manner, and have also been proposed to solve complicated computational problems. However, they are prone to suffer noise and decoherence or loss problems due to the long times involved. 
This is often problematic because some applications require many repetitions or too long times.  

Shortcuts to adiabaticity (STA) are alternative fast processes that reproduce the same final populations, or even the same final state, as the adiabatic process in a finite, shorter time. 
The expression ``shortcut to adiabaticity'' was introduced in 2010 by Chen {\it{et al.}} \cite{Chen2010} to describe protocols that speed up a quantum adiabatic process, usually, although not necessarily, through a non-adiabatic route.
There are many different approaches to design the shortcuts. For example, the counterdiabatic or transitionless tracking approach formulated by Demirplak and Rice (2003, 2005, 2008) \cite{Demirplak2003,Demirplak2005,Demirplak2008} or independently by Berry (2009) \cite{Berry2009}, based on adding counterdiabatic terms to a reference Hamiltonian $H_0$ to achieve adiabatic dynamics with respect to $H_0$. Moreover, Lewis-Riesenfeld invariants (1969) \cite{Lewis1969} were used to inverse engineer a time-dependent Hamiltonian $H(t)$ from the invariant $I(t)$.
Masuda and Nakamura (2010) developed a ``fast-forward technique'' for several manipulations \cite{Masuda2010}. There are also alternative methods that use the dynamical symmetry of the Hamiltonian or based on distributing the adiabaticity parameter homogeneously in time, or Optimal Control Theory (OCT) \cite{Torrontegui2013d}.  
In this Thesis I will not only apply these existing methods but also develop new ones.   

Since adiabatic processes are ubiquitous, the shortcuts span a broad range of applications in atomic, molecular, and optical physics, such as fast transport, splitting and expansion of ions or neutral atoms; internal population control, and state preparation (for nuclear magnetic resonance or quantum information), vibrational mode multiplexing or demultiplexing, cooling cycles, many-body state engineering or correlations microscopy \cite{Torrontegui2013d}. 
The Thesis focuses on the double well potential, which is an interesting model to study some of the most fundamental quantum effects, like interference or tunneling. Using utracold atoms it has become possible to study the double well at an unprecedented level of precision and control. This has allowed the observation of Josephson oscillations, nonlinear self-trapping and recently, second-order tunneling effects. Few-body systems are lately of much interest as they enable us to study finite-size effects for a deeper understanding of the microscopic mechanics in utracold atoms, and for the possibility of realizing operations involving a few qubits. Also, double wells for single atoms and Bose-Einstein condensates have been used for precise measurement in interferometry experiments. For trapped ions, the double well is used to implement basic operations for quantum information processing, for example, separation or recombination of ions, Fock states creation, or tunable spin-spin interactions and entanglement. 

The Thesis is divided into six chapters: The first chapter is devoted to fast splitting of matter waves. The fast-forward approach is applied to speed up the process and a two dynamical-mode model is introduced. This two-mode model will be an important test-bed model during the whole Thesis. Linear and non-linear matter waves (interacting Bose-Einstein condensates) are studied. Chapter 2 deals with an interacting few-body boson gas in a two-site potential. In particular, we investigate how to accelerate an insulator-superfluid transition and the implementation of a $1:2$ and $1:3$ beam splitter. To achieve these goals, a new STA method based on Lie transforms is worked out. In chapter 3 I present one more new STA method that uses the time dependence of a control parameter to delocalize in time the transition probability among adiabatic levels. Some general properties are described and the approach is used to speed up basic operations in three different systems: a two-mode model, interacting bosons in a double well and a few-particle system on a ring.
In chapter 4 the invariant-based inverse engineering approach is used to accelerate multiplexing or demultiplexing processes. The shortcut is designed in the two-mode model and then it is mapped into a realizable coordinate potential. Chapter 5 extends the results of the previous chapter to optical wave guides systems. Finally, chapter 6 provides a strategy based on the compensating force-approach to implement a fast bias inversion both in neutral atoms and in trapped ions. Combining this fast bias inversion with fast multiplexing and demultiplexing processes, population inversions using only trap deformations can be achieved. 

Due to the length of the manuscript and the different topics discussed, the notation is consistent within each chapter, but not necessarily throughout the Thesis.



\chapter{Engineering fast and stable splitting of matter waves}
\label{Chapter1}
\lhead{Chapter 1. \emph{Engineering fast and stable splitting of matter waves}} 
When attempting to split a coherent noninteracting atomic cloud
by bifurcating the initial trap 
into two well separated wells,
slow adiabatic following is unstable with respect to any slight trap asymmetry, and the matter wave ``collapses'' to the deepest well.
A generic fast chopping splits the wave but it also excites it.
Shortcuts to adiabaticity engineered to speed up the unperturbed adiabatic process through nonadiabatic transients provide, instead, quiet and robust balanced splitting.  
For a Bose-Einstein condensate in the mean-field limit,  the interatomic interaction makes the splitting, adiabatic or via shortcuts, more stable with respect to trap asymmetry. Simple formulas are provided to distinguish different  regimes.   
\newpage
\section{Introduction}

The splitting of a wave packet is an important operation in   
matter wave interferometry \cite{Hohenester2007,Grond2009a,Grond2009,Pezze2005}.
A strategy to improve the interferometer performance is to suppress the interaction \cite{Fattori2008,Gustavsson2008}, so let us first consider   
a non-interacting Bose-Einstein condensate.   
For this system, a complete wave splitting into two separated branches is a peculiar operation because adiabatic following, rather than robust, is intrinsically unstable with respect to a small external potential asymmetry \cite{Gea-Banacloche2002}.
The potential is assumed here to evolve from a single well to a final 
double-well where tunnelling is negligible \cite{Shin2004}. 
The ground-state wave function ``collapses'' into the final lower well (or more generally into the one that holds the lowest ground state as in \cite{Gea-Banacloche2002}) and a very slow trap potential bifurcation fails to split the wave except 
for perfectly symmetrical potentials.  
A fast bifurcation remedies this but the price is typically 
a strong excitation which is also undesired, as it produces loss of contrast in the interference patterns 
when recombining the two waves \cite{Collins2005}.  
We propose here a way around these problems
by using shortcuts to adiabaticity that speed up the adiabatic process along a nonadiabatic route \cite{Chen2010}. Wave splitting via shortcuts avoids the final excitation 
and is significantly more stable with respect to asymmetry than the adiabatic following.   
Specifically we shall use a streamlined version \cite{Torrontegui2012} of the fast-forward (FF) technique of Masuda and Nakamura \cite{Masuda2010} applied to the Gross-Pitaevskii (GP) or Schr\"odinger equations.   
There have previously been found some 
obstacles to apply the invariants-based method (quadratic-in-momentum invariants do not satisfy the required boundary conditions \cite{Torrontegui2012}) and the transitionless-driving algorithm \cite{Demirplak2003} (because of difficulties in implementing counter-diabatic terms in practice).

In Sec. \ref{FFA} we summarize the FF approach for condensates (interacting or not) 
in one dimension and its application to splitting. 
In Sec. \ref{perturbation} the effect of a small asymmetric perturbation is studied for noninteracting 
matter waves, and Sec. \ref{2mode} analyzes and interprets the results with the aid of a
moving two-mode model. Sec. \ref{interacting} studies the remarkable stability with respect to the asymmetry achieved due to interatomic
interactions in the mean-field limit, and different regimes are distinguished. Finally, Sec. \ref{discussion_splitting} discusses the 
results and open questions.         
\section{Fast-forward approach}\label{FFA}
The FF method \cite{Torrontegui2012,Masuda2010,Masuda2008} 
may be used to generate external potentials $V_{FF}$ and drive the matter wave from an initial single well to a final symmetric double-well.
The starting point 
is the  three-dimensional (3D) time-dependent GP equation,
\beq
\label{gp_e}
i\hbar \partial_t|\psi(t)\rangle=H(t)|\psi(t)\rangle,
\eeq
where $H(t)=T+G(t)+V(t)$ includes  kinetic energy $T$, external potential $V$,
and mean field potential $G$. 
We are assuming an external local potential, where ``local" means here $\langle {\bf{x}}|V(t)|{\bf{x}'}\rangle=V({\bf{x}},t)\delta({\bf{x}}-{\bf{x}'})$. The kinetic and mean field terms in the coordinate representation have the usual forms, 
\beqa
\langle{\bf{x}}|T|\psi(t)\rangle&=&\frac{-\hbar^2}{2m} \nabla^2 \psi({\bf{x}},t), \\
\langle{\bf{x}}|G(t)|\psi(t)\rangle&=&g |\psi({\bf{x}},t)|^2\psi({\bf{x}},t).  
\eeqa
The GP equation (\ref{gp_e}) is used to describe a Bose-Einstein condensate within the mean field approximation and it takes into account the atom-atom interaction through $g$, the atom-atom coupling constant. In the case of vanishing coupling constant $g=0$ the GP equation simplifies to the Schr\"odinger equation.

By solving Eq. (\ref{gp_e}) in coordinate space, $V({\bf{x}},t)$ may be written as
\beq
\label{pot1}
V({\bf{x}},t)=\frac{{i\hbar\langle {\bf{x}}| \partial_t\psi(t)\rangle-\langle{\bf{x}}|T+G(t)|\psi(t)\rangle}}{{\langle{\bf{x}}|\psi(t)\rangle}}, 
\eeq 
with $\langle{\bf{x}}|\psi(t)\rangle=\psi({\bf{x}},t)$. By introducing into Eq. (\ref{pot1}) the ansatz
\beq
\langle{\bf{x}}|\psi(t)\rangle=r({\bf{x}},t)e^{i\phi({\bf{x}},t)}, \quad r({\bf{x}},t), \phi({\bf{x}},t) \in \mathbb{R},
\eeq
we get 
\beq
\label{v_ff}
V({\bf{x}},t)=i\hbar \frac{\dot r}{r}-\hbar \dot \phi+\frac{\hbar^2}{2m} \left ( \frac{2i\nabla \phi \cdot \nabla r}{r} + i \nabla^2 \phi
-(\nabla \phi)^2 +\frac{\nabla^2 r}{r} \right ) -gr^2,
\eeq
where the dot means time derivative. The real and imaginary parts are
\beqa
{\rm{Re}}[V({\bf{x}},t)]&=&-\hbar{\dot \phi}+\frac{\hbar^2}{2m}\bigg(\frac{\nabla^2r}{r}-( \nabla \phi)^2\bigg)-g r^2,
\label{real}
\\
{\rm{Im}}[V({\bf{x}},t)]&=&\hbar\frac{\dot r}{r}+\frac{\hbar^2}{2m}\bigg(\frac{2\nabla \phi \cdot \nabla r}{r}+\nabla^2 \phi \bigg).
\label{imag}
\eeqa
Our purpose is to design a local and real potential such that an initial eigenstate of the initial Hamiltonian, $H(0)$, typically the ground state, but it could be otherwise, evolves in a time $t_f$ into the corresponding eigenstate of the final Hamiltonian, $H(t_f)$. We assume that the full Hamiltonian and the corresponding eigenstates are known at the boundary times.

By construction the potential of Eq. (\ref{v_ff}) is local. If we impose $Im[V({\bf{x}},t)]=0$, i.e.,
\beq
\label{phase_ff}
\frac{\dot r}{r}+\frac{\hbar}{2m}\bigg(\frac{2\nabla \phi \cdot \nabla r}{r}+\nabla^2 \phi \bigg)=0,
\eeq
then we get from Eq. (\ref{real}) a local and real potential.

In the inversion protocol $r({\bf{x}},t)$ is designed first,
and Eq. (\ref{phase_ff}) is solved for $\phi$ 
to get
$V_{FF}({\bf{x}},t):={\rm{Re}}[V({\bf{x}},t)]$ from Eq. (\ref{real}).  
To ensure that the initial and final states are eigenstates of the stationary GP equation we impose  
$\dot{r}=0$ at $t=0$ and $t_f$. Then Eq. (\ref{phase_ff}) has
solutions $\phi({\bf{x}},t)$ independent of $\bf{x}$ at the boundary times \cite{Torrontegui2012}.  
Using this in Eq. (\ref{real}) at $t=0$, and multiplying by $e^{i\phi(0)}$, we get
\beq
\bigg[\!-\frac{\hbar^2}{2m}\nabla^2+V({\bf{x}},0)+g |\psi({\bf{x}},0)|^2\!\bigg]\!\psi({\bf{x}},0)
\nonumber\\
\!=\!-\hbar\dot \phi(0)\psi({\bf{x}},0). 
\eeq 
The initial state $\psi({\bf{x}},0)$ is an eigenstate of the stationary GP equation with chemical potential $-\hbar\dot\phi(0)=\mu(0)$. Note that the above solution of $\phi$ (with $\dot r=0$ at boundary times) admits the addition of an arbitrary function that depends only on time and modifies the zero of energy.
A similar result is found at $t_f$.  

In the remainder of this chapter we will restrict to the one dimensional case so the potential in Eq. (\ref{pot1}) is reduced to
\beq
\label{pot2}
V(x,t)=\frac{i\hbar \langle x|\partial_t\psi(t)\rangle-\langle x|T+G|\psi(t)\rangle}{\langle x|\psi(t)\rangle}, 
\eeq
with 
$\langle{x}|T|\psi(t)\rangle=\frac{-\hbar^2}{2m} \psi''({x},t)$ and 
$\langle{x}|G(t)|\psi(t)\rangle=g_1N|\psi({x},t)|^2\psi({x},t)$.  
The primes denote derivatives with respect to $x$, $g_1$ is the effective 1D-coupling constant of the Bose-Einstein condensate, and $N$ is the number of atoms. For the
numerical examples  we consider $^{87}$Rb atoms. 
Using in Eq. (\ref{pot2}) the ansatz
$\langle{x}|\psi(t)\rangle=r({x},t)e^{i\phi({x},t)}, \quad r({x},t), \phi({x},t) \in \mathbb{R}$,
the real and imaginary parts will be
\beqa
{\rm{Re}}[V({x},t)]&=&-\hbar{\dot \phi}+\frac{\hbar^2}{2m}\bigg(\frac{r''}{r}-( \phi')^2\bigg)-g_1Nr^2,
\label{real_1}
\\
{\rm{Im}}[V({x},t)]&=&\hbar\frac{\dot r}{r}+\frac{\hbar^2}{2m}\bigg(\frac{2\phi' r'}{r}+\phi''\bigg),
\label{imag_1}
\eeqa
where the dot means time derivative.

\begin{figure}[t]
\begin{flushleft}
  \hspace*{2.3 cm}
   \includegraphics[width=0.6 \linewidth]{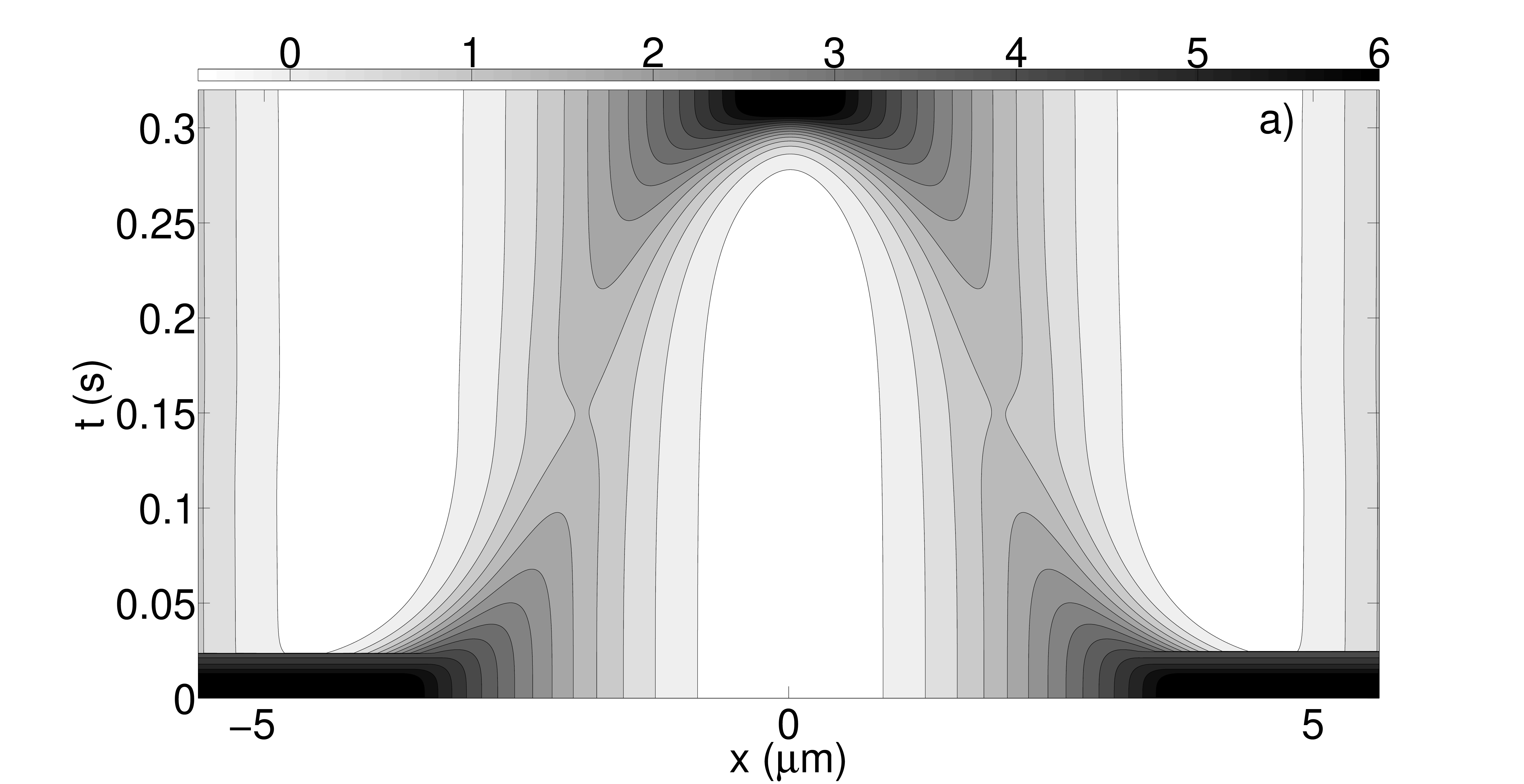}
   \end{flushleft}
   \begin{flushleft}
   \hspace*{2.3 cm}
   \includegraphics[width=0.6 \linewidth]{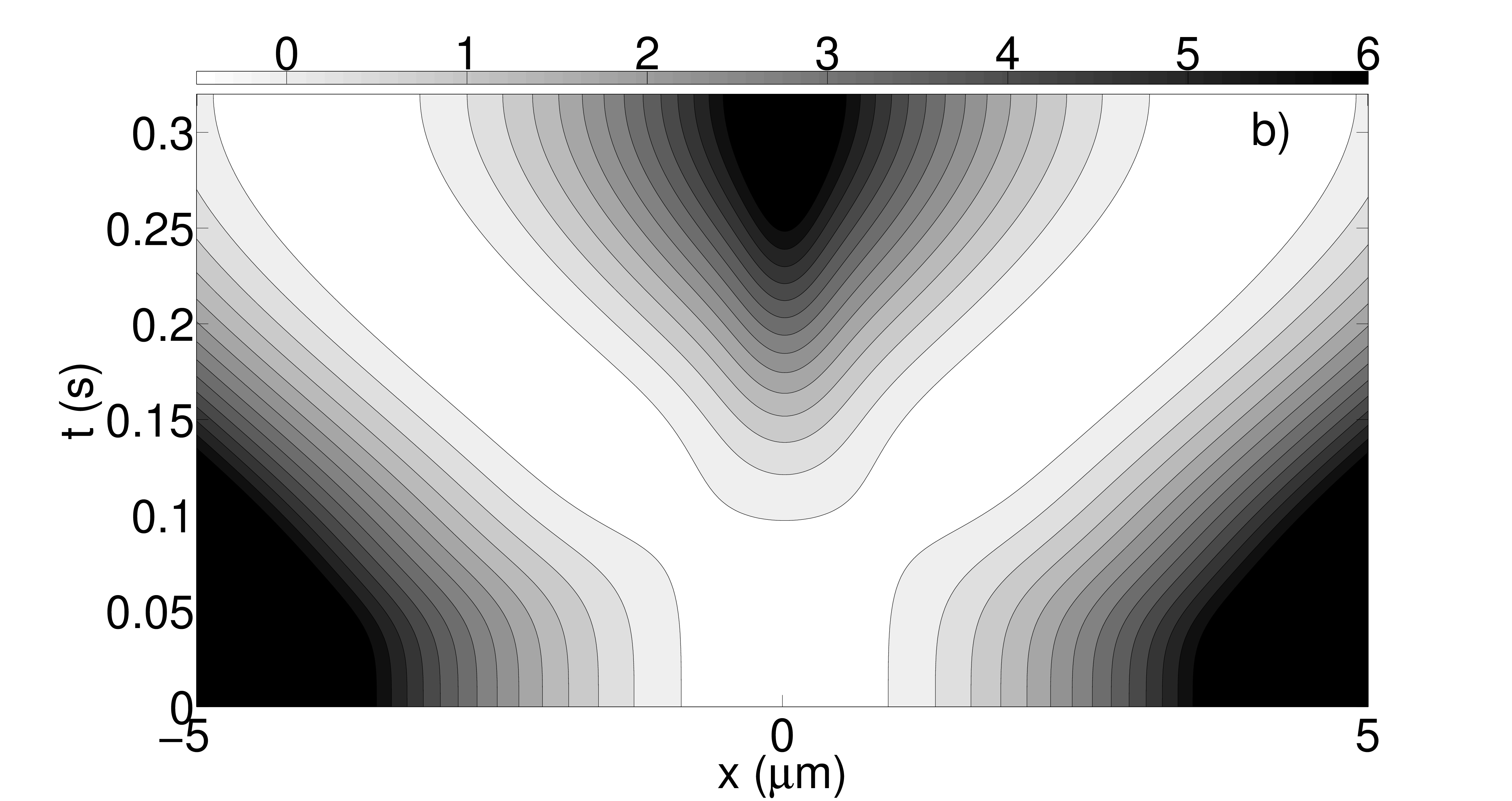}
   \end{flushleft}
\caption{Contour plot of $V_{FF}$ in units $\hbar\omega$ from Eq. (\ref{real_1}) for (a) a three-well interpolation 
and
(b) a $Y$-shaped form. 
Parameters: $\omega=780$ rad/s, and $t_f=320$ ms.}
\label{f1}
\end{figure}

In the following two sections we consider first 
$g_1=0$ and split an initial single Gaussian 
state $f(x,0)=e^{- x^2/2a_0^2}$ $(a_0=\sqrt{\hbar/m\omega})$ 
into a final double Gaussian $f(x,t_f)=e^{- (x-x_f)^2/2a_0^2}+e^{-(x+x_f)^2/2a_0^2}$. 
In previous works \cite{Torrontegui2012,Masuda2010} use has been made of the interpolation
$r(x,t)=z(t)\big\{[1-{\cal{R}}(t)]f(x,0)+ 
{\cal{R}}(t)f(x,t_f)\big\}$,
where ${\cal{R}}(t)$ is a smooth, monotonously increasing function
from 0 to 1,
and $z(t)$ is a normalization function.
This produces a triple-well potential
at intermediate times. Here we use instead the two-bump form 
$
r(x,t)=z(t)[e^{- [x-x_0(t)]^2/2a_0^2}+e^{-[x+x_0(t)]^2/2a_0^2}],
$
which generates simpler $Y$-shaped potentials (see Fig. \ref{f1}). 
We impose $\dot{x}_0(0)=\dot{x}_0(t_f)=0$, so $\dot r=0$ at the boundary times.
In the numerical examples $x_0(s)=x_f(3s^2-2 s^3)$, where $s=t/t_f$, 
and $x_f=4\, \mu$m (see e.g. \cite{Schumm2005});   
Equation (\ref{imag_1}) is solved with the initial conditions $\phi(x=0)=\frac{\partial{\phi}}{\partial x}|_{x=0}=0$.    
%
%
%
%
%
%
%
%
%
\section{Effect of the perturbation}\label{perturbation}

Assume now a perturbed Hamiltonian $H_\lambda=T+V_\lambda$ with  $V_{\lambda}=V_{FF}+\lambda\theta(x)$,
where $\theta(x)$ is the step function and $\lambda$ the potential imbalance. 
Except in the final discussion, we assume that $\lambda$ is some uncontrollable and  hard-to-avoid small perturbation, typically unknown, due to imperfections of the experimental setting.  
The adiabatic splitting becomes unstable, as we shall see, but the instability does not 
depend strongly on this particular form,  chosen for simplicity. 
It would also be found, for example,  for a linear-in-$x$ 
perturbation, a smoothed step, slightly different frequencies for the final right and left traps, or a shifted central barrier \cite{Gea-Banacloche2002}. 
In the final potential configuration, with negligible tunneling, the two wells are independent, and the 
global ground state is localized  in one of them.      
To analyze the effects of the perturbation on the wavefunction structure and on the shortcut dynamics, 
we compute several wavefunction overlaps: 
%
%
%
%
\begin{figure}[t]
 \begin{center}
   \includegraphics[width=0.5 \linewidth]{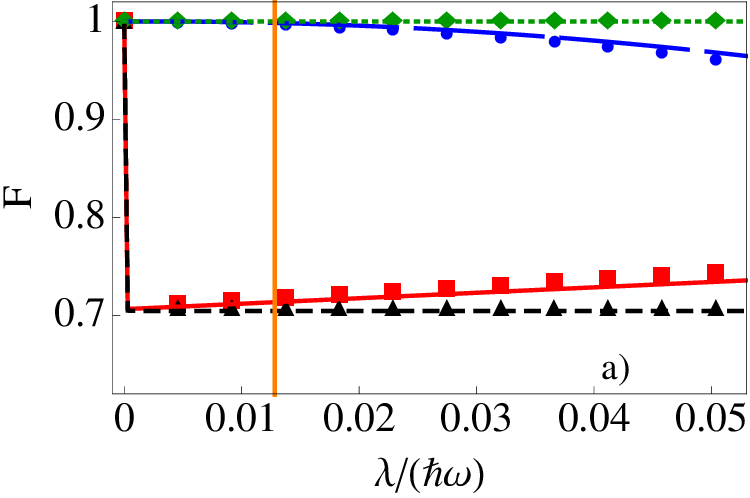}
  \includegraphics[width=0.5 \linewidth]{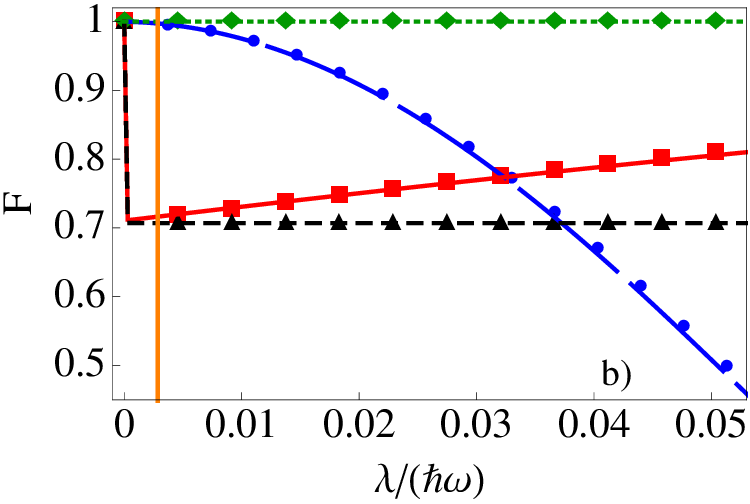}
  \includegraphics[width=0.5 \linewidth]{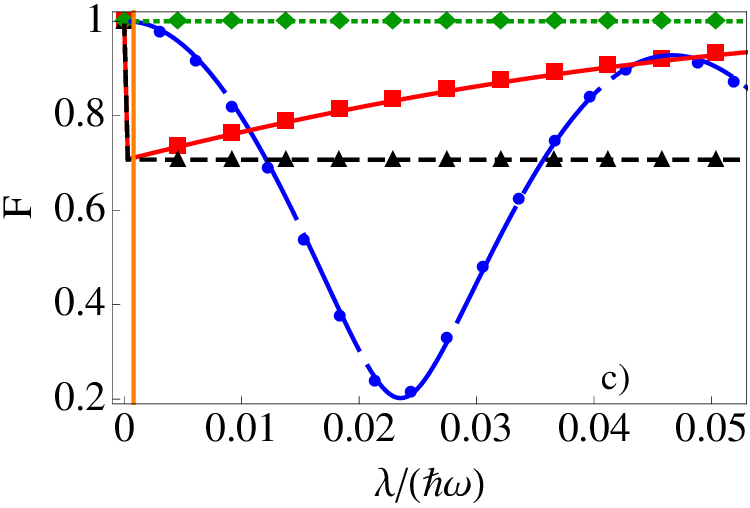}
  \end{center}
\caption{Different fidelities versus the perturbation parameter $\lambda$ for the FF approach (lines) and the two-mode model (symbols). 
$F_D^{(0)}$: (blue) long-dashed line and circles; $F_D$: (red) solid line and squares;  $F_S$: (black) short-dashed line and triangles; $F_I$: (green) dotted line and diamonds.
The vertical (orange) line is at $0.2/(t_f\omega)$. 
{(a)}  $t_f=20$ ms. {(b)} $t_f=90$ ms. {(c)} $t_f=320$ ms.
$\omega=780$ rad/s.}
\label{f2}
\end{figure}
%
%
%
%
\begin{itemize}

\item{$F_S=|\langle \psi^{-}_0(t_f)|\psi^{-}_\lambda(t_f)\rangle|$, the (black) short-dashed line in Fig. \ref{f2}, is the ``structural fidelity''   
between the (perfectly split) ground state $\psi^{-}_0(t_f)$ of the unperturbed potential $V_{FF}(t_f)$ and the final ground state $ \psi^{-}_\lambda(t_f)$ of the 
perturbed potential $V_{\lambda}$. This would be the fidelity found with the desired split state if the process were adiabatic.}
$F_S(\lambda)$ decays extremely rapidly from 1
at $\lambda=0$ to $1/\sqrt{2}$, which corresponds to the collapse 
of the ground state of the perturbed potential $V_{\lambda}$ into the deeper well. 

\item{$F_D^{(0)}=|\langle \psi^-_0(t_f)|\psi(t_f)\rangle|$, 
the (blue) long-dashed line in Fig. \ref{f2},
is the fidelity between the state dynamically evolved with $H_{\lambda}$,
$\psi(x,t_f)=\langle x|e^{iH_{\lambda}t_f/\hbar}|\psi(0)\rangle$, and $\psi^-_0(t_f)$. 
$\psi(0)=\psi^-_\lambda(0)$ is the
initial ground state with $V_{\lambda}(0)$. If  
$\psi(0)=\psi^-_0(0)$ is used instead, the results are indistinguishable; 
see the overlap $F_I=|\la \psi^-_\lambda(0)|\psi^-_0(0)\ra|\approx 1$,
[green dotted line] in Fig. \ref{f2}. 
The flat  $F_D^{(0)}(\lambda)$ at small $\lambda$, in sharp contrast to the rapid decay of $F_S(\lambda)$,  
demonstrates the robustness of the balanced splitting produced by the shortcut. Shorter process times $t_f$ make the splitting more and more stable [compare Figs. \ref{f2}(a)-\ref{f2}(c)]. ({We assume condensate lifetimes of the order of seconds; see e.g., 
\cite{Hall2007}.}) In principle, $t_f$ may be reduced arbitrarily. In practice, this reduction implies an increase in  transient energy excitation that  requires accurate 
potential engineering for higher energies \cite{Chen2010a}.
Considering that the time-averaged standard deviation of the
energy $\overline{\Delta E}$ should be limited at some value a general bound  is $t_f>h/(4\overline{\Delta E})$ \cite{Anandan1990}. For the trap frequency in the examples (780 rad/s) and setting $\overline{\Delta E}=\hbar\omega$ the bound saturates for a time $t_f=2$ ms, $10$ times shorter than our shortest time in Fig. 2.}

\item{$F_D=|\langle \psi(t_f)|\psi^{-}_\lambda(t_f)\rangle|$ [solid (red) line in Fig. \ref{f2}]  is the fidelity between the 
dynamically evolved state $\psi(t_f)$ and the final ground state $\psi^{-}_\lambda(t_f)$
for the perturbed potential. 
If the process is adiabatic, then $F_D\approx 1$. 
For very small perturbations $F_D\approx F_S$. In this regime the dynamical wave function $\psi(t_f)$ is not affected by the perturbation and becomes
$\psi^-_0(t_f)$, up to a phase factor; 
note that $F_D^{(0)}\approx 1$ there. 
We understand and quantify below this important regime as a sudden process 
in a moving-frame interaction picture.  
As $\lambda$ increases, the energies of 
the ground and excited states of $V_{\lambda}$ separate and the process 
becomes less sudden and more adiabatic. In Fig. \ref{f2}(c) for  $t_f=320$ ms and for large values of $\lambda$, $F_D$ approaches 1 again, the final evolved state collapses to one side and becomes the ground state of $V_{\lambda}$. 
For the shorter final times in Figs. \ref{f2}(a) and \ref{f2}(b), 
larger $\lambda$ values are needed so that $F_D$ approaches 1 adiabatically.}
\end{itemize}
\section{Moving two-mode model}\label{2mode}
Static two-mode models have been previously used 
to analyze splitting processes or double-well dynamics \cite{Grond2009,Javanainen1999a,Aichmayr2010}. 
Here we add the separation motion of left and right basis functions to 
provide 
analytical estimates and insight.  
In terms of a (dynamical) orthogonal bare basis
$|L(t)\rangle = \left(\scriptsize{\begin{array} {rccl} 0\\ 1 \end{array}} \right)$, $|R(t)\rangle = \left(\scriptsize{\begin{array} {rccl} 1\\ 0 \end{array}} \right)$ 
our two-mode Hamiltonian model is
\beq
\label{H_tm_FF}
H(t)=\frac{1}{2} \left ( \begin{array}{cc}
\lambda
& -\delta(t)\\
-\delta(t)& -\lambda
\end{array} \right),
\eeq 
where $\delta(t)$ is the tunneling rate \cite{Grond2009,Javanainen1999a,Aichmayr2010}.
We may consider $\lambda$ constant through a given splitting process, for the time being, and equal to the perturbative parameter that defines the asymmetry.  
A more detailed approach discussed later does not produce any significant difference.        
The instantaneous eigenvalues are
\beq
\label{eigenvalues_tm}
E^{\pm}_\lambda(t)=
\pm \frac{1}{2} \sqrt{\lambda^2+\delta^2(t)},
\eeq
and the normalized eigenstates take the form
\beq
\begin{array}{ll}
\label{eigenstates_tm}
|\psi^+_\lambda(t)\rangle = \sin{ \left ( \frac{\alpha}{2} \right ) } |L(t)\rangle-\cos {\left ( \frac{\alpha}{2} \right ) }|R(t)\rangle, 
\\
\\
|\psi^-_\lambda(t)\rangle = \cos{ \left ( \frac{\alpha}{2} \right )}|L(t)\rangle+\sin{\left ( \frac{\alpha}{2} \right )}|R(t)\rangle,
\end{array}
\eeq
where $\alpha=\alpha(t)$ is the mixing angle given by $\tan \alpha = \delta (t)/\lambda$.

The bare basis states $\left \{ |L(t)\rangle,|R(t)\rangle \right\}$ are symmetrical and orthogonal-moving left and right states. Initially they
are close  to each other  and $\delta(0)\gg\lambda$. The instantaneous eigenstates of $H$ are  the symmetric ground state 
$|\psi^{-}_{0}(0)\rangle=\frac{1}{\sqrt{2}}(|L(0)\rangle+|R(0)\rangle)$
and the antisymmetric excited state 
$|\psi^{+}_{0}(0)\rangle=\frac{1}{\sqrt{2}}(|L(0)\rangle-|R(0)\rangle)$
of the single well.
At $t_f$ we distinguish two extremes:

{\it{i)}} For $\delta(t_f)\gg\lambda$ the final eigenstates of $H$ tend to symmetric and antisymmetric splitting states $|\psi^{\mp}_{\lambda}(t_f)\rangle=\frac{1}{\sqrt{2}}(|L(t_f)\rangle\pm
|R(t_f)\rangle)$;   

{\it{ii)}} For  $\delta(t_f)\ll\lambda$ the final eigenfunctions of
$H$ collapse and become right-and left-localized states: $|\psi^{-}_{\lambda}(t_f)\rangle=|L(t_f)\rangle$ and $|\psi^{+}_{\lambda}(t_f)\rangle=|R(t_f)\rangle$.

Since $\delta(t_f)$ is set as a small number to avoid tunneling
in the final configuration, 
the transition from one to the other regime explains the sharp drop of $F_S$ at small $\lambda\approx\delta(t_f)$.    

\subsection{Moving-frame interaction picture}
We define now a moving-frame interaction-picture (IP) wave function  
$\psi^A=A^\dagger\psi^S$, where $A=\sum_{\beta=L,R} |\beta (t)\ra\la \beta(0)|$
and $\psi^S$ is the Schr\"odinger-picture wave function. 
$\psi^A$ obeys 
\beq
i\hbar\dot{\psi}^A=(H_A -K_A) \psi^A, 
\eeq
with
\beqa
H_A&=&A^\dagger H A, \\ 
K_A&=&i\hbar A^\dagger \dot{A}, 
\eeqa
but for real $\la x|R(t)\ra$ and $\la x|L(t)\ra$, the symmetry $\la x|R(t)\ra=\la -x|L(t)\ra$ makes
$K_A=0$.     
  
Inverting Eq. (\ref{eigenstates_tm}) the bare states may be written in terms of the ground and first excited states and energies. 
The
two-level model approximates the actual dynamics by first  
identifying $|\psi^{\pm}_0(t)\ra$ and $E^{\pm}_0(t)$ with the instantaneous ground and
excited states and energies of the unperturbed FF Hamiltonian.\footnote{Contrast this 
with the variational approach in \cite{Menotti2001}.}
We combine them to compute the bare basis in coordinate representation and then 
the matrix elements $\la\beta'|H_{\lambda}|\beta\ra=H_{\lambda}^{\beta'\beta}$, for $\beta\ne \beta'$. From Eq. (\ref{H_tm_FF}),  $\delta(t)=-2H_{\lambda}^{RL}=-2H_{\lambda}^{LR}$.\footnote{ 
For $\beta=\beta'$, we may consistently calculate  $\lambda'(t):=2(H_{\lambda}^{RR}-V_0)=-2(H_{\lambda}^{LL}-V_0)$, where $V_0=[E^-_\lambda(t)+E^+_\lambda(t)]/2$
is a shift to match the zero-energy point between the FF and the two-mode models. 
$\lambda'$ differs slightly from the constant $\lambda$ at short times, but    
the results of substituting $\lambda$ by $\lambda'$ are hardly distinguishable 
in the calculations, so the treatment with $\lambda$ is preferred 
for simplicity.}   
Once all matrix elements are set  
we solve the dynamics in the moving frame for the two-mode Hamiltonian. 
The initial state may be the ground state of the perturbed or unperturbed initial potential. 
The agreement  with the exact results is excellent (see the symbols of Fig. \ref{f2}),
which denotes the absence of higher excited states. 
This two-level model thus provides a powerful interpretative and control tool. 
To gain more insight we now perform further approximations.

\subsection{Sudden and adiabatic approximations}
The fidelities at low $\lambda$ may be understood with the sudden approximation
in the IP. Its validity requires 
\cite{messiah1961quantum}
\beq
\label{sudden}
t_f \ll \frac{\hbar}{\Delta \overline{H_A}}, 
\eeq
where $\Delta \overline{H_A}=[{\langle\psi(0)|\overline{H_A}^2|\psi(0)\rangle-
\langle\psi(0)|\overline{H_A}|\psi(0)\rangle^2}]^{1/2}$. 
We take $|\psi(0)\rangle=|\psi_{0}^{-}(0)\rangle$ and $\overline{H_A}=\frac{1}{t_f}\int_{0}^{t_f}dt'H_A(t')$, where
the matrix elements of $H_A(t')$ in the basis $\{|\beta(0)\ra\}$
coincide with the matrix elements of $H$ in Eq. (\ref{H_tm_FF}),
when the latter are expressed in the basis $\{|\beta(t')\ra\}$.
The condition for the sudden approximation to hold becomes 
\beq
\lambda \ll \frac{2\hbar}{t_f}.
\eeq
Vertical lines mark $\lambda=0.2\hbar/t_f$ in Fig. \ref{f2} and demonstrate that indeed this condition 
sets the range in which $F_D^{(0)}\approx 1$ so that  the fast protocol provides balanced splitting in spite of the asymmetry.    
%
%
%
%
\begin{figure}[t]
 \begin{center}
   \includegraphics[width=0.54 \linewidth]{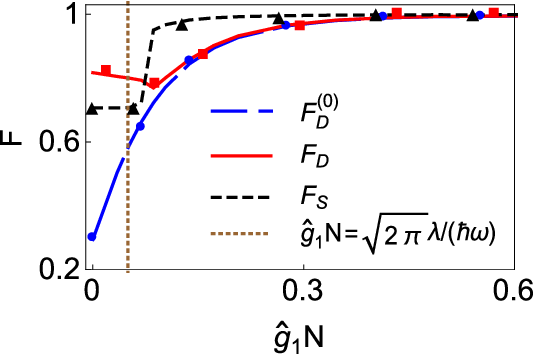}
    \end{center}
\caption{Fidelities vs dimensionless coupling constant 
for $\lambda/(\hbar \omega)=0.02$, $t_f=320$ ms, and $x_{f}=4$ $\mu$m.
Lines are the same as in Fig. \ref{f2}. Symbols are for a two-level model
like Eq. (\ref{H_tm_FF}) with the nonlinear
diagonal terms $g_2 |c_{R,L}|^2$ added, where 
$g_2=g_1\int dx |R(x)|^4=g_1\int dx |L(x)|^4$ and $|c_{R,L}|^2$ are 
populations for left and right states \cite{Ottaviani2010}.
The vertical line is at  ${\widehat{g_1}N}={\sqrt{2\pi}\lambda/\hbar\omega}$; see the Appendix \ref{Interaction versus asymmetry for adiabatic following}.} 
\label{f3}
\end{figure}
%
%
%
%

The increase in $F_D$ with increasing $\lambda$ can be explained using the adiabatic approximation. 
The adiabaticity condition is here \cite{Torrontegui2012a}
\beq
|\langle \psi^-_\lambda(t)|\partial_t\psi^+_\lambda(t)\rangle| \ll \frac{1}{\hbar}|E^-_\lambda(t)-E^+_\lambda(t)|, 
\eeq
which, using Eqs. (\ref{eigenvalues_tm}) and (\ref{eigenstates_tm}), 
takes the form
\beq
\label{adiabaticity}
\left |\frac{{\hbar\lambda \dot \delta(t)}}{{2[\lambda^2+\delta(t)^2]^{3/2}}}
\right | \ll 1.
\eeq
$^{}\\$
{\section{Interacting Bose-Einstein condensates}\label{interacting}}
We now generalize the results of the two previous sections for a condensate with interatomic interaction in the mean-field framework. 
We calculate the ground states $\chi_N(x)$ and $\chi_{\frac{N}{2}}(x)$ of a harmonic trap that holds a Bose-Einstein condensate
with $N$ and $N/2$ atoms and define   
$f(x,t)=[1-{\cal{R}}(t)]\chi_N(x)+{\cal{R}}(t)\chi_{\frac{N}{2}}(x)$, where ${\cal{R}}(t)=3(t/t_f)^2-2(t/t_f)^3$. $r(x,t)$ is 
constructed as 
\beq\label{eq11}
r(x,t)=\big\{f[x-x_0(t),t]+f[x+x_0(t),t]\big\}/z(t),
\eeq
where $z(t)$ is a normalization factor and $x_0(t)=x_f{\cal{R}}(t)$. 
We then get $V_{FF}$ from Eq. (\ref{real_1}) 
and evolve  
the initial ground state with the GP equation 
using  the perturbed  potential $V_{\lambda}(t)$.

The fidelities are shown in Fig. \ref{f3}
versus the dimensionless coupling constant $\widehat{g_1}N=g_1N/(\hbar\omega a_0)$.
Note the stabilization of $F_D^{(0)}$ towards $1$ upon increasing the interaction (this implies more stable shortcuts). $F_D$ increases too, as the dynamics tends to be more adiabatic.  
The structural fidelity jumps to $1$ around ${\widehat{g_1}N}={\sqrt{2\pi}\lambda/\hbar\omega}$
from the linear case value $1/\sqrt{2}$, i.e., 
balanced splitting by adiabatic following is robust versus trap asymmetry for  
$\widehat{g_1}N \gg\lambda/\hbar\omega$ (see the Appendix \ref{Interaction versus asymmetry for adiabatic following}).  
The extra filling of the lower well increases the nonlinear interaction there opposing the external potential imbalance. 

\begin{figure}[t]
 \begin{center}
    \includegraphics[width=0.54 \linewidth]{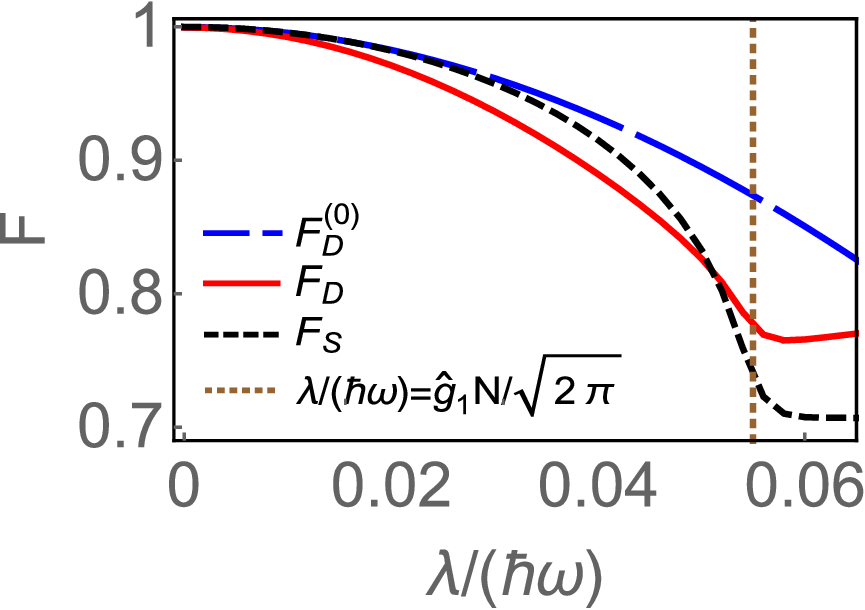}
    \end{center}
\caption{Fidelities for a Bose-Einstein condensate; lines are the same as in Fig. \ref{f2}.    
Equation (\ref{eq11}) is used to design the potential $V_{FF}$. Parameters: 
$x_{f}=4$ $\mu$m, $\omega=780$ rad/s,  $\widehat{g_1}N$=0.138,   
and $t_f=45$ ms.}
\label{f4}
\end{figure}

The two-level model may also be extended to interacting condensates with minor modifications, 
also providing an accurate description (see Fig. \ref{f3}).    
Adiabaticity fails eventually when decreasing $t_f$ and/or $g_1$,
but the shortcut provides then balanced splitting (see the example of Fig. 
\ref{f4}): for small $\lambda$, adiabatic following would be stable
(see $F_S$ and compare to the
sharp drop in Fig. \ref{f2} for linear dynamics), but the process is not 
quite adiabatic ($F_D<F_S$) for the chosen time, $t_f=45$ ms -more time would be needed. 
The shortcut is nevertheless more stable than the hypothetical adiabatic process ($F_D^{(0)}>F_S$).   
%
%
%
%
%
%

\section{Discussion}\label{discussion_splitting}
We have designed simple $Y$-shaped (position and time dependent) potential traps to fully split noninteracting matter waves rapidly without final excitation, avoiding the instability of the adiabatic approach with respect to slight trap asymmetries.  We also avoid or mitigate in this manner the decoherence and noise that affect slow adiabatic following \cite{Shin2004,Collins2005}.
The bifurcation may be experimentally implemented with optical traps created with the aid of spatial light modulators \cite{Boyer2006}. A simpler approximate approach would involve two Gaussian beams. Further manipulations, such as 
the application of differential ac Stark phase shifts could be combined with the proposed technique \cite{Shin2004}. Also, 
a differential phase among the two final wave parts will develop due to the imbalance, allowing for precision metrology 
\cite{Schumm2005,Hall2007}, without the time limitations of methods based on adiabatic splitting
\cite{Hall2007}. In addition, optimal control methods \cite{Hohenester2007,Grond2009a,Grond2009} complement  the present approach to further improve stability and/or optimize other variables  such as the transient excitation.     

A unique feature of the above application of shortcuts to adiabaticity, compared to previous
ones  \cite{Chen2010,Chen2010b,Ban2012,Ibanez2012}, is that the shortcut does not 
attempt to reproduce the result of an adiabatic following of the perturbed asymmetrical system in a shorter time. 
(The assumption has been made so far that the perturbation is 
uncontrolled and, possibly, unknown.) Instead, the shortcut reproduces the  balanced splitting of the adiabatic following corresponding to the unperturbed, perfectly symmetrical system. 
In other words, shortening the time here is not really the goal, but it means 
to achieve stability. 

Other operations may actually make positive use of 
the instability due to potential asymmetries. In particular, the ground- and first-excited-state components of the 
initial trap could be spatially separated by a controlled, slightly asymmetrical adiabatic bifurcation. 
Moreover, both states would become ground states of the right and left 
final traps, so the process may as well be used as a population inversion protocol from the excited to the ground state.  

We have also analyzed and exemplified the effect of interatomic interactions
for a condensate in the mean-field regime. 
The interaction changes the behavior of the system with respect to asymmetry,  stabilizing  dramatically balanced splitting. 
The total adiabatic collapse of the wave onto one 
of the two final separated  wells requires, in this case, a significant perturbation, 
proportional to the coupling constant.
Compared to the noninteracting case, this offers different manipulation opportunities, in particular, the possibility of considering the 
asymmetric perturbation as a known, controllable parameter, so that the imbalance between the two wells may be prepared at will.  
Examples of this type of manipulation may be found in \cite{Albiez2005,Gati2006,Esteve2008}.
Shortcuts to adiabaticity and, in particular, the FF approach may be 
readapted to that scenario by designing the fast protocol taking into account the known, controlled asymmetry. 
The emphasis would be again, as in most applications of the shortcuts,
on accelerating and reproducing the result of a slow process.          

Shortcuts to adiabaticity could play other roles in systems described by a double-well with varying parameters.
They have been applied, in particular, to speed up the generation of spin-squeezed many-body states in bosonic Josephson
junctions \cite{Julia-Diaz2012}.
Here we suggest other applications: for example, Stickney and Zozulya \cite{Stickney2002} have described a wave-function recombination instability 
due to the weak nonlinearity of the condensate. Specifically, they  consider an initially weak ground symmetric mode 
of the double-well which is exponentially amplified at the expense of an initially strong excited asymmetric mode
when the wells are recombined.  
Similarly to the instability due to asymmetry described in this chapter for noninteracting waves, 
the nonlinear instability is in fact enhanced by adiabatic following. A shortcut-to-adiabaticity strategy as the one followed in this chapter 
would stabilize the recombination. Our present  results may as well be applied to 
design Y-junctions in planar optical waveguides 
\cite{Sanz2012,Lin2012,Tseng2012}, since the equation that describes the field in the paraxial approximation 
is formally identical to the linear Schr\"odinger equation, with the longitudinal  coordinate playing the role of time. 
Finally, partial splitting, in which the final two wells 
are not completely separated and tunnelling is still allowed, may as well
be considered.


\chapter{Shortcuts to adiabaticity in three-level systems using Lie transforms}
\label{Chapter2}
\lhead{Chapter 2. \emph{STA in three-level systems using Lie transforms}} 
Sped-up protocols  that drive a system quickly to the same populations that can be reached by a slow adiabatic process 
may involve Hamiltonian terms which are difficult to realize. We use the dynamical symmetry of the Hamiltonian to 
find, by means of Lie transforms, alternative Hamiltonians  that achieve the same goals without the problematic terms.
We apply this technique to three-level systems 
(two interacting bosons in a double well, and  beam splitters with two and three output 
channels) driven by Hamiltonians that belong  
to the  four-dimensional algebra U3S3. 
\newpage
\section{Introduction}
``Shortcuts to adiabaticity'' are  manipulation protocols 
that take the system quickly to the same populations, or even the same state, 
that can be reached by a slow adiabatic process \cite{Torrontegui2013d}.  
Adiabaticity is ubiquitous in preparing a system state in atomic, molecular, and optical physics, so many applications of this concept 
have been worked out, in both theory and experiment \cite{Torrontegui2013d}.    
Some of the engineered Hamiltonians that speed up the adiabatic process in principle may involve 
terms which are difficult or impossible to realize in practice.  
In simple systems the dynamical symmetry of the Hamiltonian can be used to eliminate the problematic  terms and provide instead  feasible Hamiltonians. Examples are single particles transported or expanded by harmonic potentials \cite{Torrontegui2011,Muga2010},
or  two-level systems \cite{Chen2010b, Chen2011b, Ibanez2012}.  
In this chapter we extend this program to three-level systems whose Hamiltonians belong to a four-dimensional dynamical algebra.   
This research was motivated by a recent observation by Opatrn\'y and M\o lmer \cite{Opatrny2014}. 
Among other systems they considered two (ultracold) interacting bosons in a double well within a three-state approximation. 
Specifically, the aim was to speed up a  
transition from a ``Mott-insulator'' state with one particle in each well, to a  delocalized ``superfluid'' state. 
The reference adiabatic process  consisted in slowly turning off the interparticle interaction while increasing the tunneling
rate. To speed up this process they applied a  
method of generating shortcuts based on adding a ``counterdiabatic'' (cd) term to the original time-dependent Hamiltonian
\cite{Chen2010b,Demirplak2003,Demirplak2005,Demirplak2008,Berry2009},  but the evolution with the cd term turns out to be difficult to realize in practice \cite{Opatrny2014}.   
In this chapter we shall use the symmetry of the 
Hamiltonian (its dynamical algebra) to find an alternative shortcut  
by means of a Lie transform, namely, a unitary operator in the Lie group associated with the Lie algebra.  
Since other physical systems have the same Hamiltonian structure the results are applicable to them too. 
Specifically, the analogy between the time-dependent Schr\"odinger equation and the stationary-wave equation 
for a waveguide in the
paraxial approximation \cite{Longhi2009,Szameit2010,Longhi2011,Ornigotti2008,Rangelov2012,Chien2013} is used   
to design  short-length optical beam splitters with two and three output channels. 

In Sec. \ref{the model} we describe the theoretical model for two bosons in two wells. In Sec. \ref{CD} we summarize the counterdiabatic or transitionless tracking approach and apply it to the bosonic system. Section \ref{as} sets the approach based on unitary Lie transforms to produce alternative shortcuts. In Sec. \ref{is} we introduce the insulator-superfluid transition and  apply the shortcut designed in the previous section. In Sec. \ref{bs} we apply the technique to generate beam splitters with two and three output channels. Section \ref{discussion_lie} discusses the results and open questions. Finally, in the Appendix \ref{algebra} some features of the Lie algebra of the system are discussed.   
\section{The model}
\label{the model}
An interacting boson gas in a two-site potential  is described within the Bose-Hubbard approximation \cite{Fisher1989, Jaksch1998} by  
\beq
\label{H_BH}
H_0=\frac{U}{2}\sum^2_{j=1}n_j(n_j-1)-J(a_1a_2^\dag+a_1^\dag a_2),
\eeq
where $a_{j}$ ($a_{j}^\dag$) are the bosonic particle annihilation (creation) operators at the $j$th site and $n_j$ is the occupation number operator. The on-site interaction energy is quantified by the parameter $U$ and the hopping energy by $J$. They are assumed to be controllable functions of time.  
For two particles the Hamiltonian in the occupation number basis $|2,0\rangle=\left ( \scriptsize{\begin{array} {rcccl} 1\\ 0\\0 \end{array}} \right)$, $|1,1\rangle=\left ( \scriptsize{\begin{array} {rcccl} 0\\ 1\\0 \end{array}} \right)$, and $|0,2\rangle=\left ( \scriptsize{\begin{array} {rcccl} 0\\ 0\\1 \end{array}} \right)$ is given by \cite{Opatrny2014}
\beq
\label{H_0_Lie}
H_0=\left ( \begin{array}{ccc}
U & -\sqrt{2}J & 0\\
-\sqrt{2}J & 0 & -\sqrt{2}J \\
0 & -\sqrt{2}J & U
\end{array} \right)=UG_4-4JG_1,
\eeq
where
\beq
\label{G1_G4}
G_1=\frac{1}{2\sqrt{2}}\left ( \begin{array}{ccc}
0 & 1 & 0\\
1 & 0 & 1 \\
0 & 1 & 0
\end{array} \right), 
\, \, \,
G_4=\left ( \begin{array}{ccc}
1 & 0 & 0\\
0 & 0 & 0 \\
0 & 0 & 1
\end{array} \right).
\eeq
This Hamiltonian belongs to the vector space (Lie algebra) spanned by $G_1$, $G_4$, and 
two more generators, 
\beq
\label{generators}
G_2=\frac{1}{2\sqrt{2}}\left ( \begin{array}{ccc}
0 & -i & 0\\
i & 0 & i \\
0 & -i & 0
\end{array} \right), 
G_3=\frac{1}{4}\left ( \begin{array}{ccc}
1 & 0 & 1\\
0 & -2 & 0 \\
1 & 0 & 1
\end{array} \right), 
\eeq
with nonzero commutation relations
\beqa
\label{commutators}
[G_1,G_2]&=&iG_3, \nonumber  \\
\label{commutators}
[G_2,G_3]&=&iG_1, \nonumber  \\
\label{commutators}
[G_3,G_1]&=&iG_2, \nonumber  \\
\label{commutators}
[G_4,G_1]&=&iG_2, \nonumber  \\
\label{commutators}
[G_2,G_4]&=&iG_1.
\eeqa
This four-dimensional Lie algebra, U3S3 \cite{maccallum1999classification}, is described in more detail in the Appendix \ref{algebra}. 
To find the Hermitian basis we calculate $[G_1,G_4]$, and then all commutators of the result  
with previous elements. This operation is repeated for all operator pairs until no new linearly independent operator appears.  

To diagonalize the Hamiltonian (\ref{H_0_Lie}) it is useful to  parameterize $U$ and $J$ as \cite{Opatrny2014}
\beqa
\label{parametrization_U}
U&=&E_0 \cos\varphi,\nonumber \\
\label{parametrization_U}
J&=&\frac{E_0}{4} \sin \varphi,
\eeqa
where $E_0=E_0(t)$ and $\varphi=\varphi(t)$, so that 
\beq
\label{H_0_2}
H_0=E_0 \left ( \begin{array}{ccc}
\cos\varphi & -\frac{1}{2\sqrt{2}} \sin \varphi & 0\\
-\frac{1}{2\sqrt{2}} \sin \varphi & 0 & -\frac{1}{2\sqrt{2}} \sin \varphi \\
0 &-\frac{1}{2\sqrt{2}} \sin \varphi & \cos\varphi
\end{array} \right).
\eeq
The instantaneous eigenvalues are
\beqa
\label{eigenvalues_1}
&&E_1=\frac{E_0}{2}(\cos\varphi-1),
\\
\label{eigenvalues_2}
&&E_2=E_0\cos \varphi,
\\
\label{eigenvalues_3}
&&E_3=\frac{E_0}{2}(\cos\varphi+1),
\eeqa
corresponding to the normalized eigenstates
\beqa
\label{eigenstates_1}
&&|\phi_1\rangle= \left ( \begin{array}{c}
\frac{1}{2}\sqrt{1-\cos\varphi}\\
\frac{1}{\sqrt{2}}\sqrt{1+\cos\varphi}\\
\frac{1}{2}\sqrt{1-\cos\varphi}\\
\end{array} \right), 
\\
\label{eigenstates_2}
&&|\phi_2\rangle=\frac{1}{\sqrt{2}} \left ( \begin{array}{c}
1\\
0\\
-1\\
\end{array} \right), 
\\
\label{eigenstates_3}
&&|\phi_3\rangle= \left ( \begin{array}{c}
\frac{1}{2}\sqrt{1+\cos\varphi}\\
-\frac{1}{\sqrt{2}}\sqrt{1-\cos\varphi}\\
\frac{1}{2}\sqrt{1+\cos\varphi}\\
\end{array} \right). 
\eeqa
\section{Counterdiabatic or transitionless tracking approach}
\label{CD}
For the transitionless driving or counterdiabatic approach  formulated by Demirplak and Rice \cite{Demirplak2003,Demirplak2005,Demirplak2008} or equivalently by Berry \cite{Berry2009}, 
the starting point is a time-dependent reference Hamiltonian
\beq
\label{H_0_B}
H_0(t)=\sum_n|n_0(t)\rangle E_n^{(0)}(t)\langle n_0(t)|.
\eeq
The approximate time-dependent adiabatic solution of the dynamics with $H_0$ takes the form
\beq
\label{a_s}
|\psi_n(t)\rangle=e^{i \xi_n(t)}|n_0(t)\rangle,
\eeq
where the adiabatic phase reads
\beq
\label{a_p}
\xi_n(t)=-\frac{1}{\hbar}\int_0^t dt' E_n^{(0)} (t')+ i \int_0^t dt' \langle n_0(t')|\partial_{t'} n_0(t')\rangle.
\eeq
Defining now the unitary operator
\beq
A(t)=\sum_n e^{i\xi_n(t)}|n_0(t)\rangle \langle n_0(0)|,
\eeq
a Hamiltonian $H(t)=i\hbar \dot A A^\dag$ can be constructed to drive the system exactly along the adiabatic paths of $H_0(t)$ as 
\beqa
H(t)&=&H_0(t)+H_{cd}(t),  \nonumber
\\
H_{cd}(t)&=&i\hbar \sum_n  \left[ |\dot{n}_0(t)\rangle \langle n_0(t)|
-\langle n_0(t)|\dot{n}_0(t)\rangle |n_0(t)\rangle \langle n_0(t)| \right],
\eeqa
where $H_{cd}(t)$ is purely nondiagonal in the $\{ |n_0(t)\rangle\}$ basis and the overdot represents time derivative.  

We may change the $E_n^{(0)}(t)$, and therefore $H_0(t)$ itself, keeping the same $|n_0(t)\rangle$. We could for example make all the $E_n^{(0)}(t)$ zero, or set $\xi_n(t)=0$ \cite{Berry2009}. Taking into account this freedom the Hamiltonian for transitionless driving can be generally written as
\beq
H(t)=-\hbar \sum_{n}|n_0(t)\rangle \dot\xi_n \langle n_0(t)| + i\hbar \sum_n |\partial_t n_0(t)\rangle \langle n_0(t)|.
\eeq
Subtracting $H_{cd}(t)$, the generic $H_0$ is
\beq
H_0(t)=\sum_n |n_0(t)\rangle \left[ i\hbar \langle n_0(t)|\partial_t n_0(t) \rangle-\hbar \dot \xi_n \right] \langle n_0(t)|.
\eeq

For our system [$|n_0(t)\rangle\to|\phi_n\rangle$], the counterdiabatic term takes the form
\beq
H_{cd}=i\hbar (|\dot \phi_1\rangle \langle \phi_1|+|\dot \phi_3 \rangle \langle \phi_3|).
\eeq
Taking into account Eqs. (\ref{eigenstates_1}), (\ref{eigenstates_2}), (\ref{eigenstates_3}), and their respective time derivatives we get 
\beq
\label{H_cd}
H_{cd}=
-\hbar \dot \varphi G_2.
\eeq
Implementing this interaction is quite challenging as discussed in detail in \cite{Opatrny2014}. 
In particular, a rapid switching between $G_1$ and $G_4$, to implement  $G_2$ through 
their commutator, is not a practical option \cite{Opatrny2014}. Our goal in the following is to 
design an alternative Hamiltonian to perform the shortcut without $G_2$.     
\section{Alternative driving protocols via Lie transforms}
\label{as}
The main goal here is to define a new shortcut different from the one described by $i\hbar \partial_t \psi(t)=H(t)\psi(t)$,
where $H(t)=H_0(t)+H_{cd}(t)$. 
A wave function  $\psi_I(t)$, which represents the alternative dynamics, is related to $\psi(t)$
by a unitary operator $B(t)$, 
\beq
\label{I_state}
\psi_I(t)=B^\dag(t)\psi(t), 
\eeq
and obeys 
$i\hbar \partial_t \psi_I(t)=H_I(t)\psi_I(t)$,
%
where   
\beqa
\label{I_hamiltonian}
H_I(t)&=&B^\dag(t)[H(t)-K(t)]B(t),
\\
\label{I_K}
K(t)&=&i\hbar\dot B(t)B^\dag(t).
\eeqa
These are formally the same expressions that define an interaction picture. However, in this application the 
``interaction picture'' portrays a different physical setting from the original one \cite{Ibanez2012}. 
In other words, $H_I$ is not a mathematical 
aid to facilitate a calculation in some transformed space, 
but rather a physically realizable Hamiltonian different from $H$. Similarly, $\psi_I$ represents 
in general different dynamics from $\psi$.  
The transformation 
provides indeed an alternative shortcut   
if $B(0)=B(t_f)=1$, so that $\psi_I(t_f)=\psi(t_f)$
for a given initial state $\psi_I(0)=\psi(0)$. Moreover, if $\dot B(0)=\dot B(t_f)=0$ also the Hamiltonians coincide
at initial and final times, 
$H(0)=H_I(0)$ and $H(t_f)=H_I(t_f)$. These boundary conditions  may be relaxed 
in some cases as we shall see.   

We carry out the transformation by exponentiating a
member $G$ of the dynamical Lie algebra of the Hamiltonian, 
\beq\label{ag}
B(t)=e^{-i\alpha G},
\eeq
where $\alpha=\alpha(t)$ is a time-dependent real function to be determined.  
This type of unitary operator $B(t)$ constitutes a ``Lie transform''. Lie transforms have been used, for example, to 
develop efficient perturbative approaches that try to set the perturbation term of a
Hamiltonian in a convenient form in both classical and quantum systems \cite{Bambusi1995,Cary1981}.
   
Note that $K$ in Eq. (\ref{I_K}) becomes $-\hbar \dot{\alpha}G$ and commutes with $G$.   
Then, $H_I$, given now by 
\beqa
B^\dag (H-K) B&=&e^{i\alpha G}(H-K)e^{-i\alpha G}
\nonumber
\\
&=&H-\hbar\dot{\alpha}G+i\alpha [G,H]-\frac{\alpha^2}{2!}[G,[G,H]]
-i\frac{\alpha^3}{3!}[G,[G,[G,H]]] + \cdots, \nonumber
\\
\label{tra}
\eeqa
depends only on $G$, $H$, and its repeated commutators with $G$, 
so it stays in the algebra. 
If we can choose $G$ and $\alpha$ so that the undesired generator components in $H$  cancel out 
and the boundary conditions 
for $B$ are satisfied, the method provides a feasible, alternative shortcut.  
In the existing applications of the method \cite{Torrontegui2013d,Ibanez2012}, and in this chapter
we proceed by trial an error, testing different generators. 
In the present application we want the Hamiltonian $H_I$ to keep the structure of the original one, with  
nonvanishing components proportional to $G_1$ and $G_4$.      
We may quickly discard by inspection
$G_1$, $G_2$, and $G_3$ as candidates for $G$.       
Choosing $G\to G_4$ in Eq. (\ref{ag}), 
and substituting into Eqs. (\ref{I_hamiltonian}) and (\ref{tra}),
the series of repeated commutators may be summed up.   
$H_I$ becomes  
\beqa
\label{H_I}
H_I&=&\left ( E_0 \cos \varphi-\hbar \dot \alpha \right ) G_4
\nonumber
\\
&-& 
\left ( E_0 \sin \varphi \cos \alpha + \hbar \dot \varphi \sin \alpha \right ) G_1
\nonumber
\\
&-& 
\left ( E_0 \sin \varphi \sin \alpha - \hbar \dot \varphi \cos \alpha \right ) G_2.
\eeqa
To cancel the $G_2$ term, we choose
\beq
\label{alpha}
\alpha(t)=\arccot \left [ \frac{E_0(t)}{\hbar \dot \varphi(t)} \sin[\varphi(t)] \right].
\eeq
Substituting Eq. (\ref{alpha}) into Eq. (\ref{H_I}) we have finally 
\beqa
\label{new_HI}
H_I&=&\left [ \frac{\cos \varphi E_0^3 \sin^2 \varphi +\hbar^2 \sin \varphi \dot E_0 \dot \varphi
 +\hbar^2 E_0 \left ( 2 \cos \varphi \dot \varphi ^2 - \sin \varphi \ddot \varphi \right )}{E_0^2 \sin^2 \varphi+ \hbar^2 \dot \varphi^2} \right ] G_4 \nonumber \\
&-& \left [ E_0 \sin \varphi \sqrt{1+\frac{\hbar^2\csc^2\varphi \dot \varphi^2}{E_0^2}} \right ] G_1, 
\eeqa
which has the same structure (generators) as the reference Hamiltonian but  different time-dependent coefficients. 
\section{Insulator-Superfluid transition}
\label{is}
\begin{figure}[t]
\begin{center}
   \includegraphics[width=0.5 \linewidth]{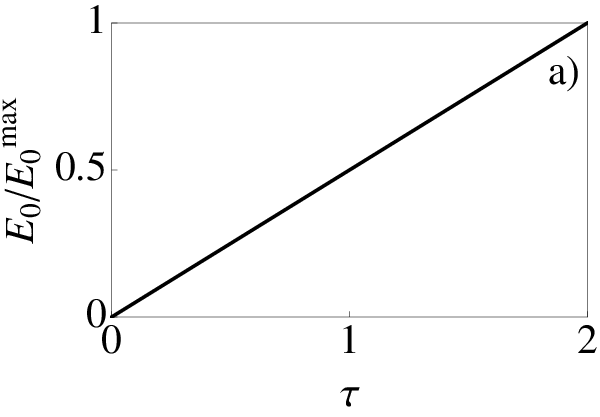}
   \includegraphics[width=0.5 \linewidth]{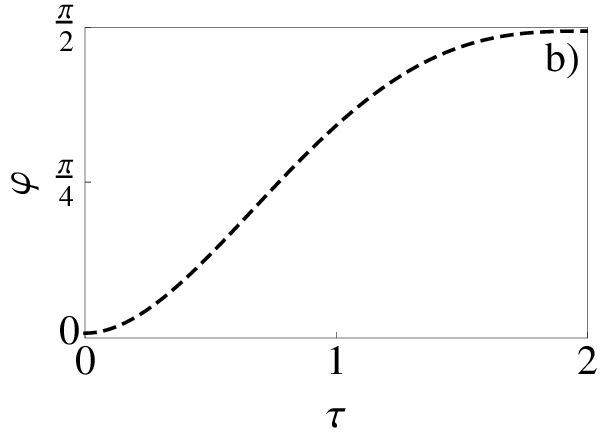}
\end{center}
\caption{\label{E0_phi_is}
Functions in $H_I(t)$: (a) $E_0(t)$ and (b) $\varphi(t)$. Parameters: $\tau=E_0^{max}t/\hbar$ where $E_0^{max}$ is the maximum value of $E_0(t)$ and $\tau_f=2$.
 }
\end{figure}

On changing the $U/J$ ratio, the system may go 
from a Mott-insulator (the two particles isolated in separate wells) to a superfluid state (in which each particle is distributed with equal probability in both wells).
From Eq. (\ref{eigenstates_1}),  the Mott-insulator ground state is  $|\phi_1\rangle=|1,1\rangle$ and in the superfluid regime the ground state becomes $|\phi_1\rangle=\frac{1}{2}|2,0\rangle+\frac{1}{\sqrt{2}}|1,1\rangle+\frac{1}{2}|0,2\rangle$.
To design a reference process (one that performs the transition when driven slowly enough)
we consider polynomial functions for $E_0(t)$ and $\varphi(t)$. 
Since we want to drive the system from $|1,1\rangle$ to $\frac{1}{2}|2,0\rangle+\frac{1}{\sqrt{2}}|1,1\rangle+\frac{1}{2}|0,2\rangle$, 
we impose in Eq. (\ref{eigenstates_1})
\beqa
\label{bc_is}
&&\varphi(0)=0, \nonumber \\
\label{bc_is}
&&\varphi(t_f) =\pi/2.
\eeqa
To have the wells isolated at $t=0$ but connected (allowing the particles to pass from one to the other) at $t=t_f$ we also set
\beqa
\label{bc_is2}
&&E_0(0)=0, \nonumber \\
\label{bc_is2}
&&E_0(t_f)\neq0,
\eeqa
so that $J(0)=U(0)=0$ and $J(t_f)\neq0$.
Moreover, for a  smooth connection with the asymptotic regimes ($t<0$, $t>t_f$) we set  
\beqa
\label{bc_is3}
&&\dot\varphi(0)=0, \nonumber \\
\label{bc_is3}
&&\dot\varphi(t_f) =0.
\eeqa
This implies that $H_{cd}(0)=H_{cd}(t_f)=0$; see Eq. (\ref{H_cd}). 
The condition  
\beq
\label{bc_is4}
\ddot \varphi(t_f)=0
\eeq
is also needed to implement alternative shortcuts, in particular, to satisfy $\dot B(t_f)=0$.
At intermediate times, we interpolate the functions as  
$E_0(t)=\sum_{j=0}^1a_jt^j$ and $\varphi(t)=\sum_{j=0}^4b_jt^j$, where the coefficients are found by solving Eqs. (\ref{bc_is}), (\ref{bc_is2}), (\ref{bc_is3}) and (\ref{bc_is4}). These functions are shown in Fig. \ref{E0_phi_is}. In this and other figures 
$\tau=E_0^{max}t/\hbar$, where $E_0^{max}$ is the maximum value of $E_0(t)$. 
%
%
%
%
%
%
%
%
%
%
%

The actual time evolution of the state
\beq
|\Psi(t)\rangle= c_1(t)|2,0\rangle + c_2(t)|1,1\rangle +c_3(t)|0,2\rangle
\eeq
is given by solving Schr\"odinger's equation with the different Hamiltonians.  
For this particular transition, $|\Psi(0)\rangle=|\phi_1(0)\rangle$ and the ideal target state is
(up to a global phase factor) $|\Psi(t_f)\rangle=|\phi_1(t_f)\rangle$. 
%
%
%
%
\begin{figure}[t]
\begin{center}
   \includegraphics[width=0.5 \linewidth]{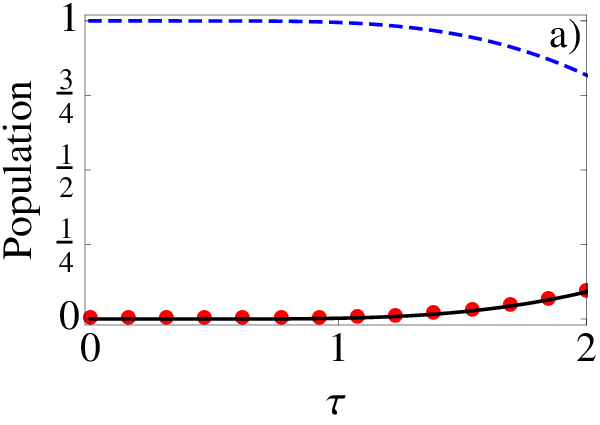}
   \includegraphics[width=0.5 \linewidth]{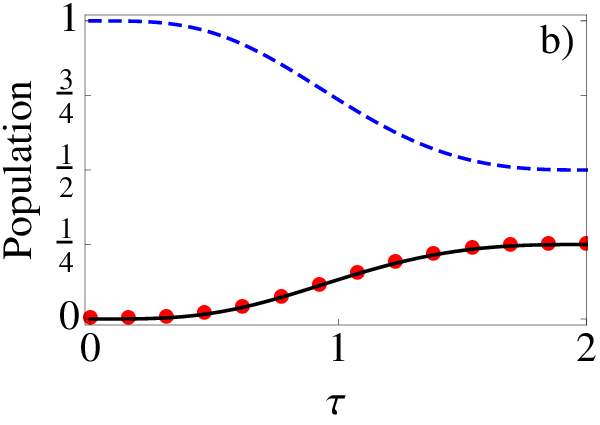}
\end{center}
\caption{\label{dynamics_is}
Bare-state populations for (a) $H_0(t)$; (b) $H(t)$ and $H_I(t)$. $|c_1(t)|^2$ (red circles), $|c_2(t)|^2$ 
(short-dashed blue line) and $|c_3(t)|^2$ (solid black line). Parameters: $\tau=E_0^{max}t/\hbar$ with $E_0^{max}$ the maximum value of $E_0(t)$, $\tau_f=2$. 
}
\end{figure}
%
%
%
%
%
%

The dynamics versus time $\tau$ is shown in Fig. \ref{dynamics_is} for $\tau_f=2$. 
For this short time $H_0(t)$ fails to drive the populations to $1/2$ and $1/4$, whereas when  
$H_{cd}(t)$ is added the intended transition occurs successfully.
As for the alternative Hamiltonian in Eq. (\ref{new_HI}),  
with $B=e^{-i\alpha G_4}$, and $\alpha$ in Eq. (\ref{alpha}),    
we find  
\beqa 
&&B(t_f)=1, \nonumber \\ 
&&\dot B(0)= \dot B(t_f)=0
\eeqa
[Eq. (\ref{bc_is4}) is necessary to have $\dot \alpha (t_f)=0$ and consequently $\dot B(t_f)=0$], 
whereas 
\beq
\label{B_is}
B(0)=\left ( \begin{array}{ccc}
e^{-i\pi/2} & 0 & 0 \\
0 & 1 & 0 \\
0 & 0 & e^{-i\pi/2} \end{array} \right )\ne 1.
\eeq
However $B^\dagger(0)|1,1\rangle=|1,1\rangle$ so $\psi^I(0)=\psi(0)$ and $H_I$ provides the desired shortcut. 

Solving numerically the dynamics for $H_I(t)$ we obtain a perfect insulator-superfluid transition [see Fig. \ref{dynamics_is}(b)]. Notice that, as $G_4$ is diagonal in the bare basis, the bare populations are the same for the dynamics driven by $H$ and $H_I$; see Fig. \ref{dynamics_is}(b).
%
%
%
%
%
%
%
%
%
\begin{figure}[t]
\begin{center}
   \includegraphics[width=0.5 \linewidth]{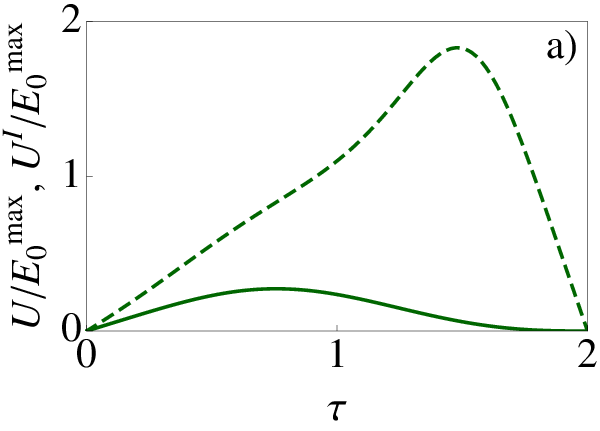}
   \includegraphics[width=0.5 \linewidth]{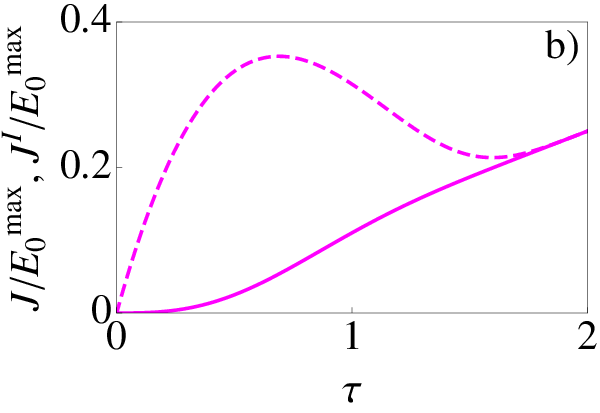}
\end{center}
\caption{\label{U_J_is}
(a) Interaction energy for the reference Hamiltonian $H_0$ (solid green line) and for $H_I$ (short-dashed green line). (b) Hopping energy for $H_0$ (solid magenta line) and $H_I$ (short-dashed magenta line). The same parameters as in Fig. \ref{E0_phi_is}.
 }
\end{figure}
%
%
%
%
%
%

In order to compare our approach 
with other protocols 
we reformulate $H_I$ as
\beq
\label{H_Ir}
H_I=\left ( \begin{array}{ccc}
U^I & -\sqrt{2}J^I & 0\\
-\sqrt{2}J^I & 0 & -\sqrt{2}J^I\\
0 & -\sqrt{2}J^I & U^I
\end{array} \right)=U^I G_4 - 4 J^I G_1.
\eeq
Comparing Eqs. (\ref{H_Ir}) and (\ref{new_HI}) we find that
\beqa
U^I&=&\frac{1} {{(E_0)}^2 \sin^2 \varphi+ \hbar^2 {(\dot \varphi)}^2} 
\left \{ \cos \varphi{(E_0)}^3 \sin^2 \varphi\right. 
\nonumber
\\
&+&\left.\hbar^2 \sin \varphi \dot E_0 \dot \varphi+\hbar^2 E_0 \left [ 2 \cos \varphi{(\dot \varphi)}^2 - \sin \varphi \ddot \varphi\right ] \right \} , 
\nonumber\\ 
J^I&=&\frac{1}{4} E_0\sin \varphi \sqrt{1+\frac{\hbar^2\csc^2\varphi {(\dot \varphi)}^2}{{(E_0)}^2}}.
\eeqa
Figure \ref{U_J_is} shows the functions $U_I$ and $J_I$. We have set $H_I(t_b)=H_0(t_b)$, for $t_b=0,t_f$, since  $H_{cd}(t_b)=0$ and 
$\dot{B}(t_b)=0$. 
In the same way as Eq. (\ref{parametrization_U}) 
we can rewrite the above energies as
\beqa
\label{new_parametrizationU}
U^I&=&E_0^I \cos\varphi^I, \nonumber \\
\label{new_parametrizationU}
J^I&=&\frac{E_0^I}{4} \sin \varphi^I,
\eeqa
where $E_0^I=E_0^I(t)$ and $\varphi^I=\varphi^I(t)$.  
The inverse transformation is 
\beqa
\label{new_parametrizationphi}
\varphi^I&=&\arctan\left ({4\frac{J^I}{U^I}}\right ), \nonumber \\
\label{new_parametrizationphi}
E_0^I&=&\frac{U^I}{\cos \varphi'}.
\eeqa
Consider a simple  protocol with $E_0(t)=E_0^M(t)=const$ and a linear $\varphi^M(t)$ from  $0$ and $\pi/2$ \cite{Opatrny2014}. 
Setting the value of $E_0^M$ so that   $\int E_0^M dt=\int E^I_0dt$, it is found that the simple protocol
needs $\tau_f=18.8$ to perform the transition with a 0.9999 fidelity.
In other words, the protocol based on $H_I$ is $9.4$ times faster according to this criterion. 
\section{Beam splitters}
\label{bs}
The three-level Hamiltonian (\ref{H_0_Lie}) describes other physical systems apart from two bosons in two wells. 
For example, it represents in the paraxial approximation, and substituting time by a longitudinal coordinate
three coupled waveguides
\cite{Longhi2009,Szameit2010,Longhi2011,Ornigotti2008,Rangelov2012,Chien2013}, 
where $J$ is controlled by waveguide separation and $U$ by the 
refractive index. In particular $J$ and $U$ may be manipulated to split an incoming wave in the central waveguide into two output channels
(corresponding to the external waveguides) or 
three output chanels   \cite{Rangelov2012,Chien2013}.
The Hamiltonian also  
represents  a single particle in a triple well \cite{Eckert2004},
where $U$ plays the role of the bias of the outer wells with respect to the central one and, $J$
the coupling coefficient between adjacent wells. The beam splitting may thus depict the evolution 
of the particle wave function from the central well either to the two outer wells 
or to  three of them with equal probabilities. 

For three-well or three-waveguide systems\footnote{The Hamiltonian (\ref{H_0_Lie}) also describes a three-level atom
under appropriate laser interactions; see \cite{Ornigotti2008}.} 
the minimal channel basis for left, center and right wave functions is $|L\rangle=\left ( \scriptsize{\begin{array} {rcccl} 1\\ 0\\0 \end{array}} \right)$, $|C\rangle=\left ( \scriptsize{\begin{array} {rcccl} 0\\ 1\\0 \end{array}} \right)$, and 
$|R\rangle=\left ( \scriptsize{\begin{array} {rcccl} 0\\ 0\\1 \end{array}} \right)$. 
\subsection{1:2 beam splitter}
%
%
%
%
%
%
\begin{figure}[t]
\begin{center}
   \includegraphics[width=0.7 \linewidth]{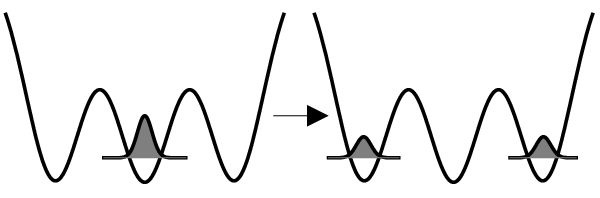}
   \end{center}
\caption{\label{lr}
Schematic representation of a $1:2$ beam splitter.}
\end{figure}
\begin{figure}[t]
\begin{center}
   \includegraphics[width=0.5 \linewidth]{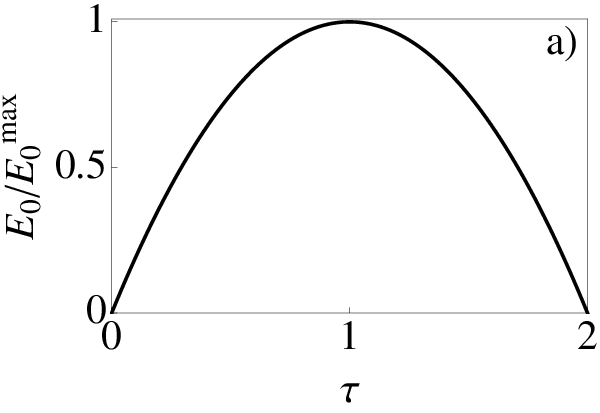}
   \includegraphics[width=0.5 \linewidth]{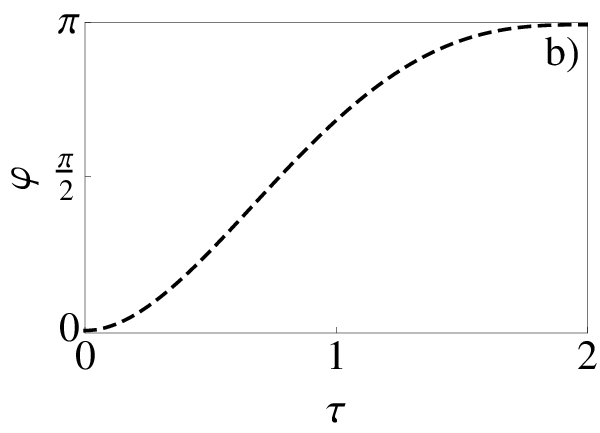}
\end{center}
\caption{\label{E0_phi_LR}
(a) $E_0(t)$ and (b) $\varphi(t)$. $\tau=E_0^{max}t/\hbar$ where $E_0^{max}$ is the maximum value of $E_0(t)$. $\tau_f=2$.
 }
\end{figure}

%
%
%
%
%
%
%
To implement a $1:2$ beam splitter (see Fig. \ref{lr}),  
the goal is to drive the eigenstate from $|\phi_1(0)\rangle=|C\rangle$ to $|\phi_1(t_f)=\frac{1}{\sqrt{2}}\left ( |L\rangle+|R\rangle \right )$. 
As in the previous section we use polynomial functions for $E_0(t)$ and $\varphi(t)$ to set a reference process.
We impose
\beqa
\label{bc_lr}
&&\varphi(0)=0, \nonumber \\
\label{bc_lr}
&&\varphi(t_f) =\pi
\eeqa
in Eq. (\ref{eigenstates_1}). 
The wells (waveguides) should be isolated at initial and final times. 
If morever all wells are at equal heights at those times we set
\beqa
\label{bc_lr2}
&&E_0(0)=E_0(t_f) =0, \nonumber \\
\label{bc_lr2}
&&E(t_f/2)\neq0
\eeqa
to satisfy $H_0(0)=H_0(t_f)=0$.
We  also impose
\beqa
\label{bc_lr3}
\dot\varphi(0)=0, \nonumber \\
\label{bc_lr3}
\dot\varphi(t_f) =\pi
\eeqa
to smooth the functions at the time boundaries and make $H_{cd}(t_b)=0$. 
In addition  
\beq
\label{bc_lr4}
\ddot \varphi(t_f)=0
\eeq
is imposed to satisfy $\dot B(t_f)=0$. 
At intermediate times  $E_0(t)=\sum_{j=0}^2a_jt^j$ and $\varphi(t)=\sum_{j=0}^4b_jt^j$, with the coefficients 
deduced from  Eqs. (\ref{bc_lr}), (\ref{bc_lr2}), (\ref{bc_lr3}) and (\ref{bc_lr4}).
These functions are shown in Fig. \ref{E0_phi_LR}.
%
%
%
%
%
%
%
%
%
%
%
%
%
%
%
%
%
\begin{figure}[t]
\begin{center}
   \includegraphics[width=0.5 \linewidth]{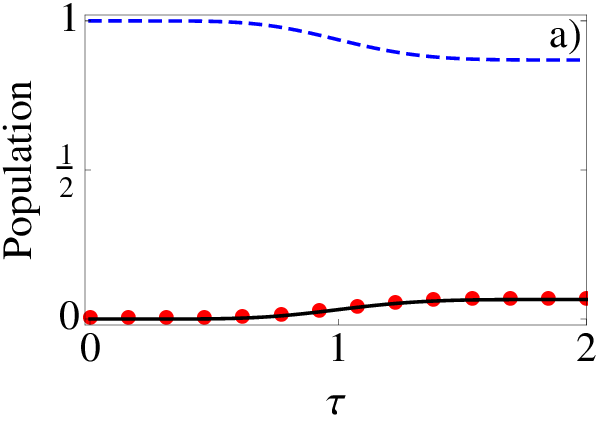}
   \includegraphics[width=0.5 \linewidth]{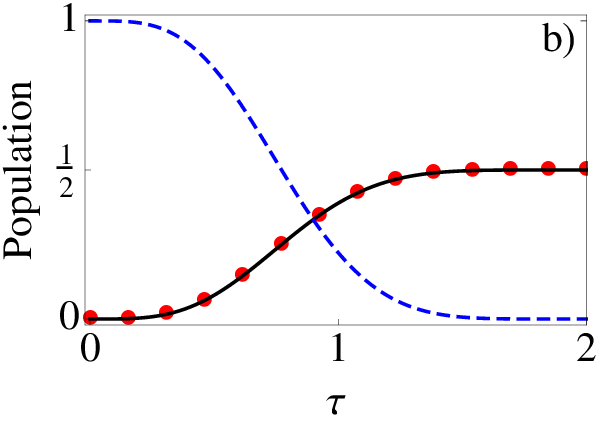}
\end{center}
\caption{\label{dynamics_LR}
Bare-state populations for (a) $H_0(t)$, and (b) $H(t)$ and $H_I(t)$. $|c_1(t)|^2$ (red circles), $|c_2(t)|^2$ 
(short-dashed blue line) and $|c_3(t)|^2$ (solid black line). Parameters: $\tau=E_0^{max}t/\hbar$ with $E_0^{max}$ the maximum value of $E_0(t)$, and $\tau_f=2$.}
\end{figure}
%
%
%
%
%
%

Figure \ref{dynamics_LR} shows the dynamics  for $\tau_f=2$. This time (corresponding to the splitter length in the optical system) is too short for the reference Hamiltonian $H_0(t)$ to drive the bare-basis populations to $0$ and $1/2$. On adding $H_{cd}(t)$ the transition occurs as desired. As in Sec. \ref{as}, we  construct an alternative shortcut $H_I(t)$ without $G_2$ 
using the transformation $B=e^{-i\alpha G_4}$. With $\alpha$ in Eq. (\ref{alpha}),  
$\dot B(0)=\dot B(t_f)=0$, whereas
\beq
\label{B_lr}
B(0)=B(t_f)=\left ( \begin{array}{ccc}
e^{-i\pi/2} & 0 & 0 \\
0 & 1 & 0 \\
0 & 0 & e^{-i\pi/2} \end{array} \right ).
\eeq
This is enough for our objective as $B^\dagger(0)|C\rangle=|C\rangle$, and $B^\dag(t_f) |\psi(t_f)\rangle=-i|\psi(t_f)\rangle$. 
%
%
%
%
%
\begin{figure}[t]
\begin{center}
   \includegraphics[width=0.5 \linewidth]{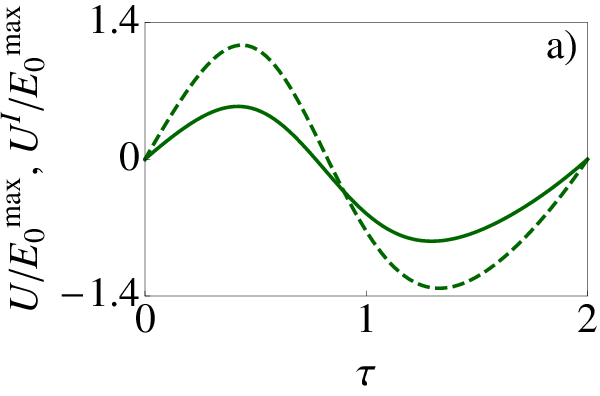}
   \includegraphics[width=0.5 \linewidth]{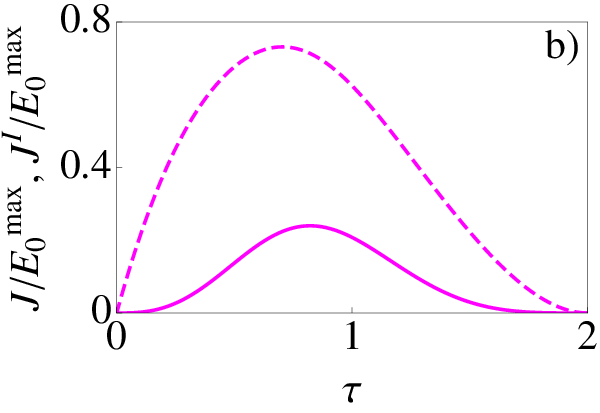}
\end{center}
\caption{\label{U_J_LR}
(a) Interaction energy for $H_0$ (solid green line) and $H_I$ (short-dashed green line). (b) Hopping energy for 
$H_0$ (solid magenta line) and $H_I$ (short-dashed magenta line). The same parameters as in Fig. \ref{E0_phi_LR}.
 }
\end{figure}
%
%
%
%
%
%

Solving numerically the dynamics for $H_I(t)$ we obtain a perfect $1:2$ beam splitting [see Figs. \ref{U_J_LR}  and \ref{dynamics_LR}(c)].

To compare the new shortcut and the simple approach with $E_0^M=const$ and $\varphi^M(t)=\frac{t}{t_f}\pi$, we 
set  
$\int E_0^M dt=\int E_0^I dt$. The constant-$E_0$ protocol needs $\tau_f \geqslant 18.6$ to achieve $0.9999$ fidelity, 
so the protocol driven by $H_I$ is $9.3$ times faster. 
\subsection{1:3 beam splitter}
%
%
%
%
%
%
\begin{figure}[t]
  \begin{center}
   \includegraphics[width=0.7 \linewidth]{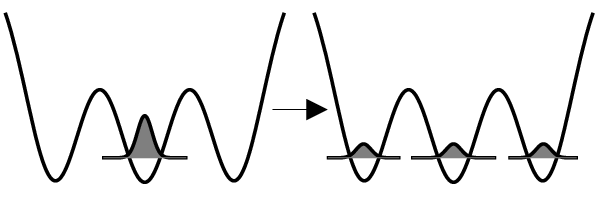}
   \end{center}
\caption{\label{lcr}
Schematic representation of the $1:3$ beam splitter.}
\end{figure}
\begin{figure}[t]
\begin{center}
   \includegraphics[width=0.5 \linewidth]{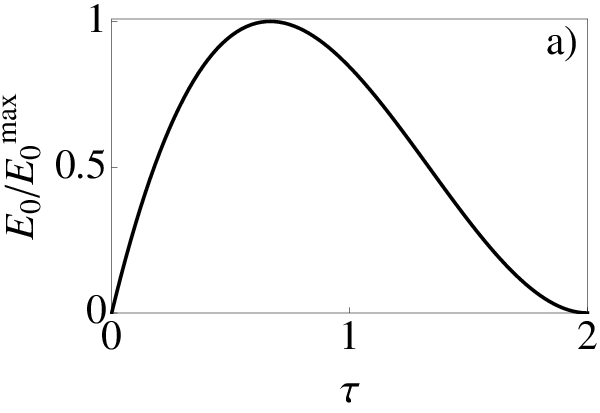}
   \includegraphics[width=0.5 \linewidth]{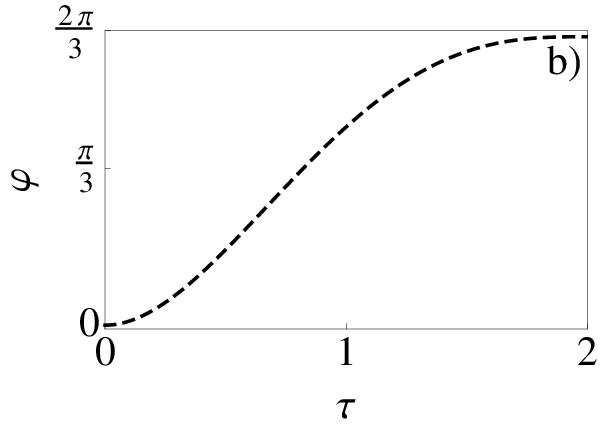}
\end{center}
\caption{\label{E0_phi_LCR}
(a) $E_0(t)$ and (b) $\varphi(t)$. $\tau=E_0^{max}t/\hbar$, where $E_0^{max}$ is the maximum value of $E_0(t)$. $\tau_f=2$.
 }
\end{figure}
\begin{figure}[t]
\begin{center}
   \includegraphics[width=0.5 \linewidth]{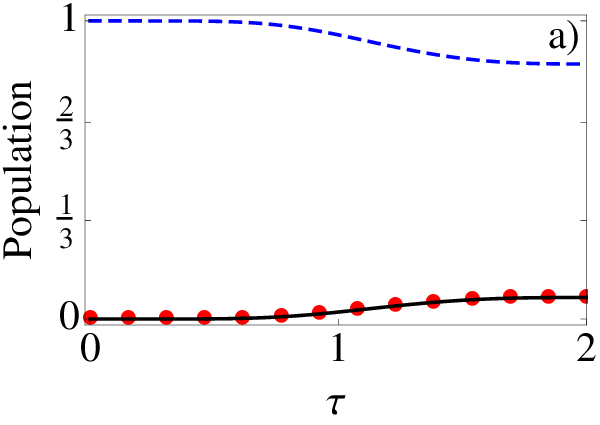}
   \includegraphics[width=0.5 \linewidth]{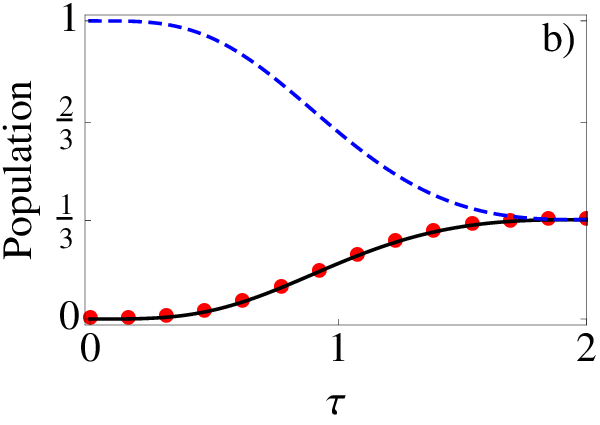}
\end{center}
\caption{\label{dynamics_LCR}
Bare-state populations for (a) $H_0(t)$, and (b) $H(t)$ and $H_I(t)$. $|c_1(t)|^2$ (red circles), $|c_2(t)|^2$ 
(short-dashed blue line) and $|c_3(t)|^2$ (solid black line). Parameters: $\tau=E_0^{max}t/\hbar$ with $E_0^{max}$ the maximum value of $E_0(t)$, and $\tau_f=2$.}
\end{figure}
\begin{figure}[t]
\begin{center}
   \includegraphics[width=0.5 \linewidth]{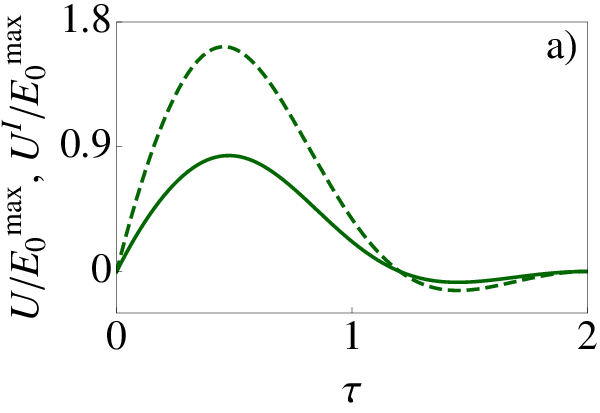}
   \includegraphics[width=0.5 \linewidth]{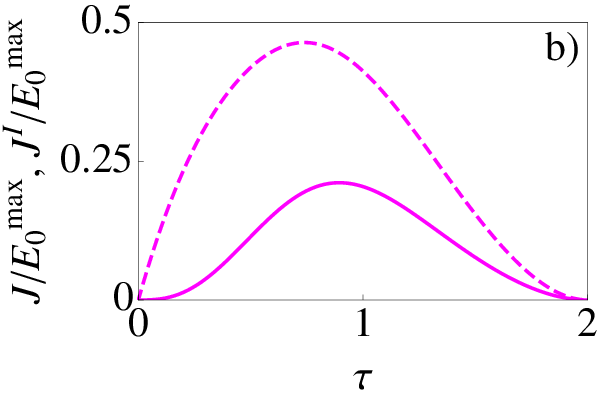}
\end{center}
\caption{\label{U_J_LCR}
(a) Interaction energy for $H_0$ (solid green line) and $H_I$ (short-dashed green line). (b) Hopping energy for
$H_0$ (solid magenta line) and $H_I$ (short-dashed magenta line). The same parameters as in Fig. \ref{E0_phi_LCR}
 }
\end{figure}

%
%
%
%
%
%
%
%
We also describe briefly a $1:3$ beam splitter; see Figs. \ref{lcr}-\ref{U_J_LCR}.   
The aim is to drive the system from $|\phi_1(0)\rangle=|C\rangle$ to equal populations in 
$|L\rangle$, $|C\rangle$, and $|R\rangle$. To design a reference protocol we use polynomial interpolation for $E_0(t)$ and $\varphi(t)$ (see Fig. \ref{E0_phi_LCR}), 
with the same boundary conditions as for the $1:2$ splitter but with $\varphi(t_f) =0.60817\pi=\arccos(-1/3)$ and the additional condition $\dot E_0(t_f)=0$ [to satisfy $U^I(t_f)=U(t_f)$ so that $H_I(t_f)=H_0(t_f)$].  The Lie  transform may be applied as before on the protocol with the counterdiabatic correction; see  Fig. \ref{dynamics_LCR}(b). 
%
%
%
%
%
%
%
%
%
%
%
%

A simple protocol
with $E_0^M$ and $\varphi(t)=\frac{t}{t_f} 0.60817\pi$
needs  $\tau_f=22$, if  $\int E_0^M dt=\int E_0^I dt$, for a 0.9999 fidelity, 
so the protocol based on $H_I$ is 11 times faster. 
%
%
%
%
%
%
%
%
%
%
%
%
%
%
%
%
%
%
%
%
%
%

\section{Discussion}
\label{discussion_lie}
We started with shortcuts to adiabaticity for three-level systems with U3S3 symmetry 
(a four-dimensional Lie algebra) 
that include Hamiltonian terms which are difficult to implement in the laboratory.  
Alternative shortcuts without them have then been found by means of Lie transforms.  
These transformations are formally equivalent to IP transformations.
However the resulting IP Hamiltonian and state represent a different
physical process from the original  (Schr\"odinger) Hamiltonian and dynamics.      
We have set shortcuts for different physical systems. For two particles in two wells we have implemented 
a fast insulator-superfluid transition. For coupled waveguides, or a particle in a triple well we have implemented 
fast beam splitting with one input channel and two or three output channels. In all cases the IP Hamiltonian  
involves only two realizable terms (generators).  

A related method  has been worked out in \cite{Torrontegui2014}. 
Both approaches rely on Lie algebraic methods 
and aim at constructing shortcuts to adiabaticity. However, we do not use dynamical invariants
explicitly in the current approach,  whereas the bottom-up approach in 
\cite{Torrontegui2014} engineers the Hamiltonian by making explicit use of     
its relation to dynamical invariants. In contrast, we start here from an existing, known shortcut
--for example the one generated by a counterdiabatic method;  
then, a Lie transform is applied to generate alternative, 
feasible, or more convenient shortcuts, as in \cite{Ibanez2012}.    
A connection between the transformation method and dynamical invariants is sketched briefly in the
Appendix \ref{algebra} but it deserves a separate study.  We note that the dynamics of all our examples 
takes place in a degenerate eigenspace of an algebraic invariant  
which  is not proportional to the unit matrix and commutes with all members of the algebra. 
The degeneracy is required to produce nontrivial dynamics, so identifying degenerate subspaces 
of nontrivial invariants, as well as the conditions allowing the cancellation of certain generators will be instrumental in 
finding further applications in systems described by other Lie algebras.    

Optimal control theory (OCT) offers an alternative way to generate fast dynamics \cite{Doria2011,Lapert2012}. 
In this chapter no optimization has been attempted, but  the combination of shortcut-to-adiabaticity
techniques offering multiple exact protocols with perfect fidelity, such as the one based on Lie transforms,  
and OCT, has been shown to be fruitful \cite{Stefanatos2010,Chen2011a,Torrontegui2012b}.
OCT may select among the protocols generated the ones 
that optimize a physically significant variable  \cite{Stefanatos2010,Chen2011a,Torrontegui2012b}.   

Within the scope of the  algebra U3S3, other physical systems that could be treated are 
in quantum optics (three-level atoms) 
 \cite{Chen2012, Bergmann1998}, nanostructures (triple wells or dots) \cite{Kiselev2013}, optics 
(mode converters) \cite{Lin2012,Tseng2012}, or Bose-Einstein condensates in an accelerated optical lattice
\cite{Dou2014}.


\chapter{Fast quasi-adiabatic dynamics}
\label{Chapter3}
\lhead{Chapter 3. \emph{Fast quasi-adiabatic dynamics}} 
We work out the theory and applications 
of a fast quasi-adiabatic approach to speed up slow adiabatic
manipulations of quantum systems  by driving a control parameter
as near to the adiabatic limit as possible over the entire protocol duration. 
We find characteristic time scales, such as the minimal time to achieve fidelity $1$, 
and the optimality of the approach within the iterative superadiabatic sequence.
Specifically, we show that the population inversion in a two-level system, the splitting and cotunneling 
of two-interacting bosons, and the stirring of a Tonks-Girardeau gas on a ring to achieve mesoscopic superpositions
of many-body rotating and nonrotating states can be significantly speeded up.    
\newpage
\section{Introduction}

Developing  technologies based on delicate quantum coherences of  
atomic systems is a major scientific and technical challenge due to pervasive noise-induced and manipulation errors.  
Shortening the process  below characteristic decoherence times provides
a way out to avoid the effects of noise, but the protocol (time dependence of control parameters) 
should still be robust with respect to offsets of the external driving parameters. 
Shortcuts to adiabaticity (STA) are a set of techniques to reduce the duration of slow adiabatic processes, 
minimizing noise effects while keeping or enhancing robustness \cite{Chen2010,Ruschhaupt2012a,Torrontegui2013d}.  
There are  different approaches but, as we have already discussed in the previous chapter,  
they are not always easy to implement in practice, because of 
the need to control many variables, or the difficulty to realize certain
terms added to the original  Hamiltonian to speed up the adiabatic dynamics. 
Here we  work out the theory and present several applications of a simple, but effective, fast quasi-adiabatic (FAQUAD) approach that  
engineers the time dependence of a single control parameter $\lambda(t)$, 
without changing the structure of the original Hamiltonian,  $H[\lambda(t)]$,  
to perform a process as quickly as possible while making 
it as adiabatic as possible at all times. 
The two goals are contradictory so a compromise is needed. 

\section{The method}\label{FAQUAD method}
We impose that the standard adiabaticity parameter \cite{schiffquantum} is constant throughout
the process, and consistent with 
the boundary conditions  (BC) of $\lambda(t)$ at $t=0$ and $t=t_f$.   

In the simplest scenario we assume that the adiabatic process driven by changing $\lambda(t)$ involves a passage through at least one avoided crossing.
While real systems are in general multilevel, only the two quasicrossing levels (say $E_1$, $E_2$) in the
instantaneous basis $\{|\phi_j\ra\}$ need to be considered under the adiabaticity condition \cite{schiffquantum}, 
\beq
\hbar \left |\frac{ \langle \phi_1(t)|\partial_t \phi_2(t)\rangle}{ E_1(t)-E_2(t)} \right |\ll1.
\eeq
(More levels can be taken into account if necessary.)  
We then impose 
\beq
\label{f_adiabatic}
\hbar \left |\frac{ \langle \phi_1(t)|\partial_t \phi_2(t)\rangle}{ E_1(t)-E_2(t)} \right |=
\hbar \left |\frac{ \langle \phi_1(t)|\frac{\partial H}{\partial t}| \phi_2(t)\rangle}{ [E_1(t)-E_2(t)]^2} \right | =c,
\eeq 
and as $\lambda=\lambda(t)$ and $t=t(\lambda)$ we apply the chain rule to write    
\beq
\label{d_e}
\dot \lambda\!=\!\mp\frac{c}{\hbar}Ê\!\left | \frac{ E_1(\lambda)-E_2(\lambda)}{ \langle \phi_1(\lambda)|\partial_\lambda \phi_2(\lambda)\rangle} \right |\!
\!=\!\mp\frac{c}{\hbar}\!\left | \frac{ [E_1(\lambda)-E_2(\lambda)]^2}{ \langle \phi_1(\lambda)|\frac{\partial H}{\partial \lambda}|\phi_2(\lambda)\rangle} \right |\!,
\eeq
where the overdot is a time derivative and $\mp$ applies to a monotonous decrease or increase of $\lambda(t)$.  
Equation (\ref{d_e}) must be solved with the BC 
$\lambda(0)$ and  $\lambda(t_f)$, which fixes $c$ and the integration constant. 
The corresponding FAQUAD solution, $\lambda_{F}(t)$,
changes quickly when the transitions among instantaneous  eigenstates are unlikely and slowly otherwise. 
An equation equivalent to Eq. (\ref{d_e}) has been applied to specific
models \cite{Kastberg1995,Daems2007,Daems2008,Torrontegui2012a,Bowler2012,Martinez-Garaot2013}, for example,
the two-level system \cite{Daems2008} and three-level lambda systems
\cite{Daems2007}.

In this chapter, we derive 
important properties of FAQUAD including characteristic time
scales, such as the minimal time to achieve fidelity $1$, and its optimality within the iterative superadiabatic sequence. 
We  also apply FAQUAD to several physical systems for which other shortcut techniques are difficult or impossible
to implement, 
including a process for creating a collective superposition state between rotating
and nonrotating atoms on a ring.

The FAQUAD strategy belongs to a family of processes that use the
time dependence of a control parameter to delocalize  in time the transition
probability among adiabatic levels. 
In the parallel adiabatic transfer technique \cite{Guerin2002,Guerin2011} the level
gap is required to be constant, which prevents it from being applicable when 
the initial and final gaps are different [see the Tonks-Girardeau (TG) gas
example below]. The uniform adiabatic (UA) method developed in \cite{Quan2010}
relies on a comparison of {\it transition} and {\it relaxation} time scales
and predicts (in a notation consistent with the one used in the work)
\beq
\dot \lambda
=\mp\frac{c_{UA}}{\hbar}\left | \frac{ [E_1(\lambda)-E_2(\lambda)]^2}{\partial [E_1(\lambda)-E_2(\lambda)]/\partial \lambda} \right |.
\eeq
Furthermore, the local adiabaticity (LA) approach \cite{Roland2002,Richerme2013} predicts an equation similar to Eq.~(\ref{d_e}), however without the factor $\langle \phi_1(\lambda)|\frac{\partial H}{\partial \lambda}| \phi_2(\lambda)\rangle$. This leads to a different constant, $c_{LA}$, and 
time dependence of the parameter, $\lambda_{LA}(t)$, and therefore different minimal times as illustrated below.      
Note that in \cite{Roland2002} Eq. (\ref{d_e}) is also written down but not applied
as such.
\subsection{General Properties}\label{g_p}
We rewrite Eq. (\ref{d_e}) in terms of  $s=t/t_f$ and  define $\tilde{\lambda}(s)\coloneqq\lambda(s t_f)$ so that  
\beq
\dot \lambda(t)=\tilde{\lambda}' \frac{1}{t_f}, 
\eeq
where the prime is the derivative with respect to $s$. We get 
\beq
\label{deltap}
\tilde{\lambda}'=\mp\frac{\tilde{c}}{\hbar} \left|  \frac{{E}_1-{E}_2}{ \langle {\phi}_1|\partial_{\tilde{\lambda}}\phi_2\rangle} \right |_{\tilde{\lambda}}, 
\eeq
with 
\beq
\tilde{c}=c t_f=\mp\hbar\int_{\tiny{\tilde{\lambda}(0)}}^{\tiny{\tilde{\lambda}(1)}} \frac{d{\tilde{\lambda}}}{\big|  \frac{{E}_1-{E}_2}{ \langle {\phi}_1|\partial_{\tilde{\lambda}}\phi_2\rangle} \big|_{\tilde{\lambda}}}.
\eeq
It is thus enough to solve the FAQUAD protocol once, i.e., using Eq. (\ref{deltap}) we get $\tilde{\lambda}_F(s)$ and $\tilde{c}$ to satisfy  
$\tilde{\lambda}(s=0)$ and $\tilde{\lambda}(s=1)$, and then 
adapt (scale) the result for each $t_f$, as $\lambda_F(t=s t_f)=\tilde{\lambda}_F(s)$, and $c=\tilde{c}/t_f$. 
Similarly, the gap 
\beq
\omega_{12}(t)=\frac{E_1(t)-E_2(t)}{\hbar}
\eeq
is given in terms of a universal gap function $\tilde{\omega}_{12}[\tilde{\lambda}_F(s)]$ as 
$\omega_{12}(t)=\tilde{\omega}_{12}[\tilde{\lambda}_F(t/t_f)]$.   
Depending on $\tilde{c}$, a large time $t_f$ 
might be necessary to make the process fully adiabatic (i.e., with a small enough $c$) but, surprisingly, much shorter times for which the 
process is not fully adiabatic also lead to the desired results.      

Since the system is nearly adiabatic, this is explained by adiabatic perturbation theory.    
In the adiabatic basis the wave function is expanded as \cite{schiffquantum,Ibanez2014}
\beq
|\Psi(t)\rangle=\sum_n g_n(t) e^{i\beta_n (t)}|\phi_n(t)\rangle,
\eeq
where 
\beq
\beta_n(t)=-\frac{1}{\hbar}\int_0^t E_n(t')dt'+i\int_0^t\langle\phi_n(t')|\dot \phi_n(t')\rangle dt'.
\eeq
From 
\beq
i\hbar|\dot \Psi(t)\rangle=H(t)|\Psi(t)\rangle
\eeq
we get, choosing
$\langle\phi_n(t)|\dot \phi_k(t)\rangle$ to be real (in particular $\langle\phi_n(t)|\dot \phi_n(t)\rangle=0$), 
\beq
\label{ge}
\dot g_n(t)=-\sum_{k\neq n} e^{iW_{nk}(t)}\langle\phi_n(t)|\dot \phi_k(t)\rangle g_k(t),
\eeq
where 
\beq
W_{nk}(t)=\int_0^t \omega_{nk}(t')dt'
\eeq
is a dynamical-gap phase and 
\beq
\omega_{nk}(t)\coloneqq\frac{E_n(t)-E_k(t)}{\hbar}.
\eeq
Integrating, 
\beq
g_n(t)\!-\!g_n(0)\!=\! \!-\!\sum_{k\neq n}\!\int_0^t\!\!e^{iW_{nk}(t')}\!\langle\phi_n(t')|\dot \phi_k(t')\rangle g_k(t') dt',
\eeq
which is still exact. 
Assuming that the initial state is $|\phi_m(0)\rangle$ and approximating $g_k(t')=\delta_{km}$
one finds to first order, for $n\neq m$,
\beq
\label{perturbation_sol}
g_n^{(1)}(t)=-\int_0^t\langle\phi_n(t')|\dot \phi_m(t')\rangle e^{iW_{nm}(t')}dt',
\eeq
which should satisfy $|g_n(t)|\ll1$ for an adiabatic evolution.
In FAQUAD, setting $n=2$, $m=1$ and neglecting transitions to further states,   
$
\langle\phi_2(t)|\dot \phi_1(t)\rangle=c r\omega_{21}(t),
$
with 
$
r={\rm{sgn}}[\langle\phi_2(t)|\dot \phi_1(t)\rangle \omega_{21}], 
$
so we find (higher-order corrections are also explicit) 
\beqa
g_2^{(1)}(t)\!=\!-{r}\!\int_0^t\!\!\! c \omega_{21}(t') e^{iW_{21}(t')}dt' 
\!=\!ic{r}
(e^{iW_{21}(t)}\!-\!1).
\label{fa_perturbation_sol}
\eeqa
Note the scaling  $W_{21}(t_f)=t_f \Phi_{21}$ where $\Phi_{21}=\int_0^1 \tilde{\omega}_{21}(s) ds$, 
and $\tilde{\omega}_{21}(s)=\omega_{21}(s t_f)$.    
The oscillation period for the final population with FAQUAD
is 
$
T=\frac{2\pi}{\Phi_{12}}, 
$
which is also a good estimate of the minimal (final) time to pass through the avoided crossing with fidelity $1$ [since $g_2^{(1)}(T)=0$]. 
The upper envelope for the  probability of level 2 is $4\tilde{c}^2/t_f^2$.  
The period, envelope, and Eq. (\ref{fa_perturbation_sol}) are important general results
of this work.  
The oscillation is due to a   
quantum interference:   
$g_2^{(1)}(t_f)$ results from the sum of paths where 
the jump at time $t'$ from $1$ to $2$ has an amplitude  $c\omega_{21}(t')$.  
$e^{iW_{21}(t')}$ represents the dynamical phases before and after the jump, as 
\beq
e^{iW_{21}(t')}= e^{\frac{-i}{\hbar}\!\int_0^{t'}\!dt''E_2(t'')} e^{\frac{-i}{\hbar}\!\int_{t'}^{t_f}\!dt''E_2(t'')} e^{\frac{i}{\hbar}\!\int_{0}^{t_f}\!dt''E_2(t'')},
\eeq
where the last exponential is a phase factor independent of $t'$.  

To illustrate these general properties, we will first examine the two-level model, a
paradigmatic test bed. Then, to show the power of FAQUAD,
we will apply it to more complicated atomic systems.
\section{Population inversion}
\begin{figure}[t]
\begin{center}
\includegraphics[width=0.46 \linewidth]{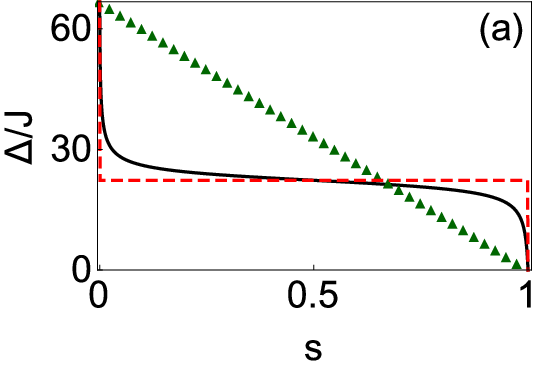}
\includegraphics[width=0.46 \linewidth]{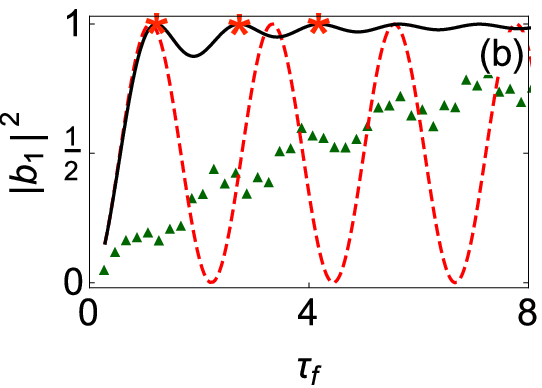}
\includegraphics[width=0.46 \linewidth]{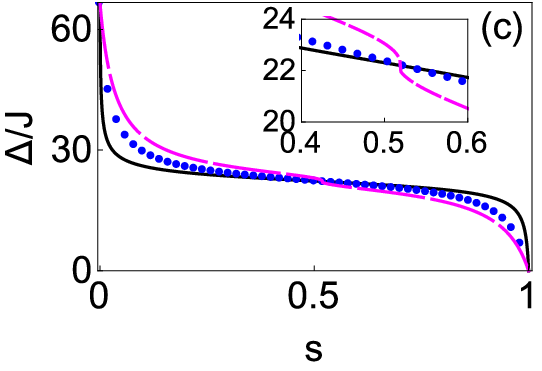}
\includegraphics[width=0.46 \linewidth]{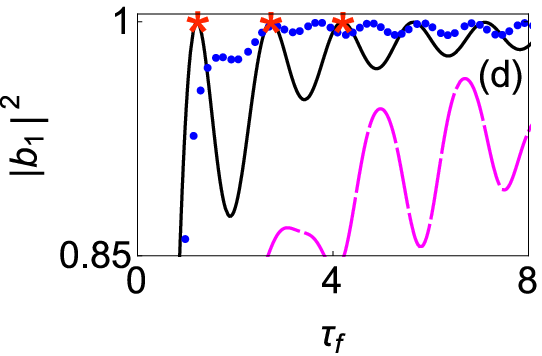}
\end{center}
\caption{\label{finalt_tmm}
(a) Bias vs $s$ for linear-in-time bias (green triangles), $\pi$ pulse (short-dashed red line), and 
FAQUAD (solid black line).  
(b) Final ground-state population $|b_1(t_f)|^2$ vs   
$\tau_f=Jt_f/\hbar$ for linear-in-time bias (green triangles), $\pi$ pulse
(short-dashed red line), and FAQUAD (solid black line).
(c) Bias vs $s$ for FAQUAD (solid black line), LA approach (blue dots), and UA approach  (long-dashed magenta line).  
The inset amplifies the kink of the UA approach.
(d) $|b_1(t_f)|^2$ vs $\tau_f=Jt_f/\hbar$ for FAQUAD (solid black line), 
LA approach (blue dots), and UA approach  (long-dashed magenta line). The
stars in (b) and (d) correspond to integer multiples of the characteristic
FAQUAD time scale $2\pi/\Phi_{12}$.
$\Delta(0)/J=66.7$, $U/J=22.3$.}
\end{figure}

Consider first a  two-mode model
with a single avoided crossing.
In the bare basis, $|1\rangle=\left (\scriptsize{\begin{array} {rccl} 1\\0 \end{array}}\right)$ and $|2\rangle=\left ( \scriptsize{\begin{array} {rccl} 0\\1 \end{array}} \right)$,  
the time-dependent state is 
$
|\Psi(t)\rangle=b_1(t)|1\rangle+b_2(t)|2\rangle
$
and 
\beq
\label{H_0_tmm}
H=\left(\scriptsize{\begin{array}{cc}
0 & -\sqrt{2}J
\\
-\sqrt{2}J & U-\Delta
\end{array}} \right),
\eeq
where the bias $\Delta=\Delta(t)$ is the control parameter, and $U>0$ and $J>0$ are constant.  
The instantaneous eigenvalues are
\beqa
\label{eigenvalues_1_tmm}
&&E_1=\frac{1}{2}(U-\Delta-P),
\\
\label{eigenvalues_2_tmm}
&&E_2=\frac{1}{2}(U-\Delta+P),
\eeqa
where $P=P(t)=\sqrt{8J^2+U^2-2U\Delta(t)+\Delta^2(t)}$,
and the normalized eigenstates are
\beqa
\label{eigenstate_1_tmm}
|\phi_1\rangle&=&\frac{1}{\sqrt{1+\frac{(U-\Delta+P)^2}{8J^2}}} \left (\!\! \begin{array}{c}
\frac{1}{2\sqrt{2}J}(U-\Delta+P)
\\
1
\\
\end{array}\!\! \right)\!, 
\\
\label{eigenstate_2_tmm}
|\phi_2\rangle&=&\frac{1}{\sqrt{1+\frac{(U-\Delta-P)^2}{8J^2}}}  \left (\!\! \begin{array}{c}
\frac{1}{2\sqrt{2}J}(U-\Delta-P)
\\
1
\\
\end{array}\!\! \right)\!. \nonumber 
\\
\eeqa
The goal is to drive the eigenstate from $|\phi_1(0)\rangle=|2\rangle$ to $|\phi_1(t_f)\rangle=|1\rangle$. To design the reference adiabatic protocol 
we impose on $\Delta(t)$ the BC
$\Delta(0)\gg U,J$ and $\Delta(t_f) =0$.
The FAQUAD protocol is shown in Fig. \ref{finalt_tmm}(a) compared to a linear-in-time $\Delta(t)$ and a 
constant $\Delta=U$.  
%
%
%
%
%
%
%
%
%
%
%
%
%
%
The final ground-state populations $|b_1(t_f)|^2$ versus dimensionless final time $\tau_f=Jt_f/\hbar$ are shown in Fig. \ref{finalt_tmm}(b).
Since the dressed states are essentially pure bare states at initial and final times, their populations in bare and dressed state bases coincide at these times.  
For $\Delta=U$ between 0 and $t_f$, 
``Rabi oscillations'' (we use a terminology appropriate for quantum optics but of course the two-level model is more broadly applicable) 
occur [see Fig. \ref{finalt_tmm}(b)]. The conditions for a $\pi$ pulse 
or multiple $\pi$ pulses are met periodically over $t_f$, alternated with times 
where the probability drops to zero because of 
destructive interference among two dressed states superposed with equal weights. 
By contrast the 
FAQUAD process is dominated by one dressed state and the influence of the transitions to the other one is minimized, because they are small in amplitude, and because at certain times they completely cancel each other out by 
destructive interference. The time interval between population maxima for
FAQUAD is $2\pi/\Phi_{1,2}$ [also shown in Figs. \ref{finalt_tmm}(b) and \ref{finalt_tmm}(d) by  stars],
i.e., it is not governed by the Rabi frequency. The first maximum is at a small $t_f$ similar to the one for the $\pi$ pulse, but broader. 
The FAQUAD maxima are more stable with respect to errors in $\Delta$ as $t_f$ increases, whereas the flat-pulse maxima decrease their stability.  
Figure \ref{finalt_tmm}(b) also shows the poorer results of the linear ramp for $\Delta(t)$. 

FAQUAD is compared to the LA and UA approaches in Figs. \ref{finalt_tmm}(c) and
\ref{finalt_tmm}(d). It provides  shortcuts at smaller process times (it achieves $0.9998$ 
probability three times faster than LA) 
and an analytically predictable behavior
via the perturbation theory analysis.  
Let us now consider more complicated atomic systems where FAQUAD can be applied whereas other 
STA techniques cannot.  
\section{Interacting bosons in a double well}
\begin{figure}[t]
\begin{center}
  (a)\includegraphics[width=0.6 \linewidth]{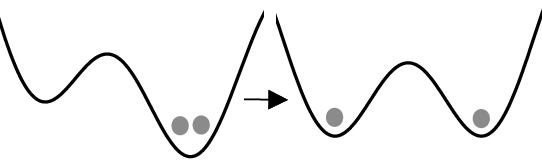}
   (b)\includegraphics[width=0.6 \linewidth]{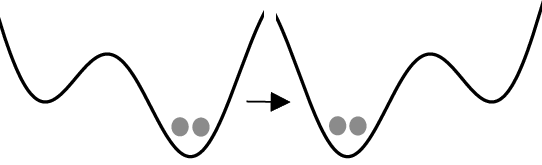}
   \end{center}
\caption{\label{tunneling}
(a) Schematic representation of splitting from $|0,2\ra$ to $|1,1\ra$. (b) Cotunneling from
$|0,2\ra$ to $|2,0\ra$.}
\end{figure}

Pairs of interacting bosons in a double-well potential may be manipulated to implement universal quantum logic gates for quantum computation 
or to observe fundamental phenomena such as cotunneling  of two atoms
\cite{Anderlini2007,Folling2007}.
We shall speed up two processes: the splitting of the two particles from one to the two separate wells, 
and cotunneling (see Fig. \ref{tunneling}). 
The boson dynamics in a double well with tight lateral confinement 
is described by a two-site Bose-Hubbard Hamiltonian\footnote{This Hamiltonian is similar to the one in the previous chapter [Eq.(\ref{H_0_Lie})] but we add two diagonal terms that make the potential asymmetric.} \cite{Folling2007}.
The Hamiltonian in the occupation number basis $|2,0\rangle=\left (\scriptsize{\begin{array} {rcccl} 1\\ 0\\0 \end{array}} \right)$, $|1,1\rangle=\left ( \scriptsize{\begin{array} {rcccl} 0\\ 1\\0 \end{array}} \right)$, and $|0,2\rangle=\left ( \scriptsize{\begin{array} {rcccl} 0\\ 0\\1 \end{array}} \right)$ is   
\beq
\label{H_0}
H=\left ( \begin{array}{ccc}
U+\Delta & -\sqrt{2}J & 0\\
-\sqrt{2}J & 0 & -\sqrt{2}J \\
0 & -\sqrt{2}J & U-\Delta
\end{array} \right),
\eeq
where the bias $\Delta=\Delta(t)$ is the control function, $J$ is the hopping energy, and $U$ the interaction energy. 
We write the time-dependent states as 
$|\Psi(t)\rangle=c_1(t)|2,0\rangle+c_2(t)|1,1\rangle+c_3(t)|0,2\rangle$.
%
%
%
%
%
%
%
%
%
%
%
%
%
%
%
Adiabatic processes that change $\Delta(t)$ slowly, keeping the $U/J$ ratio constant,
are possible to implement splitting or cotunelling. 
Speeding them up by a ``counterdiabatic'' approach  is 
not possible in practice because of the need to apply new terms in the Hamiltonian which are difficult to implement. 
Alternative techniques, like the one introduced in Chapter \ref{Chapter2}, could not be applied \cite{Opatrny2014} or are  
cumbersome \cite{Torrontegui2014,Martinez-Garaot2014} because of the relatively large algebra involved.  
The FAQUAD approach provides a viable way out.  

%
%
%
%
%
%
%
%
\begin{figure}[t]
\begin{center}
   \includegraphics[width=0.5 \linewidth]{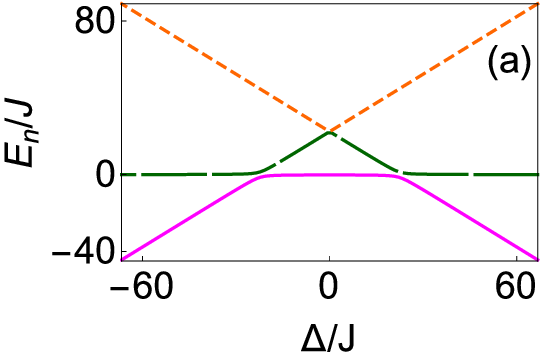}
   \includegraphics[width=0.5 \linewidth]{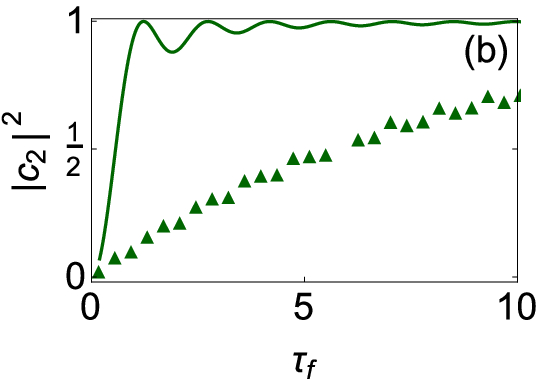}
   \end{center}
\caption{\label{levels_ap}
(a) Energy levels vs $\Delta$. For $n=1,2,3$: $E_1$ (solid magenta line), $E_2$ (long-dashed green line),
and $E_3$ (short-dashed orange line).  
$U/J= 22.3$. (b) $|c_2|^2$ vs $\tau_f$ for linear-in-time bias (green triangles) and FAQUAD (solid green line). 
$\Delta(0)/J=100$, $U/J=33.45$, and $\tau_f=Jt_f/\hbar$.}
\end{figure}
%
%
%
%
%
%
- In a splitting process  $\Delta(0)\gg U,J$ and  $\Delta(t_f) =0$ 
[see Fig. \ref{tunneling}(a)].  
The initial ground state is $|\phi_1\rangle=|0,2\rangle$ and the final ground state $|\phi_1\rangle=|1,1\rangle$. 
Figure \ref{levels_ap}(a) shows the dependence of the three eigenenergies
with $\Delta$. 
$\Delta_{F}(t)$ is very similar to the result for the two-level system in Fig. \ref{finalt_tmm}(a). 
%
%
%
%
%
%
%
%
\begin{figure}[t]
\begin{center}
   \includegraphics[width=0.5 \linewidth]{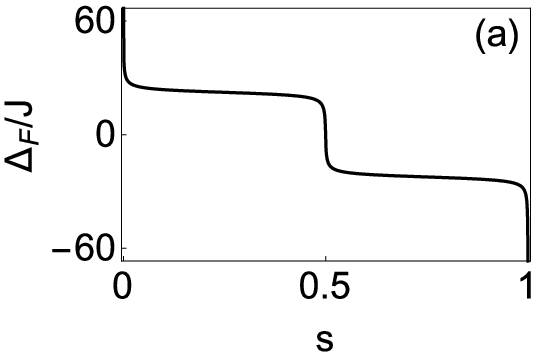}
   \includegraphics[width=0.5 \linewidth]{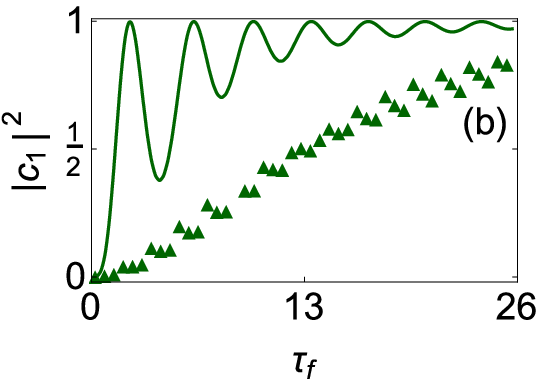}
\end{center}
\caption{\label{cotu}
(a) Time dependence of the bias with FAQUAD. 
(b) $|c_1|^2$ vs $\tau_f$ for linear-in-time bias (green triangles) and FAQUAD (solid green line). 
$\Delta(0)/J=66.7$, $U/J=22.3$, and $\tau_f=Jt_f/\hbar$.}
\end{figure}
%
%
%
%
%
%
The results of FAQUAD and the linear protocol are compared in Fig. \ref{levels_ap}(b). The probability of the first peak for FAQUAD, 0.998
at $\tau_f=1.2$, is achieved with the linear ramp for $\tau_f=43$.  

- In a speeded-up cotunneling, shown in Fig. \ref{tunneling}(b), the goal is to 
drive the system fast from $|\phi_1(0)\rangle=|0,2\rangle$ to $|\phi_1(t_f)\rangle=|2,0\rangle$ intermediated by $|1,1\ra$ 
[the Hamiltonian  
(\ref{H_0}) 
does not  connect $|2,0\rangle$ and $|0,2\rangle$ directly]. We impose $\Delta(0)\gg U,J$ and $\Delta(t_f) =-\Delta(0)$ to have $|0,2\rangle$ and $|2,0\rangle$ as the ground states at initial and final times, respectively.
The energy levels versus $\Delta$ are depicted in  
Fig. \ref{levels_ap}(a) for repulsive interaction ($U>0$). 
Figure \ref{cotu}(a) shows the FAQUAD trajectory for $\Delta(t)$
for the repulsive strong-interaction regime, $U/J=22.3$.
Figure \ref{cotu}(b) depicts the final probabilities of the bare state $|2,0\ra$ for FAQUAD and a linear protocol
that needs about $\tau_f=65$ to achieve the value of the first peak of the FAQUAD method
($|c_1|^2=0.998$ at $\tau_f=2.3$).  
The minima in the FAQUAD probability go in this case below the lower envelope $1-4\tilde{c}^2/t_f^2$
predicted by perturbation theory. The reason is a leak through the narrow avoided crossing at 
$\Delta=0$ from the second to the third energy level 
[see Fig. \ref{levels_ap}(a)]. The leak occurs at total process times in which the first avoided crossing produces a minimum 
of the ground-state probability.          
\section{Collective superpositions of rotating and nonrotating atoms on a ring}
Creating a macroscopic or mesoscopic superposition of a many-particle system is a difficult task and of interest for research in quantum information, quantum metrology and fundamental aspects of quantum mechanics. However, it was recently proposed that a low-dimensional gas of interacting bosons in the TG limit \cite{Girardeau1960} placed on a ring can be perturbed in such a way, that a robust superposition of two angular momentum states can be achieved. This perturbation corresponds to the introduction of a narrow potential, which is then accelerated to a certain value to spin up the gas \cite{Hallwood2010}.
%
%
%
%
%
\begin{figure}[t]
\begin{center}
   \includegraphics[width=0.5 \linewidth]{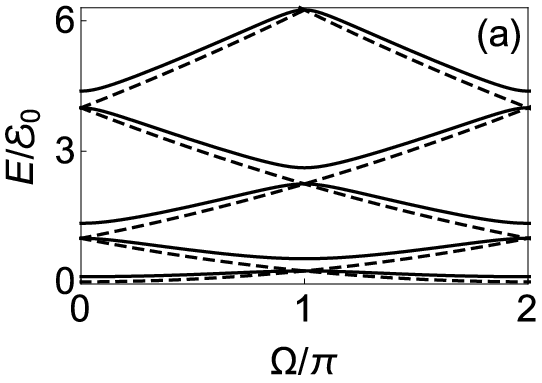}
   \includegraphics[width=0.5 \linewidth]{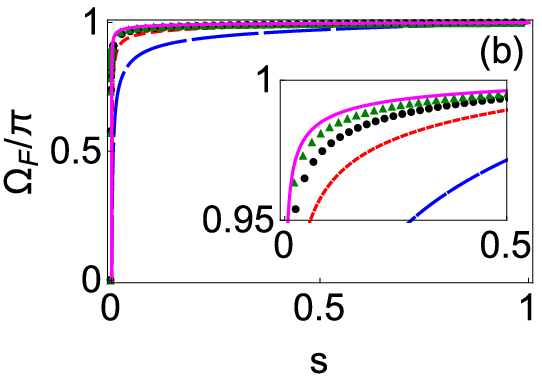}
  \end{center}
  \caption{\label{spoon}
(a) Single-particle energy levels for $U_0=0$ (dashed lines) and $U_0ML/\hbar^2=4$ (solid lines) in units of ${\cal{E}}_0=2\pi^2\hbar^2/(ML^2)$. The ordering is $E_1(n=0)<E_2(n=1)<E_3(n=-1)<E_4(n=2)<E_5(n=-2)<...$. (b) $\Omega_F(s)$ for $N=1,3,5,7,9$, from the bottom up to the top.}
\end{figure}
%
%
%
%
%

For a single particle this is described by
\beq
i\hbar \partial_t \psi (x,t)\!=\!\bigg\{-\frac{\hbar^2}{2M}\frac{\partial^2}{\partial x^2}\!+\!U_0\delta[x-x_0(t)]\bigg\} \psi(x,t),
\eeq
where the stirrer is represented by a $\delta$ function of strength $U_0$ and periodic BC are assumed. In a comoving frame one can then define $y=x-x_0(t)$ and the Hamiltonian is 
\beq
H=\frac{1}{2M} \big[\hat P_y - \hbar\Omega(t)/L \big]^2+U_0\delta(y),
\eeq
where $L$ is the ring perimeter, $\hbar\Omega(t)=M\dot x_0$ and $\hat P_y=-i\hbar\partial/\partial y$. The instantaneous energy eigenvalues are 
\beq
E(n)=\frac{2\hbar^2\pi^2}{L^2M}\alpha_n^2,
\eeq
and the $\alpha_n$ are solutions of 
\beq
\frac{4\pi\hbar^2\alpha_n}{MLU_0}=\cot(\pi \alpha_n- \Omega/2)+ \cot(\pi \alpha_n+\Omega/2).
\eeq
For $U_0\rightarrow 0$, the $\alpha_n$ tend to $n-\Omega/(2\pi)$, with $n=0,\pm1,\pm2,\dots$, 
where the different signs are for clockwise or counterclockwise rotation in the laboratory frame, and the $n$th eigenstates are plane waves with momentum $n\hbar 2\pi/L$. For $0<\Omega<\pi$ the energies in the moving frame increase for $n\leqslant 0$ and decrease for $n>0$. For $U_0=0$ the spectrum shows degeneracies at $\Omega=0,\pi$, which turn into avoided crossings once the stirrer couples different angular momentum eigenstates, as shown in Fig. \ref{spoon}(a). Adiabatically increasing the stirring frequency from $\Omega=0$ to $\pi$ then, allows us to drive the system into a superposition of two angular momentum states, and for a TG gas with an odd number of particles $N$ it can be shown that the ground state at $\Omega=\pi$ corresponds to macroscopic superposition between states with angular momentum zero and $N\hbar$.
%
%
%
%
%
%
\begin{figure}[t]
 \begin{center}
   \includegraphics[width=0.5 \linewidth]{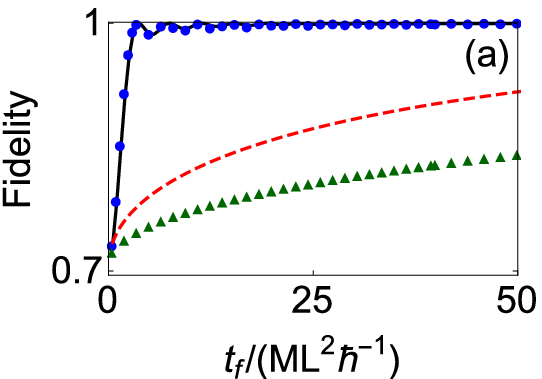}
  \includegraphics[width=0.5 \linewidth]{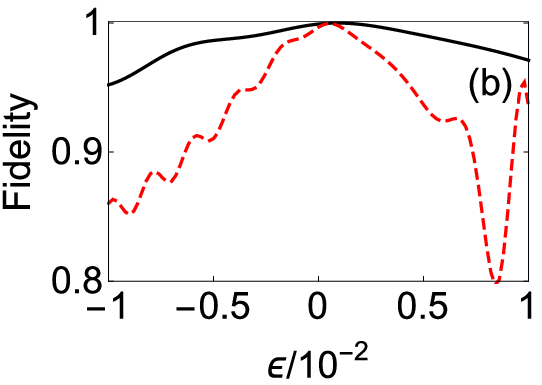}
  \end{center}
  \caption{\label{spoon2}
(a) Fidelity $|\la\Psi_{TG}(t_f)|\Phi_{TG}\ra|$  for $N=3$ [FAQUAD (solid black line) and linear $\Omega(t)$ (short-dashed red line)] and $N=9$ [FAQUAD (blue circles) and linear $\Omega(t)$ (green triangles)].  $\Psi_{TG}(t_f)$ is the time-evolved TG state starting from the ground state for $\Omega=0$, and $\Phi_{TG}$ is the ground state of the TG gas at $\Omega=\pi$. (b) Fidelity $|\la\Psi_{TG}(t_f)|\Phi_{TG}\ra|$ vs $\epsilon$ if FAQUAD is applied following a {\it wrong} $\Omega_e(t)=\Omega_F(t)(1+\epsilon)$ for $N=3$ (solid black line) and $N=9$ (short-dashed red line). Here $U_0ML/\hbar^2=0.5$.}
\end{figure}
%
%
%
%
%
%

To design an optimal $\Omega(t)$ for the TG gas, we note that the fidelity depends mostly on leakage from the highest occupied levels. This can be seen by considering the time evolved TG gas state $\Psi_{TG}(x_1,x_2,\ldots, x_N)$ defined by
\beq
\Psi_{TG} = \frac1{\sqrt{N!}} \prod_{i<j}\textnormal{sgn}(x_i-x_j) \sum_{\mu \in P} \epsilon_\mu \psi_{\mu_1}(x_1)\cdots\psi_{\mu_N}(x_N),
\eeq
where $P$ represents the set of all permutations of $\{0,1,\ldots,N-1\}$, $\epsilon_\mu$ is the antisymmetric tensor of the permutation $\mu$, and $\psi_i$ are the one-particle orbitals. Assuming that the system is isolated and contains only $N$ eigenvectors $\phi_j$, the orbitals can be expressed as $\psi_i=\sum_j U_{ij} \phi_j$ 
with $U$ some unitary operator. If we now compare $\Psi_{TG}$ to the ground state  $\Phi_{TG}$ of the TG gas at the final $\Omega$, we can calculate the fidelity $F=|\la \Phi_{TG} | \Psi_{TG} \ra |$ as
\begin{eqnarray}
F &=&\frac1{N!}   \left|  \sum_{\nu, \mu} \epsilon_\nu \epsilon_\mu  \la \phi_{\nu_1} |  \psi_{\mu_1} \ra \cdots  \la \phi_{\nu_1} |  \psi_{\mu_1} \ra  \right| \nonumber \\
&=&  \frac1{N!}  \left|  \sum_{\nu, \mu} \epsilon_\nu \epsilon_\mu U_{\mu_1,\nu_1} \cdots U_{\mu_N,\nu_N}  \right| \nonumber \\ 
&=&  \left|    \det(U)\right| =1,
\end{eqnarray}
since $U$ is unitary. Of course, in reality the system we consider contains more than $N$ eigenvectors and the fidelity does not remain 1, but this argument shows that leaking between two occupied states does not influence the fidelity of a TG gas at all; only leaks into modes above the Fermi level do, such as with nonzero mixing terms $U_{N,N+1}$. We should therefore optimize  $\Omega_F(s)$ for the avoided crossing of the highest occupied level as shown in  Fig. \ref{spoon}(b). The corresponding final-state fidelities for $N=3$ and $9$ with respect to the exact ground states clearly outperform the ones for the linear ramp [see Fig. \ref{spoon2}(a)]. The linear ramp fidelity deteriorates as $N$ increases whereas, remarkably, the fidelity of the FAQUAD protocol stays constant. The effect of an error of the form $\Omega_e(t)=\Omega_F(t)(1+\epsilon)$ is shown in Fig. \ref{spoon2}(b).

\section{Discussion} 
The FAQUAD approach to speed up adiabatic manipulations of quantum systems achieves significant time shortenings by distributing homogeneously the
adiabaticity parameter along the process while satisfying the boundary conditions of the
control parameter.
We have derived general time scales and    
we have demonstrated its applicability in  different systems, in particular where 
other approaches are not available,  and expect a broad range of applications in quantum, optical, and mechanical
systems,  due to the ubiquity of adiabatic methods. 

A natural extension is to attempt a scheme similar to Eq. (\ref{f_adiabatic}) in a superadiabatic 
rather than an adiabatic frame \cite{Ibanez2013}. 
The set of nested frames is described in detail in \cite{Ibanez2013}. A brief summary is provided here. 
Let us start with a Schr\"odinger picture Hamiltonian $H_0(t)$ and corresponding wave function $\psi_0(t)$. Defining the unitary operator 
\beq
A_0(t)=\sum_n|\phi_n(t)\rangle\langle n| 
\eeq
with $|\phi_n(t)\rangle$ the adiabatic basis in Eqs. (\ref{eigenstate_1_tmm}) and 
(\ref{eigenstate_2_tmm}), and $|n\rangle$ the bare basis, the Hamiltonian that governs the dynamics of the 
interaction picture state $A_0^\dagger \psi_0$ is 
\beq
\label{H_1}
H_1(t)=A_0^{\dag}(H_0-K_0)A_0,
\eeq
where $K_0=i\hbar\dot{A_0} A_0^{\dag}$.
For the two level model, taking into account Eq. (\ref{H_0_tmm}) in Eq. (\ref{H_1}) we get
\beq
H_1=\left(\scriptsize{\begin{array}{cc}
\frac{1}{2}(U-P-\Delta) & i\frac{\sqrt{2}J\hbar\dot\Delta}{P^2}
\\
-i\frac{\sqrt{2}J\hbar\dot\Delta}{P^2} & \frac{1}{2}(U+P-\Delta)
\end{array}} \right),
\eeq
with instantaneous eigenvalues
\beqa
\label{eigenvalues_1_tmm_sa}
&&E_1^{(1)}=\frac{1}{2}\left (U-\Delta-\frac{\sqrt{P^6+8J^2\hbar^2\dot\Delta^2}}{P^2}\right ),
\\
\label{eigenvalues_2_tmm_sa}
&&E_2^{(1)}=\frac{1}{2}\left (U-\Delta+\frac{\sqrt{P^6+8J^2\hbar^2\dot\Delta^2}}{P^2}\right ),
\eeqa
and normalized eigenstates
\beqa
\label{eigenstate_1_tmm_sa}
|\phi_1^{(1)}\rangle&=&\left (\!\! \begin{array}{c}
\frac{-2iJ\hbar\dot \Delta}{(P^6+8J^2\hbar^2\dot\Delta^2)^{1/4}\sqrt{-P^3+\sqrt{P^6+8J^2\hbar^2\dot\Delta^2}}}
\\[2.6 ex]
\frac{\sqrt{-P^3+\sqrt{P^6+8J^2\hbar^2\dot\Delta^2}}}{\sqrt{2}(P^6+8J^2\hbar^2\dot\Delta^2)^{1/4}}
\\
\end{array}\!\! \right)\!\!, 
\\
\label{eigenstate_2_tmm_sa}
|\phi_2^{(1)}\rangle&=& \left (\!\! \begin{array}{c}
\frac{2iJ\hbar\dot \Delta}{(P^6+8J^2\hbar^2\dot\Delta^2)^{1/4}\sqrt{P^3+\sqrt{P^6+8J^2\hbar^2\dot\Delta^2}}}
\\[2.6 ex]
\frac{\sqrt{P^3+\sqrt{P^6+8J^2\hbar^2\dot\Delta^2}}}{\sqrt{2}(P^6+8J^2\hbar^2\dot\Delta^2)^{1/4}}
\\
\end{array}\!\! \right)\!.
\eeqa
The first superadiabatic frame is defined by the unitary operator 
\beq
A_1(t)=\sum_n|\phi_n^{(1)}(t)\rangle \langle n|.
\eeq
The state $A_1^\dagger \psi_1$ is governed by the 
Hamiltonian 
\beq
H_2(t)=A_1^{\dag}(H_1-K_1)A_1,
\eeq
where $K_1=i\hbar\dot{A_1}A_1^{\dag}$.

Note that superadiabaticity, i.e., the possibility to neglect $K_1$, does not necessarily imply adiabaticity,  which amounts to neglecting $K_0$. 
Also,  a shortcut to superadiabaticity 
is only a STA if the superadiabatic states $|\phi_n^{(1)}\rangle$ coincide, up to phase factors, 
with the eigenstates of $H_0$,  $|\phi_n\rangle$, at boundary times. 
This will imply additional boundary conditions on the control parameter. 
The equation that substitutes Eq. (\ref{f_adiabatic}) 
for the lowest superadiabatic scheme beyond the adiabatic level is
\beq
\label{f_adiabatic_1}
\hbar \left | \frac{\langle\phi_1^{(1)}(t)|\partial_t\phi_2^{(1)}(t)\rangle}{E_1^{(1)}-E_2^{(2)}} \right |=c.
\eeq
Using Eqs. (\ref{eigenvalues_1_tmm_sa}), (\ref{eigenvalues_2_tmm_sa}), (\ref{eigenstate_1_tmm_sa}) and (\ref{eigenstate_2_tmm_sa}) in (\ref{f_adiabatic_1}), we get a second-order differential equation for $\Delta$:
\beq
\label{f_adiabatic_sa}
\frac{\sqrt{2}J\hbar^2P^4(-3\dot P\dot \Delta+P\ddot\Delta)}{(P^6+8J^2\hbar^2\dot\Delta^2)^{3/2}}=c.
\eeq
To satisfy $|\phi_1^{(1)}(0)\rangle=|\phi_1(0)\rangle=|2\rangle$ and  $|\phi_1^{(1)}(t_f)\rangle=|\phi_1(t_f)\rangle=|1\rangle$ (up to phase factors)
we have to impose
four boundary conditions,
\beqa
&&\Delta(0)\gg U,J , \, \, \, \Delta(t_f)=0, \nonumber
\\
&&\dot\Delta(0)=0 , \, \, \, \dot\Delta(t_f)\gg \Delta(0), U, J,
\eeqa
that cannot be satisfied with two integration constants plus the $c$. The mismatch between number 
of conditions and free parameters actually gets worse when increasing the order of superadiabaticity in further iterations. 
In the second superadiabatic frame defined by the unitary operator 
\beq
A_2(t)=\sum_n|\phi_n^{(2)}(t)\rangle\langle n|, 
\eeq
due to the $K_2=i\hbar\dot A_2A_2^{\dag}$ term, second-order derivatives of the control parameter appear 
in the superadiabatic eigenstates, so 
the number of boundary conditions necessary to satisfy $|\phi_1^{(2)}(0)\rangle=|2\rangle$ and $|\phi_1^{(2)}(t_f)\rangle=|1\rangle$ (up to phase factors) increases to $6$. 
Moreover, the differential equation resulting from applying the FAQUAD concept in the second superadiabatic basis is of third order in $\Delta$. Once again, the differential equation cannot satisfy the six boundary conditions with three integration constants plus the $c$.
In general, as the order of the iteration increases, the number of boundary conditions to satisfy grows as $2n+2$, where $n$ is the order of the iteration, while the order of the differential equation increases as $n+1$. 
Hence, the adiabatic frame is in fact optimal to apply the FAQUAD concept within the series of iterative superadiabatic frames, as it is the only one for which the number of conditions equals the number of free parameters available.


\chapter{Vibrational mode multiplexing of ultracold atoms}
\label{Chapter4}
\lhead{Chapter 4. \emph{Vibrational mode multiplexing of ultracold atoms}} 
Sending multiple messages on qubits encoded in different 
vibrational modes of cold atoms or ions along a transmission waveguide
requires us to  merge first and then separate the modes at input and output ends.  
Similarly, different qubits can be stored in the modes of a trap and be separated later.     
We design the fast splitting of a harmonic trap into an asymmetric  double well so that  the initial ground vibrational state
becomes the ground state of one of two final wells, and the initial first excited state 
becomes the ground state of the other well.   
This might be done adiabatically by slowly  deforming the trap. We speed up the process by  
inverse engineering a  double-function trap using  dynamical invariants.   
The separation (demultiplexing) followed by an inversion of the asymmetric bias 
and then by the reverse process (multiplexing)  provides a population inversion protocol 
based solely on trap reshaping.    
\newpage
\section{Introduction} 

One of the main goals of atomic physics is to achieve an exhaustive 
control of atomic states and dynamics \cite{cohen2011advances}. The ultracold domain is particularly 
suitable for this aim as it provides a rich scenario of quantum states and phenomena. 
Atom optics and atomtronics \cite{Seaman2007} intend to manipulate cold atoms in circuits and devices 
for applications  in metrology, quantum information, or fundamental science. These devices are frequently
inspired by electronics (e.g., the atom diode \cite{Ruschhaupt2004,Raizen2005}, the transistor \cite{Seaman2007}, atom chips \cite{reichel2011atom}), 
or optics
(e.g., beam splitters \cite{Gattobigio2012}, or multiplexing \cite{Burns1975,Riesen2012}).   
%
%
%
%
\begin{figure}[t]
\begin{center}
\includegraphics[width=0.99 \linewidth]{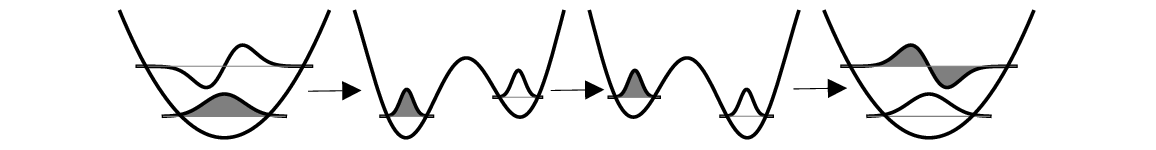}
\end{center}
\caption{\label{adiabatic1}
Population inversion using trap deformations in three steps:  
demultiplexing, bias inversion, and multiplexing.}
\end{figure}
%
%
%
%

In this chapter we shall focus on a cold-atom realization of multiplexing, a basic process in modern telecommunications.    
Multiplexing is the transmission of different messages 
via a single physical medium. 
A multiplexer combines signals from several emitters into a single medium, whereas a demultiplexer
performs the reverse operation. 
The concept of multiplexing is relevant for quantum information processing (for its use in quantum repeaters,
see \cite{Collins2007,Sangouard2011}, or for trapped ions \cite{Wineland1998}).  
We envision here optical or magnetic waveguides for atoms holding several transverse orthogonal modes \cite{Lizuain2006,Leanhardt2002,Guerin2006,Gattobigio2009}. 
If the qubit is encoded in the internal state of the atom, several qubits may be carried out simultaneously by different modes.   
To develop such a quantum-information  architecture, fast  multiplexers or demultiplexers that could  join
the modes from different waveguides into one guide, or separate them, are needed.
We shall discuss trap  designs for demultiplexing since the multiplexer would simply operate in reverse. 
For a proof of principle, we propose the  simplified setting of a single initial harmonic trap 
for noninteracting cold atoms whose first two 
eigenstates will be separated, as in the first step of Fig. \ref{adiabatic1}, into two different wells.   
In a complete demultiplexing process, the final wells should be independent, with negligible tunneling.   
The challenge  is to design the splitting   (a) without final excitation of higher vibrational levels, (b) 
in a short time, and (c) with a 
realizable trap potential.     
Condition (a) may be achieved by an adiabatic asymmetric splitting \cite{Gea-Banacloche2002,Torrontegui2013e} in
which, for  moderate bias compared to the vibrational quanta, the initial ground state becomes the ground state of the well with the 
lowest energy, and the excited state becomes the ground state of the other well (see Chapter \ref{Chapter1}). 
This adiabatic approach generally fails to satisfy the condition (b), which we shall implement 
applying  STA \cite{Chen2010,Chen2011b,Torrontegui2013d,Bowler2012}. As for (c), we shall make use of a simple two-level model for the shortcut design,
and then map it to a realistic potential
recently 
implemented to  realize an atomic Josephson junction 
\cite{Gati2006}.
Finally, several applications, such as separation of multiple modes, population inversion, or controlled excitation,  will be discussed.

\section{Slow adiabatic and fast quasi-adiabatic processes}

Suppose that a harmonic potential evolves adiabatically into two well-separated and asymmetric wells as in 
the first step  of Fig. \ref{adiabatic1}. 
To accelerate the dynamics we shall use the moving two-level approximation presented in Chapter \ref{Chapter1} (Section \ref{2mode}). This moving two-level approximation is
based on a process where  
a symmetrical potential evolves from an initial harmonic trap  to a final double well.  
Then, we construct a time-dependent
orthogonal bare basis 
$|L(t)\rangle = \left(\scriptsize{\begin{array} {rccl} 0\\ 1 \end{array}} \right)$, $|R(t)\rangle = \left(\scriptsize{\begin{array} {rccl} 1\\ 0 \end{array}} \right)$ of left and right states, obtained by a
linear combination of the instantaneous ground and first excited
states. 
An approximate two-mode Hamiltonian model for a generally  {\it{asymmetrical}} process is written in this basis as 
\beq
\label{H_tm}
H_{2\times2}(t)=\frac{\hbar}{2} \left ( \begin{array}{cc}
\lambda(t)
& -\delta(t)\\
-\delta(t)& -\lambda(t)
\end{array} \right),
\eeq
where, for the double well  configuration, $\delta(t)$ is  the tunneling rate, and $\hbar\lambda(t)$ the relative gap, or bias, between the two wells. 
Note that, this Hamiltonian is just the Hamiltonian in Eq. (\ref{H_tm_FF}) multiplied by a factor $\hbar$.
For the initial harmonic potential at $t=0$, $\lambda(0)=0$ and $\delta(0)=\omega_0$.  
The instantaneous eigenvalues are
\beq
E^{\pm}_\lambda(t)=
\pm \frac{\hbar}{2} \sqrt{\lambda^2(t)+\delta^2(t)},
\eeq
and the normalized eigenstates 
\beqa
\label{eigenstates_tls}
|\psi^+_\lambda(t)\rangle &=& \sin{ \left ( \frac{\alpha}{2} \right ) } |L(t)\rangle-\cos {\left ( \frac{\alpha}{2} \right ) }|R(t)\rangle, \nonumber
\\
|\psi^-_\lambda(t)\rangle &=& \cos{ \left ( \frac{\alpha}{2} \right )}|L(t)\rangle+\sin{\left ( \frac{\alpha}{2} \right )}|R(t)\rangle,
\eeqa
where the mixing angle $\alpha=\alpha(t)$ is given by $\tan \alpha = \delta (t)/\lambda(t)$. 
The boundary conditions on $\lambda(t)$ and $\delta(t)$ are
\beqa
\label{bc_d}
&&\delta(0)=\omega_0, \nonumber \\   
\label{bc_d}
&&\lambda(0)=0,\nonumber \\
\label{bc_d}
&&\delta(t_f)=0, \nonumber \\ 
\label{bc_d}  
&&\lambda(t_f)=\lambda_f, 
\eeqa
which correspond, at time $t=0$, to a harmonic well, and at time $t_f$ to two independent wells
with asymmetry bias $\hbar\lambda_f$.   

To design a FAQUAD process,
we shall first assume the simplifying conditions: $\lambda(t)=\lambda$ constant and
$\lambda/\delta(0)\ll 1$. Thus, $\alpha(0)\approx\pi/2$ and the initial eigenstates essentially coincide with the ground and 
first excited states of the harmonic oscillator.
As we have seen in Chapter \ref{Chapter1}, for a constant $\lambda$, the adiabaticity condition reads \cite{Torrontegui2013e}
\beq
\left |\frac{\lambda \dot \delta(t)}{2[\lambda^2+\delta(t)^2]^{3/2}} \right |\ll1.
\eeq
To get the FAQUAD solution we proceed as in Chapter \ref{Chapter3}. Therefore, 
imposing a constant value $c$ for the adiabaticity parameter 
and using the boundary conditions 
for $\delta$ in 
Eq. (\ref{bc_d}), we  fix the integration constant 
and the value of $c$,
\beq
c=\frac{\omega_{0}}{2 \lambda \sqrt{\omega_{0}^2+\lambda^2}\, t_f}.
\eeq
The solution of the differential equation for $\delta(t)$ takes finally the form 
\beq
\label{delta_fa}
\delta_{fa}(t)=\frac{\omega_0\lambda(t_f-t)}{\sqrt{{\lambda^2t_f^2+\omega_0^2t(2t_f-t) }}}.
\eeq
Although this protocol can be work for shorter times for which the process is not fully adiabatic, the FAQUAD approach is limited by 
\beq
t_f=\frac{2\pi}{\phi_{12}},
\eeq
where $\phi_{12}=\int_0^1\tilde\omega_{12}(s)ds$, and $\tilde\omega_{12}(s)=\omega_{12}(st_f)$ [see Chapter \ref{Chapter3}, Subsection \ref{g_p}].

We shall now work out an alternative, faster protocol
based on invariants, in which the boundary conditions
on $\lambda(t)$ and $\delta(t)$  
will be exactly satisfied.     

\section{Invariant-based inverse engineering}\label{lr}

\subsection{Lewis-Riesenfeld invariants}
The Lewis-Riesenfeld \cite{Lewis1969} theory is applicable to a quantum system that evolves with a time-dependent Hermitian Hamiltonian $H(t)$, which supports a Hermitian dynamical invariant $I(t)$ satisfying
\beq
\label{inv_prop}
i\hbar\frac{\partial I(t)}{\partial t}-[H(t),I(t)]=0.
\eeq

Therefore, its expectation values for an arbitrary solution of the time-dependent Schr\"odinger equation 
\beq
i\hbar \frac{\partial}{\partial t}|\Psi (t)\rangle=H(t)|\Psi (t)\rangle
\eeq
do not depend on time. $I(t)$ can be used to expand $|\Psi(t)\rangle$ as a superposition of ``dynamical modes'' $|\psi_n(t)\rangle$,
\beqa
&&|\Psi(t) \rangle=\sum_n c_n|\psi_n(t)\rangle, \nonumber \\
&&|\psi_n(t)\rangle=e^{i\alpha_n(t)}|\phi_n(t)\rangle,
\eeqa
where $n=0,1,\dots$; $c_n$ are time-independent amplitudes, and $|\phi_n(t)\rangle$ are orthonormal eigenvectors of the invariant $I(t)$, 
\beq
I(t)=\sum_n |\phi_n(t)\rangle \lambda_n \langle \phi_n(t)|.
\eeq

The $\lambda_n$ are real constants, and the Lewis-Riesenfeld phases are defined as \cite{Lewis1969}
\beq
\alpha_n(t)=\frac{1}{\hbar}\int_0^t \bigg \langle \phi_n(t') \bigg | i\hbar \frac{\partial}{\partial t'}- H(t') \bigg | \phi_n(t') \bigg \rangle dt'.
\eeq
We use, for simplicity, a notation for a discrete spectrum of $I(t)$ but the generalization to a continuum or mixed spectrum is straightforward. We also assume a non-degenerate spectrum.

\subsection{Inverse engineering}

Supose that we want to drive the system from an initial Hamiltonian $H(0)$ to a final one $H(t_f)$, in such a way that the populations in the initial and final instantaneous bases are the same, but admitting transitions at intermediate times. To inverse engineer a time-dependent Hamiltonian $H(t)$ and achieve this goal, we may first define the invariant through its eigenvalues and eigenvectors. The Lewis-Riesenfeld phases $\alpha_n(t)$ may also be chosen as arbitrary functions to write down the time-dependent unitary evolution operator $U$
\beq
U=\sum_n e^{i\alpha_n(t)}|\phi_n(t)\rangle \langle \phi_n(0)|.
\eeq

$U$ obeys $i\hbar \dot U=H(t)U$, where the dot means time derivative. Solving formally this equation for $H(t)=i\hbar \dot U U^{\dag}$, we get
\beq
\label{hamiltonian_inv}
H(t)=-\hbar \sum_n|\phi_n(t)\rangle \dot \alpha_n \langle \phi_n(t)|+i\hbar \sum_n |\partial_t\phi_n(t)\rangle \langle \phi_n(t)|.
\eeq

According to Eq. (\ref{hamiltonian_inv}), for a given invariant there are many possible Hamiltonians corresponding to different choices of phase functions $\alpha_n(t)$. In general $I(0)$ does not commute with $H(0)$, so the eigenstates of $I(0)$, $|\phi_n(0)\rangle$, do not coincide with the eigenstates of $H(0)$. $H(t_f)$ does not necessarily commute with $I(t_f)$ either. If we impose $[I(0),H(0)]=0$ and $[I(t_f),H(t_f)]=0$, the eigenstates will coincide, which guarantees a state transfer without final excitations. In typical applications, the Hamiltonians $H(0)$ and $H(t_f)$ are given, and set the initial and final configurations of the external parameters. Then we define $I(t)$ and its eigenvectors accordingly, so that the commutation relations are obeyed at the boundary times and, finally, $H(t)$ is designed via Eq. (\ref{hamiltonian_inv}). While the $\alpha_n(t)$ may be taken as fully free time-dependent phases in principle, they may also be constrained by a pre-imposed or assumed structure of $H(t)$. 

We will focus now on the two-level system, so for the Hamiltonian in Eq. (\ref{H_tm}),
there is a dynamical invariant $I(t)$  of the form \cite{Chen2011b}
\beq
\label{inv}
I(t)=\frac{\hbar}{2}\Omega_0 \left ( \begin{array}{cc}
\cos \theta(t)
& \sin \theta(t)e^{i\varphi(t)}\\
\sin \theta(t)e^{-i\varphi(t)}& -\cos \theta(t)
\end{array} \right),
\eeq
where $\varphi(t)$ and $\theta(t)$ are auxiliary (azymuthal and polar) angles, and
$\Omega_0$ is an arbitrary constant with units of frequency. 
The role of the invariant is therefore to drive the initial eigenstates 
of $H_{2\times2}(0)$  to the eigenstates of $H_{2\times2}(t_f)$. In  our application this implies a unitary mapping
from the first two eigenstates of the harmonic oscillator to the ground states of the left and right final wells.      

From the invariance property (\ref{inv_prop}), choosing $H(t)$ as the Hamiltonian in (\ref{H_tm}), it follows that 
\beqa
\label{inv_de}
\delta(t)&=&-\dot\theta(t)/\sin{\varphi(t)}, \nonumber
\\ 
\lambda(t)&=&-\delta(t)\cot{\theta(t)}\cos{\varphi(t)}-\dot \varphi(t).
\eeqa
The commutativity of $I(t)$ and $H_{2\times2}(t)$ at boundary times $t_b=0,t_f$ imposes  the conditions
\beqa
\label{con_0}
&&\lambda(t_b) \sin[\theta(t_b)]e^{i\varphi(t_b)}+\delta(t_b)\cos[\theta(t_b)] = 0, \nonumber
\\
&&\lambda(t_b) \sin[\theta(t_b)]e^{-i\varphi(t_b)}+\delta(t_b)\cos[\theta(t_b)] = 0, \nonumber
\\
&&\delta(t_b)\sin[\theta(t_b)]\sin[\varphi(t_b)]=0.
\eeqa
Taking into account Eq. (\ref{bc_d}),
we get from Eq. (\ref{con_0}),
\beqa
\label{b_c}
&&\theta(0)=\pi/2,\nonumber \\
\label{b_c}
&&\varphi(0)=\pi,\nonumber \\
\label{b_c}
&&\theta(t_f)=0,\nonumber \\
\label{b_c}
&&\dot\theta(t_f)=0.
\eeqa
These conditions lead to indeterminacies in Eq. (\ref{inv_de}). To resolve them we apply l'H\^ opital's rule repeatedly and find  
additional boundary conditions,
\beqa
\label{b_c_2}
&&\dot\theta(0)=\ddot\theta(0)=\dot{\varphi}(0)=0,\nonumber \\
\label{b_c_2}
&&\dddot\theta(0)=-\omega_0\dot\lambda(0), \nonumber \\
\label{b_c_2}
&&\ddot\varphi(0)=-\dot\lambda(0),\nonumber \\
\label{b_c_2}
&&\varphi(t_f)=\pi/2,\nonumber \\
\label{b_c_2}
&&\dot\varphi(t_f)=-\frac{\lambda_f}{3},
\eeqa
with $\dot\lambda(0)\neq 0$.
At intermediate times, we interpolate the angles assuming a polynomial ansatz, 
$\theta(t)=\sum_{j=0}^5 a_j t^j$ and 
$\varphi(t)=\sum_{j=0}^4 b_j t^j$, where the
coefficients are found by solving the equations for the boundary conditions. 
Thus, we obtain  the Hamiltonian functions $\delta^{\rm{inv}}(t)$ 
and $\lambda^{\rm{inv}}(t)$ from Eq. (\ref{inv_de}).  
%
%
%
%
%
\begin{figure}[t]
\begin{center}
\includegraphics[width=0.5 \linewidth]{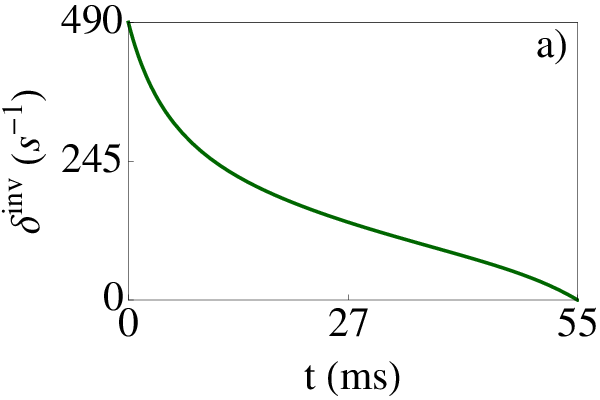}
\includegraphics[width=0.5 \linewidth]{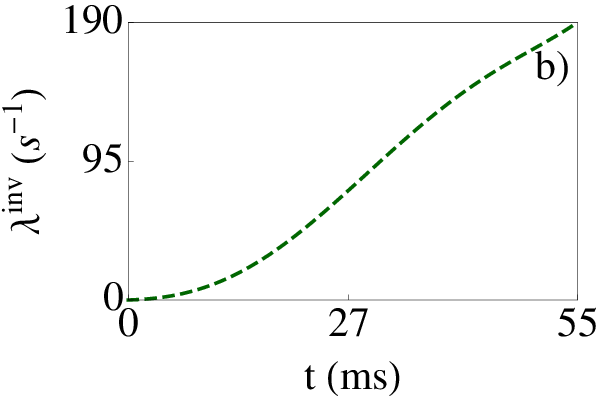}
\end{center}
\caption{\label{deltainv_10}
(a) $\delta^{\rm{inv}}(t)$ and (b) $\lambda^{\rm{inv}}(t)$.
$\delta(0)=2\pi\times 78$ Hz, 
$\lambda_f=190$ s$^{-1}$, $\dot\lambda(0)=190$ s$^{-2}$, and $t_f=55$ ms.}
\end{figure}
%
%
%
%
%
Figure \ref{deltainv_10} 
provides an example of parameter trajectories.

\section{Mapping to coordinate space\label{mapp}}
\begin{figure}[t]
\begin{center}
\includegraphics[width=0.5 \linewidth]{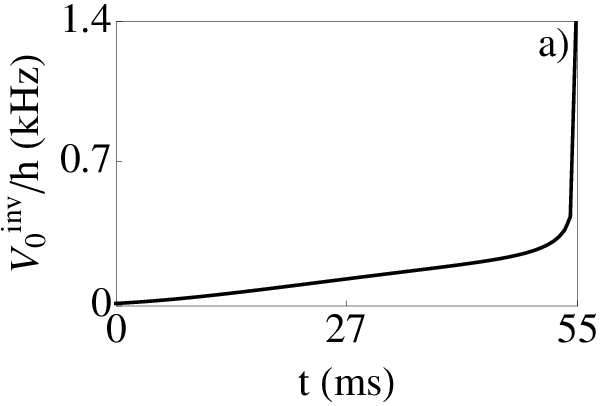}
\includegraphics[width=0.5 \linewidth]{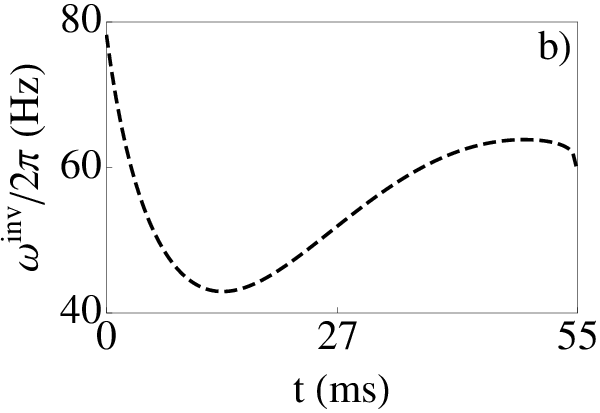}
\end{center}
\caption{\label{v0inv_10}
Lattice height $V_0$, and trap frequency $\omega/(2\pi)$ using invariant-based engineering and mapping.    
$\Delta x=200$ nm.  
}
\end{figure}

Our purpose now is to map the $2\times2$ Hamiltonian into a
realizable potential, 
\beq
\label{V_Oberthaler_1}
V(x,t)=\frac{1}{2}m\omega^2x^2+V_0\cos^2 \left [ \frac{\pi(x-\Delta x)}{d_l} \right ]. 
\eeq
This form has already been implemented \cite{Gati2006} with optical dipole potentials, combining a harmonic confinement 
due to a crossed beam dipole trap with a periodic light shift potential provided by the interference pattern of two mutually coherent laser beams.  
The control parameters are in principle the frequency $\omega$, the displacement $\Delta x$ of the optical lattice relative 
to the center of the harmonic well, the amplitude $V_0$ and the lattice constant $d_l$,  
but in the following examples we fix $d_l$ and $\Delta x$;  the other two parameters offer enough flexibility and 
are easier to control as time-dependent functions.     
To perform the mapping, we minimize numerically 
\beq
F[V_0(t),\omega(t)]=[\delta^{\rm{id}}(t)-\delta(t)]^2+[\lambda^{\rm{id}}(t)-\lambda(t)]^2,
\eeq
using the simplex method.  
The functions $\delta^{\rm{id}}(t)$ and $\lambda^{\rm{id}}(t)$ are designed according to the shortcuts 
discussed before, 
and (following the same procedure as in Chapter \ref{Chapter1})
$\delta(t)$ and $\lambda(t)$ are computed  
as 
\beqa
\delta(t)&=&-\frac{2}{\hbar}\langle L(t)|H|R(t) \rangle =-\frac{2}{\hbar}\langle R(t)|H|L(t) \rangle,
\\
\lambda(t)&=&\frac{2}{\hbar}\langle R(t)|H-\Lambda|R(t) \rangle =-\frac{2}{\hbar}\langle L(t)|H-\Lambda|L(t) \rangle,
\eeqa
where $H=H(V_0(t),\omega(t); \Delta x, d_l)=-\frac{\hbar^2}{2m}\frac{\partial^2}{\partial x^2}+V$ is the full Hamiltonian in coordinate space 
with a kinetic energy term and the potential (\ref{V_Oberthaler_1}) and $\Lambda(t)=[E_{\lambda}^-(t)+E_{\lambda}^+(t)]/2$ is a shift 
defined from the first two levels $E_\mp$ of $H$ to match the zero-energy point between the coordinate and the two-level system.  
Finally,
\beqa 
|R(t)\ra&=&(|g(t)\ra+|e(t)\ra)/2^{1/2}, \nonumber \\
|L(t)\ra&=&(|g(t)\ra-|e(t)\ra)/2^{1/2} 
\eeqa
form the base, where  
$|g(t)\ra$ is the ground state and  $|e(t)\ra$ the first excited state  
of the symmetrical Hamiltonian   $H_0(V_0(t),\omega(t); \Delta x=0, d_l)$, defined as $H$ but with $\Delta x=0$, 
which we diagonalize numerically.       
In our calculations, $\delta(t)$ and $\lambda(t)$ become indistinguishable from their 
ideal counterparts.   Figure \ref{v0inv_10}  depicts   $V_0(t)$ and $\omega(t)$ for the  parameters   
of Fig. \ref{deltainv_10}. We use $^{87}$Rb atoms  and a lattice spacing $d_l=5.18$ $\mu$m. 
The sharp final increase of $V_0(t)$  makes the two wells totally independent, but    
for most applications this strict condition may be relaxed to avoid  intrawell excitations. 
%
%
%
%
%
%
%
%
%
%

%
%
%
%
%
\begin{figure}[t]
\begin{center}
\includegraphics[width=0.5 \linewidth]{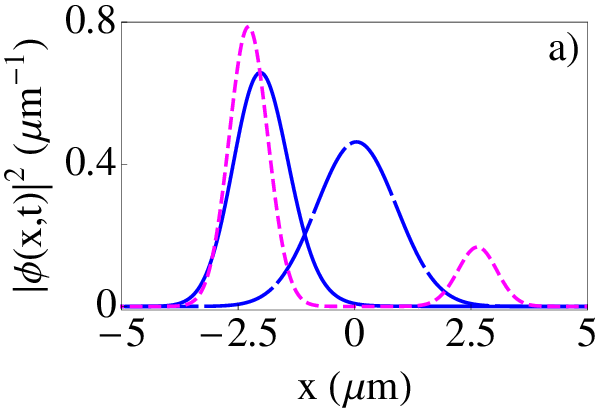}
\includegraphics[width=0.5 \linewidth]{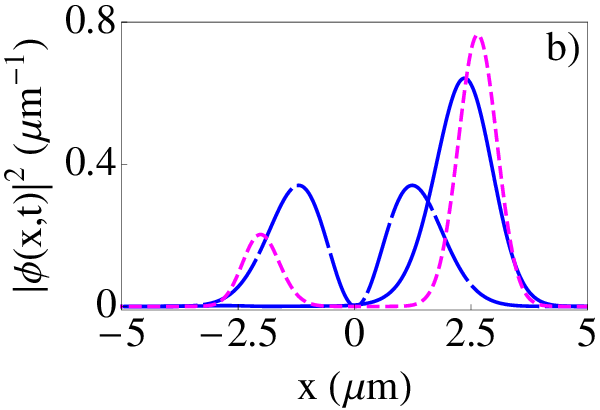}
\end{center}
\caption{\label{psiinv_10}
(a): Ground state at $t=0$ (long-dashed,  blue line); final state with the shortcut  (solid, blue line, indistinguishable 
from the ground state of the final trap); final state with linear ramp for $V_0(t)$
and $\omega=2\pi\times78$ Hz (short-dashed,  magenta line).  
(b): Same as (a) for the first excited state. 
Parameters like in Fig. \ref{v0inv_10} at $t=53$ ms. 
The linear ramp for $V_0(t)$ ends in the same value 
used for the shortcut.} 
\end{figure}
%
%
%

\begin{figure}[t]
\begin{center}
\includegraphics[width=0.5 \linewidth]{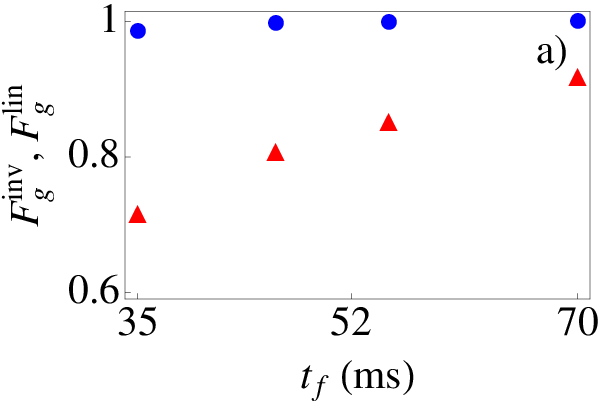}
\includegraphics[width=0.5 \linewidth]{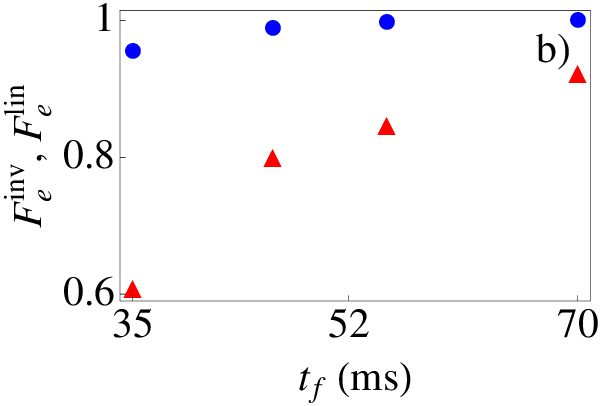}
\end{center}
\caption{\label{Fidelity}
Fidelities with respect to the final ground state starting at the ground state (a) and with respect to the final first excited state starting at the excited state (b) versus final time $t_f$, 
via shortcuts  ($F_{g}^{inv}$ and $F_{e}^{inv}$, blue circles), or  linear ramping  of $V_0(t)$ ($F_{g}^{lin}$ and $F_{e}^{lin}$, red triangles). 
The fidelity is computed at 2 ms less than the nominal $t_f$. Other parameters as in Figs. \ref{deltainv_10}, \ref{v0inv_10}, and \ref{psiinv_10}.
}
\end{figure}

\begin{figure}[t]
\begin{center}
\includegraphics[width=0.5 \linewidth]{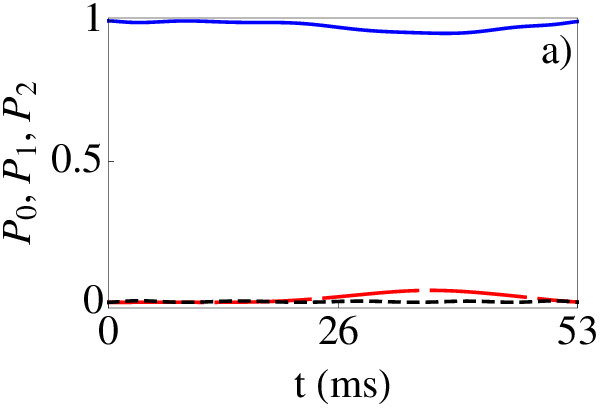}
\includegraphics[width=0.5 \linewidth]{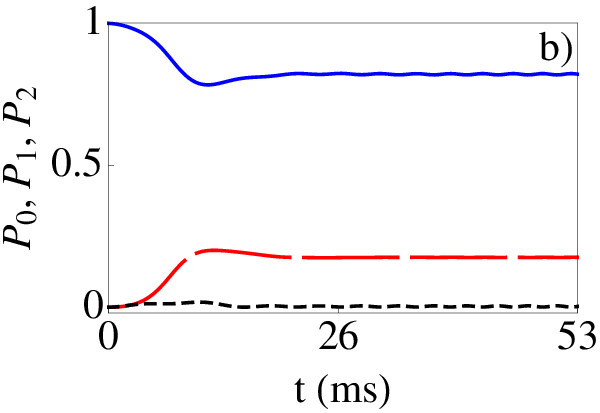}
\end{center}
\caption{\label{pinv_10}
Populations of the states for the shortcuts (a) and the linear ramp for $V_0(t)$ (b). 
Ground state ($P_0$, solid blue line); first excited state ($P_1$, long-dashed red line); second excited state ($P_2$, short-dashed black line). 
Parameters as in Fig. \ref{psiinv_10} (a).
}
\end{figure}

Figure \ref{psiinv_10} demonstrates perfect transfer for the ground (a) and the excited state (b)
using the very same protocol in both cases, the one depicted in Figs. \ref{deltainv_10} and Ê\ref{v0inv_10}.  
(Thanks to the superposition principle, the
same protocol would produce a perfect demultiplexing for any linear
combination of the ground and excited states.) Initial and final states are represented, solving the Schr\"odinger equation with the potential (\ref{V_Oberthaler_1}).
We stop the process $2$ ms before the nominal time $t_f$, as the fidelity
reaches a stable maximum there, and a further increase of $V_0$ is not required.    
We also include the  results for the protocol in which $\omega$
is kept constant and $V_0(t)$ is a linear ramp (with the same durations as the  shortcut protocols). For    
this linear protocol the final state includes a significant  density in the ``wrong'' well.   
This simple linear-$V_0$  approach needs $t_f\gtrsim0.7$ s  to become adiabatic and produce  the same fidelity, 
0.9997,  found for a shortcut protocol ten times faster, $t_f=0.07$ s, the rightmost point in Fig. \ref{Fidelity} (a).    
Figure \ref{pinv_10} compares  the populations   
in the instantaneous basis of the (full, coordinate-space) Hamiltonian for the shortcut and the linear protocols
when the system starts in the ground state, corresponding to  Fig. \ref{psiinv_10} (a). 
The shortcut protocol implies a transient exchange between ground and (first) excited levels but finally 
takes the system to the desired ground state. In contrast to the linear protocol, the excitation is permanent,
leading to a poor final fidelity. 
%
%
%
%
%
%
%
%
%
%
%
%
%

In the two-level model, 
$t_f$ may be reduced arbitrarily,  but in the coordinate space Hamiltonian, levels 0-1 will only be ``independent'' as long as 
higher levels are not excited.  These excitations are the limiting factor to shorten the times further with the 
current mapping scheme. Some guidance is provided 
by the Anandan-Aharonov relation  $t_f>h/(4 \overline{\Delta E})$, where $\overline{\Delta E}$ is the time average of the
standard deviation \cite{Anandan1990}.   

\section{Discussion}

Vibrational multiplexing may be combined with internal-state multiplexing \cite{Wineland1998} to provide 
a plethora  of possible operations. Motivated by the prospective use of multiplexing or demultiplexing  for quantum information 
processing,  we have applied shortcuts to adiabaticity techniques to speed up the spatial separation of vibrational modes 
of a harmonic trap.  
A  similar approach 
would separate $n$ modes  into $n$ wells so as to deliver more information into different processing sites. 
The number of modes that could be separated 
will depend on the asymmetric bias in relation to other potential parameters:  
the bias among the extreme wells should not exceed the
vibrational quanta in the final wells. 
The bias determines possible speeds, too, as smaller biases generally imply longer times.  
           
Chapter \ref{Chapter1} dealt also with splitting operations and shortcuts to adiabaticity, 
but the objective was the opposite to our aim here. 
Since adiabatic following from a harmonic trap to an asymmetric  double well collapses the ground state wave to one of the two wells, 
a  FF technique  \cite{Masuda2010,Torrontegui2012}  was applied to {\em avoid} the collapse and achieve perfect, 
balanced  splitting, as required, e.g., for matter-wave interferometry.  
The idea was that for a fast nonadiabatic shortcut, the perturbative effect of the asymmetry becomes negligible. 
The stabilizing effect of interactions was also
characterized within a mean-field treatment.  
In the present chapter, the objective is to send each mode of the initial harmonic trap as fast as possible 
to a different final well, so we needed  a different methodology.  
Instead of FF, which demands an arbitrary control of the potential function 
in position and time, we have restricted the potential to a form with a few controllable
parameters (in practice we have let only two of them evolve in time). Inverse engineering of the Hamiltonian is carried
out for a two-level model using invariants of motion, and the resulting (analytical) Hamiltonian is then mapped 
to real space.  The
discrete Hamiltonian
is useful  as it provides a simple picture to understand and design the dynamics at will.  
The method provides also a good basis to apply OCT, which complements 
invariant-based engineering (see. e.g. \cite{Chen2011a})  
by selecting among the 
fidelity-$1$ protocols
according to other physical requisites.  
As for interactions and nonlinearities,  they will generally spoil a clean multiplexing or demultiplexing
processes, so we have only examined linear dynamics here.    
     
An application of the demultiplexing schemes discussed in this work is the population inversion of the first two levels of the harmonic trap 
without making use of internal state excitations \cite{Bouchoule1999}.  This is useful to avoid decoherence effects induced by decay, 
or for species without an appropriate (isolated two-level)  structure. 
The scheme is based on the three steps shown in Fig. \ref{adiabatic1}. 
A mechanical excitation of the ground state level into the first excited state of a fixed  anharmonic potential 
was implemented by shaking the trap along a trajectory calculated with an OCT algorithm \cite{Bucker2013}.  
Our proposed approach relies instead on a smooth potential deformation. This type of inversion could be applied to interacting
Bose-Einstein condensates
as long as the initial states are pure ground or excited levels. 
The production of twin-atom beams from the excited state is an outstanding application \cite{Bucker2011}. 

Asymmetric double wells may also be used for other state-control operations such as preparing nonequilibrium Fock states 
through a ladder excitation process. The vibrational number may be increased by one at every step. 
Each excitation would start and finish with  demultiplexing and multiplexing 
operations from the harmonic oscillator to the double well and vice versa,  as described in the main text. 
Between them  the two wells are independent and 
their height or width can be adjusted to produce the desired level ordering. 
For an even-to-odd vibrational number transition, 
this requires an inversion of the bias, as in Fig. \ref{adiabatic1};
transitions from odd to even levels  are performed by deepening the left well until the initially occupied level on the right well  
surpasses one of the levels in the left  well. The steps may be repeated until a given Fock state is reached.  
Operating in reverse mode, a given excited state could be taken down to the ground state, as in sideband cooling, 
just with trap deformations, as we will see in Chapter \ref{Chapter6}. 

Open questions left for future work include optimizing 
the robustness of parameter trajectories versus noise and perturbations \cite{Ruschhaupt2012a}, 
or finding time bounds in terms of average energies, similar to the ones 
for harmonic trap expansions \cite{Chen2010a} or transport
\cite{Torrontegui2011}. 
The present results may also be applied 
for  optical waveguide design, as we will see in the next chapter, or to two-dimensional systems as a way to generate vortices.


\chapter{Compact and high conversion efficiency mode-sorting asymmetric Y junction using shortcuts to adiabaticity}
\label{Chapter5}
\lhead{Chapter 5. \emph{Compact and high conversion efficiency mode-sorting asymmetric Y junction using shortcuts to adiabaticity}} 
We propose a compact and high conversion efficiency asymmetric Y junction mode multiplexer/demultiplexer for applications in on-chip mode-division multiplexing. 
Traditionally, mode-sorting is achieved by adiabatically separating the arms of a Y junction.
We shorten the device length using invariant-based inverse engineering 
and achieve better conversion efficiency than the adiabatic device.
\newpage
\section{Introduction}

As optical communications over single-mode optical waveguides are quickly approaching their capacity limits, using multiple spatial modes in optical transmission systems to increase information capacity has attracted lots of attention\cite{Stuart2000,Wang2012}. In mode-division multiplexing (MDM) systems \cite{Berdague1982}, the multiple propagating modes in the system provide the extra degrees of freedom for potential capacity increase. However, to avoid intermodal dispersion, one needs to be able to excite and detect the spatial modes independently in MDM systems. So far, most of the efforts for the multiplexing or demultiplexing in MDM systems are focused on fiber-based systems, but there is also interest in realizing integrated multimode systems \cite{Liu2012, Chen2013, Driscoll2013}. In integrated optical waveguides, the asymmetric Y junction can be designed to function as a mode sorter \cite{Burns1975, Love2012, Riesen2012}. The asymmetric Y junction has a two-modes stem and two diverging single-mode arms with different widths.  When the fundamental (second) mode of the stem propagates through the junction, it evolves into the fundamental mode of the wider (narrower) output arm, and vice versa. The mode sorting behavior can be attributed to the fact that a mode would evolve into the mode of the output arm with the closest effective index \cite{Burns1975}. However, this smooth evolution can only occur when the variation at the junction is slow enough, such that the evolution is adiabatic, reducing the coupling between the local eigenmodes (supermodes) of the structure. However, the adiabatic criterion often leads to a small branching angle between the arms, and thus, a long device length to achieve the desired arm separation. The challenge in the integrated mode-sorting Y junction multiplexer or demultiplexer design is thus to reduce the device lengths while minimizing the cross talk between the arms. 

So far the efforts have been focused on optimizing the device length without violating the adiabatic criterion \cite{Love1996}. There have also been attempts to find the optimal shape function that minimizes the coupling between the supermodes \cite{Sun2009}. These approaches are based on the adiabatic approximation, and a well-known criterion for mode-sorting operation of the asymmetric Y junction is given by the mode conversion factor (MCF) as \cite{Burns1975}
\begin{equation}
\label{mcf}
\text{MCF}=\frac{|\beta_A-\beta_B|}{\theta\gamma_{AB}},
\end{equation}
where $\theta$ is the branching angle of the Y junction arms, $\beta_A$ and $\beta_B$ are the propagation constants of the modes supported by single mode arms A and B, and 
\beq
\gamma_{AB}=0.5\sqrt{(\beta_A+\beta_B)^2-(2k_0n)^2}
\eeq
with $n$ the cladding refractive index and $k_0$ the free-space wavenumber. When the MCF is larger (smaller) than 0.43, an asymmetric Y junction acts as a mode sorter (power divider). For a given material system $n$, branching waveguides dimensions $\beta_A$ and $\beta_B$, and branch separation $D$, the required device length $L=D/\theta$ is limited by $\theta$, obtained from Eq. (\ref{mcf}). Moreover, as long as there is finite coupling between the supermodes in the adiabatic evolution, the conversion efficiency will only be unity at specific operating points \cite{Syms1992, Paloczi2004}. 

Recently, many coherent quantum phenomena have been exploited to implement light manipulation in waveguide structures based on the analogies between quantum mechanics and wave optics \cite{Longhi2009}. At the same time, the development in new ways to manipulate quantum systems with high-fidelity and in a short interaction time using STA \cite{Torrontegui2013d} has inspired the design of a family of novel coupled-wave devices \cite{Tseng2012,Lin2012, Tseng2013, Tseng2013a, Yeih2014, Chien2013}. In particular, the invariant-based inverse engineering approach \cite{Chen2011b, Martinez-Garaot2013}, introduced in Chapter \ref{Chapter4}, provides a versatile tool for the design of fast and robust waveguide couplers \cite{Tseng2013a}, in which the system dynamics are described using the eigenstates of the invariant $I$ corresponding to the system Hamiltonian $H$. While previous works \cite{Tseng2012,Lin2012,Yeih2014,Chien2013} have focused on grating-assisted mode conversion in multimode waveguides, in this chapter, we apply the STA to design short asymmetric Y junction mode multiplexer or demultiplexer beyond the adiabatic limit.

\section{The model}

%
%
%
%
%
\begin{figure}[t]
\centering\includegraphics[width=0.7 \linewidth]{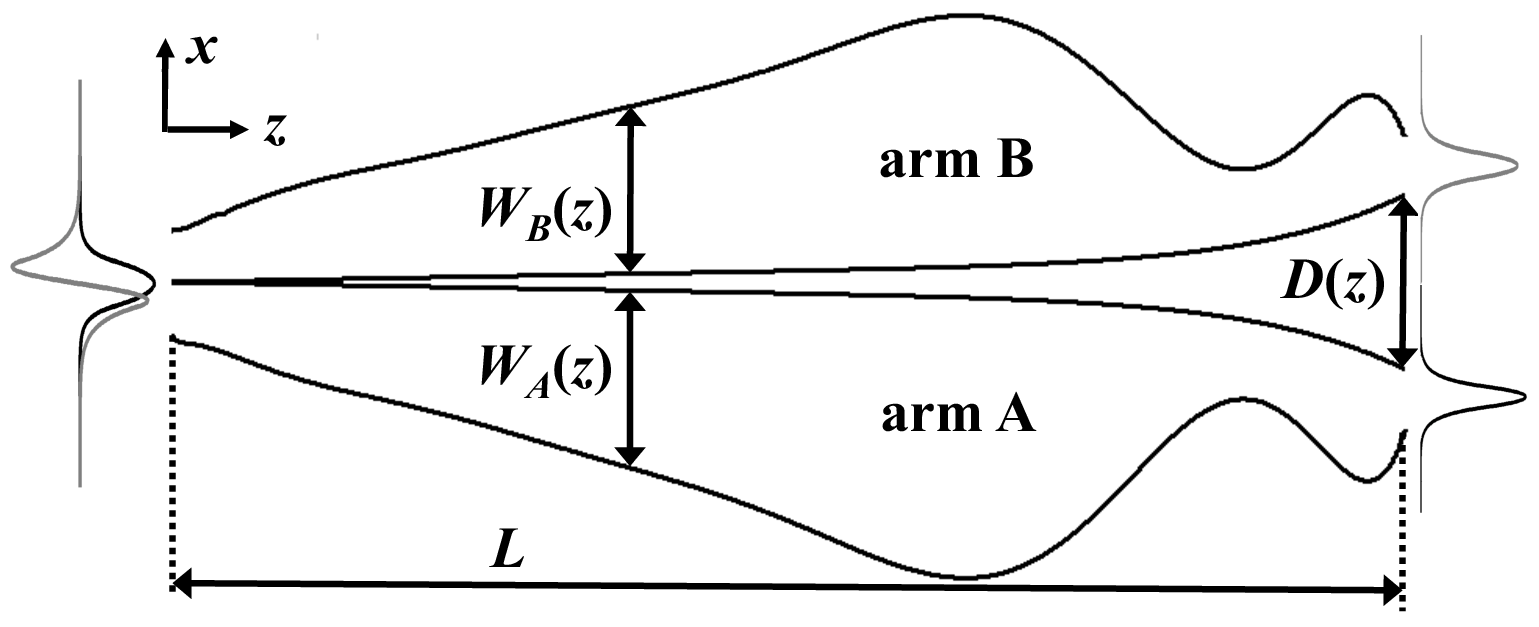}
\caption{Schematic of the asymmetric Y junction.}
\label{scheme_mode_sorting}
\end{figure}
%
%
%
%
%
We consider the asymmetric Y junction shown schematically in Fig. \ref{scheme_mode_sorting},  in which a two-modes stem waveguide evolves to two single-mode waveguides $A$ (wider) and $B$ (narrower) in a length $L$. The evolution of the fundamental modal amplitudes in waveguides $A$ and $B$ can be described by the coupled mode equations as
\begin{equation}
\label{cmt}
\frac{d}{dz}\left[
\begin{array}{c}
A\\
B
\end{array}
\right]=-i\left[
\begin{array}{cc}
\lambda(z) & -\delta(z)\\
-\delta(z) & -\lambda(z)
\end{array}
\right]\left[
\begin{array}{c}
A\\
B
\end{array}
\right]=-iH\left[
\begin{array}{c}
A\\
B
\end{array}
\right],  
\end{equation}
where $\delta$ (real) is the coupling coefficient, and $\lambda=(\beta_B-\beta_A)/2$ describes the mismatch. For the two-modes stem waveguide at $z=0$, $\lambda(0)=0$ and $\delta(0)=\omega_0$.  Solving for the eigenvectors of $H$, we find two adiabatic supermodes, 
\beqa
|a_A\rangle&=&\sin{\alpha}|\Psi_1\rangle+\cos{\alpha}|\Psi_2\rangle, \nonumber
\\ 
|a_B\rangle&=&\cos{\alpha}|\Psi_1\rangle-\sin{\alpha}|\Psi_2\rangle, 
\eeqa
where $|\Psi_1\rangle\equiv$$0\choose 1$, $|\Psi_2\rangle\equiv$$1\choose 0$, and $\alpha=(1/2)\tan^{-1}(\delta/\lambda)$. $\lambda$ and $\delta$ are related to the branch geometry, which is yet to be specified. We impose the boundary conditions
\beqa
\label{bc}
&&\delta(0)=\omega_0, \nonumber \\
&&\lambda(0)=0, \nonumber \\ 
&&\delta(L)=0, \nonumber \\
&&\lambda(L)=\lambda_L,
\eeqa
such that the structure corresponds to a two-modes stem waveguide at $z=0$ and two single-mode waveguides at $z=L$. For the conventional adiabatic Y junction design, the goal is to design the evolution of $\lambda$ and $\delta$ through the device geometry such that the coupling between $|a_A\rangle$ and $|a_B\rangle$ are minimized. When the adiabatic criterion is not satisfied, and there is a finite coupling between $|a_A\rangle$ and $|a_B\rangle$, the mode-sorting performance will deteriorate.  In the following, we use the invariant-based inverse engineering approach to design a protocol in which the mode-sorting is achieved at a shorter length than required by the adiabatic criterion. 

\subsection{Invariant-based inverse engineering}
Replacing the spatial variation $z$ with the temporal variation $t$, Eq. (\ref{cmt}) is equivalent to the time-dependent Schr\"{o}dinger equation ($\hbar=1$) describing the interaction dynamics of a two-state system, and $H$ is the Hamiltonian.  

As we have seen in the previous chapter, associated with $H$ there are Hermitian
dynamical invariants $I(t)$, fulfilling 
\beq
\label{inva}
\partial_t I+\frac{1}{i}[I,H]=0, 
\eeq
so that their expectation values remain
constant. $I$ can be written as (where $t$ is replaced by $z$ and hereafter)\cite{Chen2011b}
\begin{equation}
\label{I}
 I(z)=\frac{1}{2}
\left(
                \begin{array}{cc}
                  \cos\theta & \sin\theta e^{i\varphi} \\
                  \sin\theta e^{-i\varphi} & -\cos\theta
                \end{array}
\right),
\end{equation}
where $\theta\equiv\theta(z)$
and $ \varphi\equiv\varphi(z)$ are $z$-dependent angles. 
The eigenstates of the invariant $I(z)$ satisfy
$I(z)|\phi_{n}(z)\rangle=\lambda_{n}|\phi_{n}(z)\rangle$, and they can be
written as
\beqa
\label{phiz} 
&&|\phi_{+}(z)\rangle=\left(\begin{array}{c}
                      \cos\frac{\theta}{2}e^{-i\varphi} \\
                      \sin\frac{\theta}{2}
                    \end{array}
                    \right), \nonumber \\
&&|\phi_{-}(z)\rangle=\left(\begin{array}{c}
                      \sin\frac{\theta}{2} \\
                      -\cos\frac{\theta}{2}e^{i\varphi}
                    \end{array}
                    \right).
\eeqa
An invariant $I(z)$ of $H(z)$ satisfies $i\hbar\partial_z(I(z)|\Psi(z)\rangle)=H(z)(I(z)|\Psi(z)\rangle) $\cite{Lewis1969}. According to the Lewis-Riesenfeld theory, the state of the system can be written as 
\beq
|\Psi(z)\rangle=\Sigma_{n}c_{n}e^{i\gamma_{n}(z)}|\phi_{n}(z)\rangle,
\eeq
where the $c_{n}$ 
are z-independent amplitudes, and the $\gamma_{n}(z)$ are
Lewis-Riesenfeld phases. The z-independent $c_n$ implies that the system state will follow the eigenstate of the invariant exactly without mutual coupling. 

To engineer the Hamiltonian $H(z)$ such that  the mode sorting is exact, we will proceed as in Chapter \ref{Chapter4} (Sec. \ref{lr}). We design the invariant first and then obtain the Hamiltonian from it. Applying the boundary conditions $[H(z), I(z)]=0$ at $z=0$ and $z=L$ such that the eigenvectors of $H(z)$ and $I(z)$ coincide at the input and output, the invariant will drive the input eigenstates of $H(z)$ to the output eigenstates of  $H(z)$ exactly. Using the
invariance condition (\ref{inva}), we find 
\beqa
\label{delta}
&&\delta(z)=-\dot{\theta} (z)/ \sin\varphi (z), \\
\label{lambda}
&&\lambda(z)=-\delta(z)\cot\theta(z)\cos\varphi(z)-\dot{\varphi}(z).
\eeqa
Using the commutativity of $H(z)$ and $I(z)$  at the input and output and Eq. (\ref{bc}), we obtain
\beqa
\label{bc2}
&&\theta(0)=\pi/2, \nonumber \\
&&\varphi(0)=\pi, \nonumber \\
&&\theta(L)=0, \nonumber \\ 
&&\dot{\theta}(L)=0.
\eeqa
These conditions lead to indeterminacies in Eqs. (\ref{delta}) and (\ref{lambda}), so we apply l'H\^{o}pital's rule repeatedly and find the additional boundary conditions \cite{Martinez-Garaot2013}
\beqa
\label{bc3}
&&\dot{\theta}(0)=\ddot{\theta}(0)=\dot{\varphi}(0)=0, \nonumber \\
&&\dddot{\theta}(0)=-\omega_0\dot{\lambda}(0), \nonumber \\
&&\ddot{\varphi}(0)=-\dot{\lambda}(0), \nonumber \\
&&\varphi(L)=\pi/2, \nonumber \\
&&\dot{\varphi}(L)=-\lambda_L/3,
\eeqa
with $\dot{\lambda}(0)\neq0$. With the boundary conditions in Eq. (\ref{bc3}), the evolution of the invariant parameters $\theta(z)$ and $\varphi(z)$ can be obtained through interpolation, assuming a polynomial ansatz (see Chapter \ref{Chapter4}). Then, the Hamiltonian functions $\delta$ and $\lambda$ can be obtained from Eqs. (\ref{delta}) and (\ref{lambda}). We finally  use the simplex-based mapping method described in Sec. \ref{mapp} to map the designed Hamiltonian to a realizable waveguide geometry. Device performance will be related to the choice of the interpolation ansatz. It is beyond the scope of this chapter to categorize or evaluate the various ansatz that are possible; rather, we focus on the polynomial ansatz to demonstrate the device concept. 

\section{Numerical results}
%
%
%
%
%
\begin{figure}[t]
\centering\includegraphics[width=0.7 \linewidth]{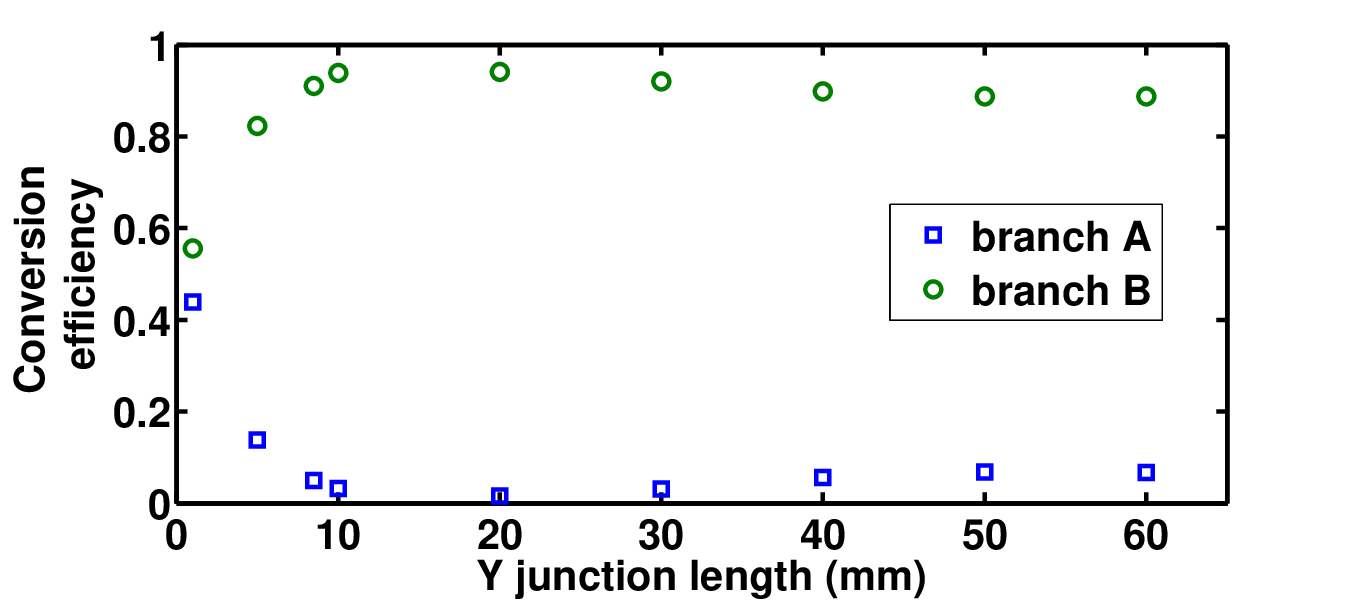}
\caption{Conversion efficiencies of a linearly separating Y junction using the second mode as the input for different device lengths}
\label{c_e}
\end{figure}
%
%
%
%
%
%

Now we illustrate the design of a compact mode-sorting asymmetric Y junction in a conventional planar integrated optics platform, and perform beam propagation method (BPM) simulations to verify the designs. The scalar 2D BPM code used in the simulations solves the scalar and paraxial wave equation using the finite difference scheme with the transparent boundary condition. We choose a polymer channel waveguide structure for beam propagation simulations. The design parameters are chosen as follows: 3 $\mu$m thick SiO$_2$ ($n=$1.46) on a Si ($n=3.48$) wafer is used for the bottom cladding layer, the core consists of a 2.4 $\mu$m layer of BCB ($n=1.53$), and the upper cladding is epoxy ($n=1.50$). The device is simulated at 1.55 $\mu$m input wavelength and the TE polarization. Subsequent analysis are performed on the 2D structure obtained using the effective index method. For the Y junction input and outputs, we choose an input stem waveguide width of 5.8 $\mu$m supporting two modes, and the output single-mode waveguides widths are $W_A(L)$=3.5 $\mu$m and $W_B(L)$=3.29 $\mu$m. We target a final waveguide separation $D(L)$ of 10 $\mu$m so that the coupling between the output branches is negligible. Substituting the corresponding waveguide parameters into Eq. (\ref{mcf}), MCF=0.1277/$\theta$ (with $\theta$ in degrees) indicating the device is a mode sorter for $\theta<0.3^{\circ}$. For a conventional linearly separating adiabatic Y junction, this corresponds to a device length of larger than 2 mm to achieve a final separation $D(L)$ of 10 $\mu$m. In Fig. \ref{c_e}, we show the simulated fractional power in the fundamental modes (conversion efficiency) of waveguides A and B using the second mode as the input for different device lengths. When the device length is greater than 10 mm, the mode-sorting characteristics are well established. The transition from power divider to mode sorter at around 2 mm predicted by MCF calculations is also evident. We also observe that the conversion efficiency starts to fall and will oscillate when the length keeps increasing beyond 10 mm, as a result of finite coupling between the supermodes \cite{Sun2009}.  
%
%
%
%
\begin{figure}[t]
\centering\includegraphics[width=0.7 \linewidth]{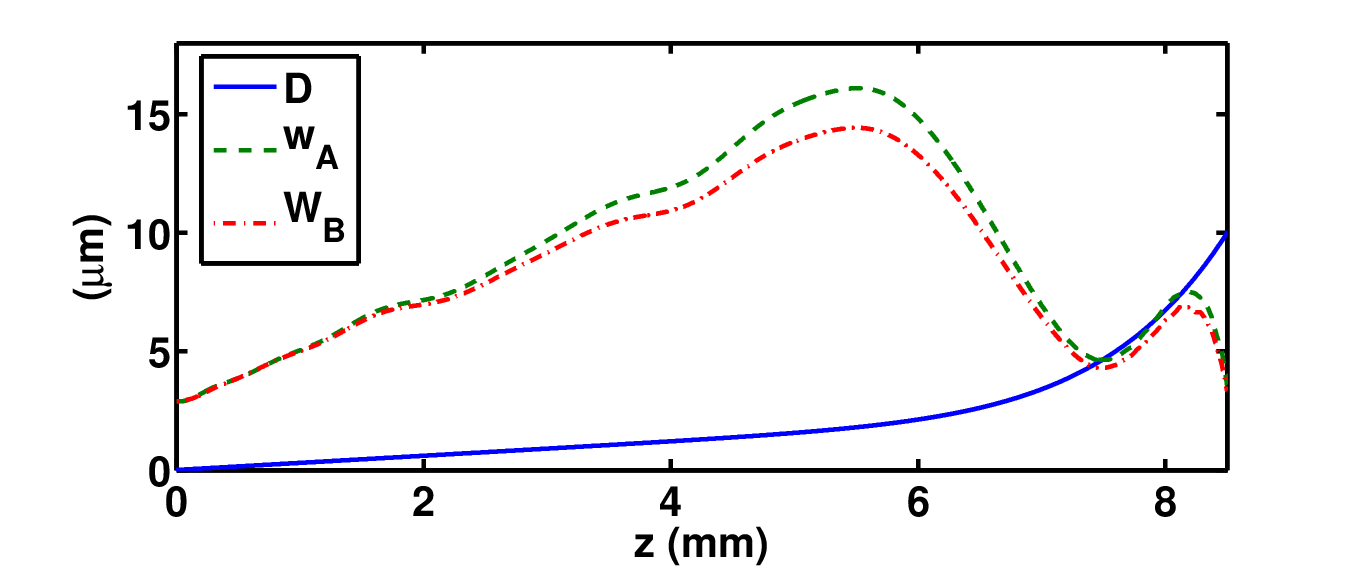}
\caption{Parameters for the invariant-based Y junction.}
\label{parameters}
\end{figure}
%
%
%
%
%
%
%
%
%
%
\begin{figure}[t]
\centering\includegraphics[width=0.7 \linewidth]{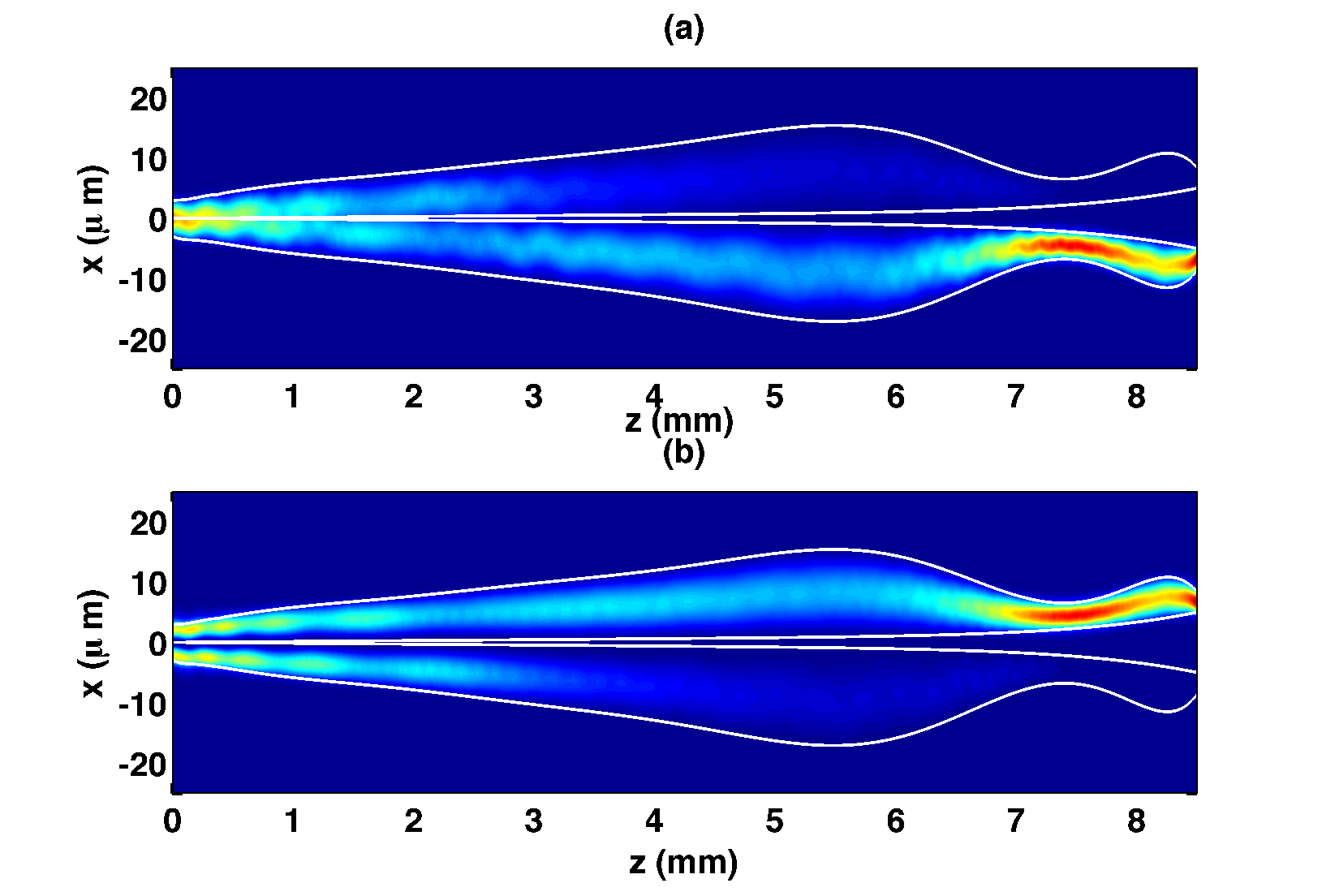}
\caption{Mode-sorting operation of the invariant-based Y junction. Input (a) fundamental mode (b) second mode.}
\label{mode_sorting}
\end{figure}
%
%
%
%
%

\begin{figure}[t]
\centering\includegraphics[width=0.7 \linewidth]{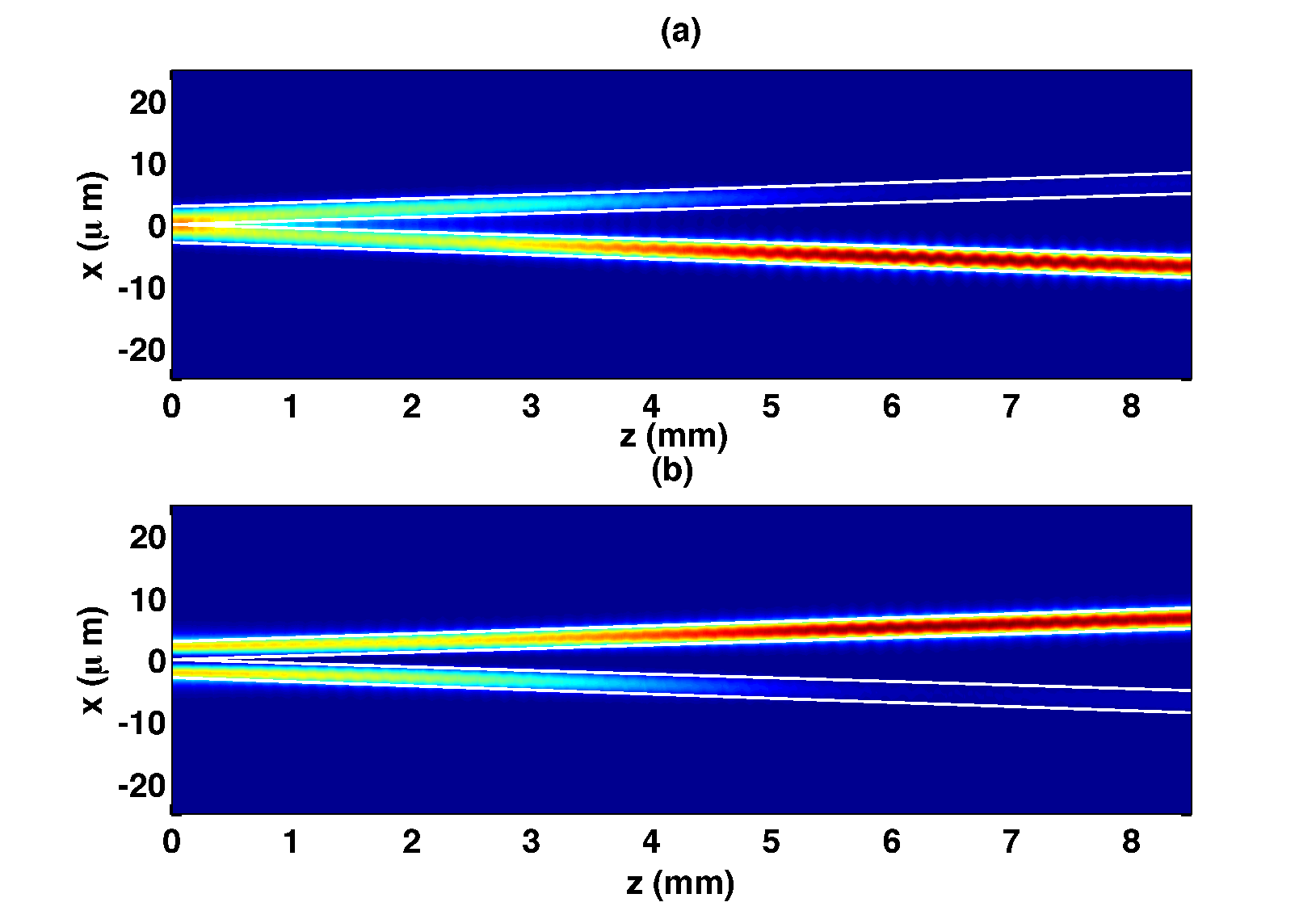}
\caption{Mode-sorting operation of the linearly separating Y junction. Input (a) fundamental mode (b) second mode.}
\label{mode_sorting_linear}
\end{figure}

\begin{figure}[t]
\centering\includegraphics[width=0.7 \linewidth]{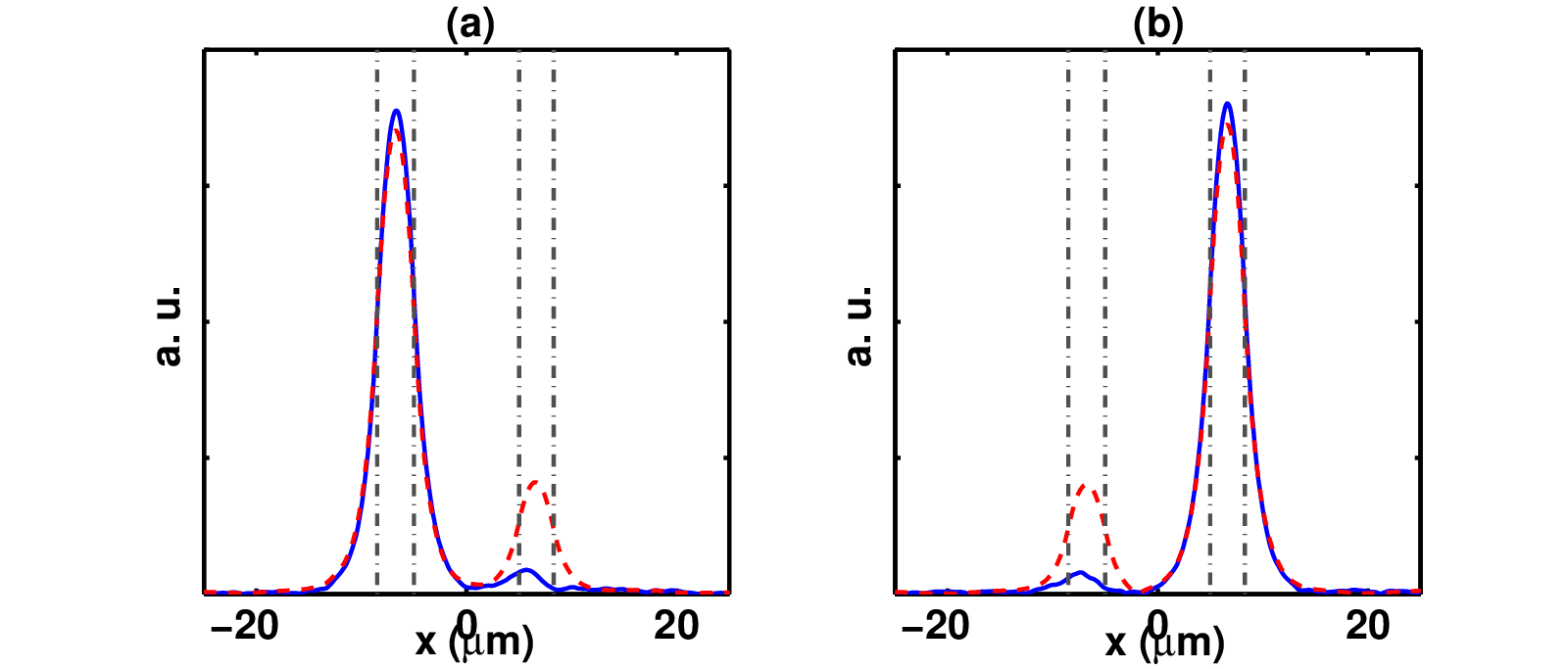}
\caption{Output field profile of the Y junctions. Solid: invariant-based. Dashed: linearly separating. Dash-dotted: waveguide walls. Input (a) fundamental mode (b) second mode.}
\label{output_field}
\end{figure}

\begin{figure}[t]
\centering\includegraphics[width=0.7 \linewidth]{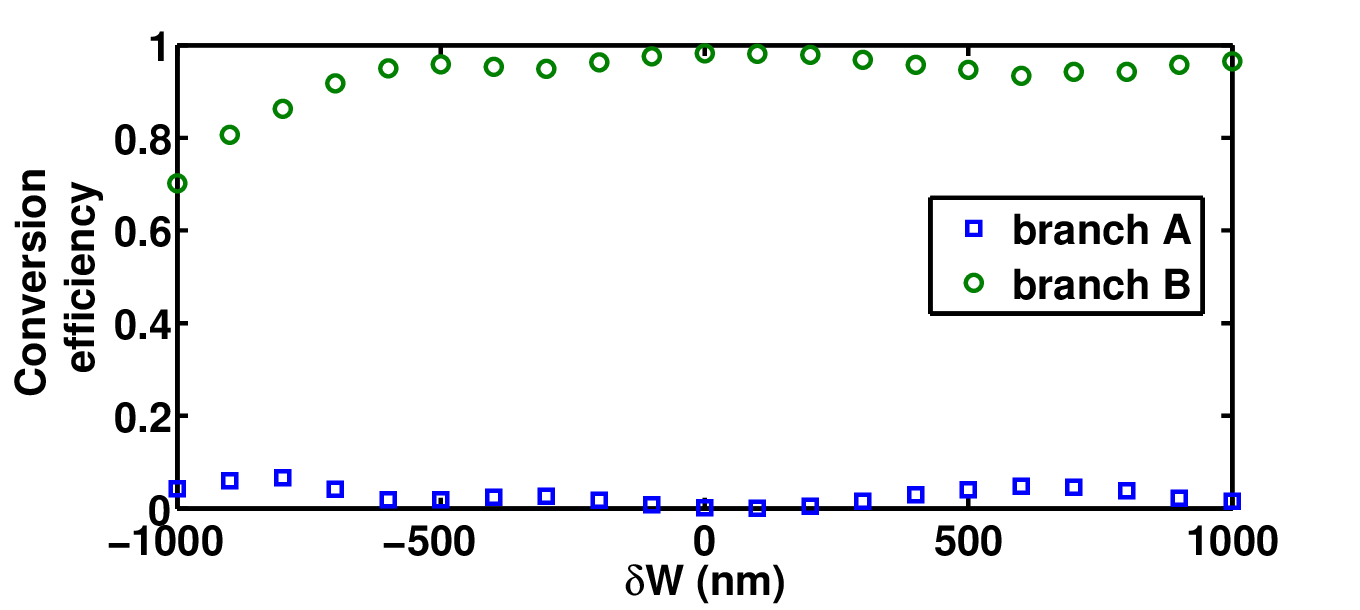}
\caption{Conversion efficiencies as a function of width variation using the second mode as the input.}
\label{tolerance}
\end{figure}

For the invariant-based design, the boundary conditions in Eq. (\ref{bc}) are fixed by the waveguide parameters at the device input and output. To map the Hamiltonian to the waveguide parameters of the Y junction, we choose the widths of waveguides $W_A(z)$ and $W_B(z)$ and the separation $D(z)$ shown in Fig. \ref{scheme_mode_sorting} as the free parameters in the simplex search. The resulting parameters are shown in Fig. \ref{parameters} for a $L=8.5$ mm device.  The corresponding Y junction geometry is shown in Fig. \ref{mode_sorting}. In Fig. \ref{mode_sorting}(a), the BPM results show that the fundamental mode has evolved to waveguide A at the output. And the evolution of the second mode to waveguide B is shown in Fig. \ref{mode_sorting}(b). We also show the BPM results for the linearly separating adiabatic Y junction of the same length in Fig. \ref{mode_sorting_linear}. In Fig. \ref{output_field}, we compare the output field of the invariant-based mode sorter and the linear mode sorter, both at a length of 8.5 mm. The conversion efficiency of the invariant-based design is calculated to be 0.98 for both modes while the linearly separating design is 0.92 for both modes. The insertion loss of the invariant-based design are 0.267 dB and 0.185 dB for the fundamental and the second modes, respectively, and 0.481 dB and 0.604 dB for the linearly separating design. The higher insertion loss of the linearly separating design can be attributed to coupling into the radiation modes. On the other hand, the evolution of the invariant-based design should follow the eigenstates of the invariant exactly without coupling into the radiation modes. The observed loss can be attributed to small coupling into the radiation modes because the ideal protocol is only approximately mapped to the coordinate space model in the simplex-based mapping \cite{Martinez-Garaot2013}. This also results in the conversion efficiency being less than 1. Although the width of the invariant-based design is larger than the linearly separating design, we note from Fig. \ref{c_e} that the conversion efficiency of the linearly separating design would not reach 0.98 even when the length of the junction is increased to 60 mm. As a result, the invariant-based design can achieve high conversion efficiency with a more compact device footprint. The fabrication tolerance is studied by adding width variations $\delta W$ to $W_A$ and $W_B$ while keeping $D$ unchanged in the simulations. The resulting conversion efficiencies for different $\delta W$ using the second mode as the input is shown in Fig. \ref{tolerance}, indicating that the proposed device has a large fabrication tolerance better than 1000 nm. 

%
%
%
%
%
%
%
%
%
%
%
%
%
%
%
%
%
%
%
%
%
%
%
%
%
%
%
%
%
%

\section{Conclusions}

In conclusion, we demonstrated that the invariant-based inverse engineering approach can be applied successfully to asymmetric Y junction design. By describing the system dynamics using the dynamical invariants, the system evolution can be engineered to achieve mode sorting in a short distance. The compact design exhibits a higher conversion efficiency than the conventional adiabatic design at a shorter device length.


\chapter{Fast bias inversion of a double well without residual particle excitation}
\label{Chapter6}
\lhead{Chapter 6. \emph{Fast bias inversion of a double well without residual particle excitation}} 
We design  fast bias inversions of an asymmetric double well so that the lowest states in each well stay in the same well 
they started,  
free from residual motional excitation.
This cannot be done adiabatically, and a sudden bias switch produces 
in general motional excitation. The residual excitation is suppressed by complementing a 
predetermined fast bias change with  a
linear ramp whose  time-dependent slope 
compensates for the displacement of the wells. The process, combined with vibrational multiplexing
and demultiplexing, can produce vibrational state inversion without exciting internal states, 
just by deforming the trap.    
\newpage
\section{Introduction}

\begin{figure}[t]
\begin{center}
\hspace*{-.3cm}\includegraphics[width=0.95 \linewidth]{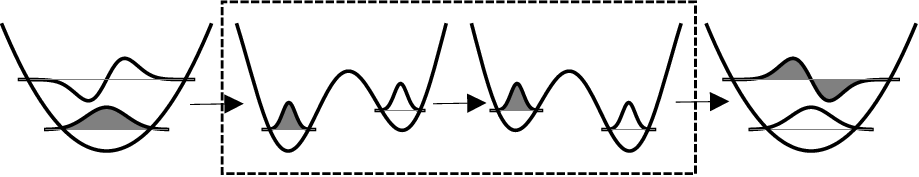}
\end{center}
\caption{\label{general_scheme}
Schematic  representation of demultiplexing (left arrow), bias inversion (framed in dashed line, central arrow), and multiplexing (right arrow).
The densities of two one-atom eigenstates are represented in all potentials. In the harmonic potentials 
(unframed potentials on the left and right charts) the states are the ground state and first excited state. 
In the two central charts with tilted double wells the states are the lowest for each well. 
The color (white or gray) indicates how they would evolve sequentially
following  the fast protocol described in the text. For example, the gray state is initially the ground state of the harmonic oscillator, then 
it becomes the lowest state of the left well, and remains being the lowest state of that well after the bias inversion. In the last step it 
becomes the first excited state
of the final harmonic oscillator.}
\end{figure}

In Chapter \ref{Chapter4} a protocol to realize fast vibrational state multiplexing or demultiplexing  of ultra cold atoms was introduced.  By a properly designed time-dependent potential deformation between a harmonic trap and a biased double well, the states of a single atom in a harmonic trap can be dynamically mapped into states
localized at each well  (vibrational demultiplexing; see the left arrow in Fig. \ref{general_scheme}), or vice versa (multiplexing; see the right arrow in Fig. \ref{general_scheme}), faster than adiabatically and without residual  excitation at the final time. 
It was suggested that these processes may be combined with a bias inversion to produce (i) state inversions, from the ground to the first excited state of the harmonic trap and vice versa, 
based on trap deformations only 
(see the evolution of gray and white states in Fig. \ref{general_scheme}), or (ii) to produce vibrationally excited Fock states from an initial ground state \cite{Martinez-Garaot2013}.  
These are basic operations to implement quantum
information processing  and fundamental studies. 
Thus the possibility to perform them without exciting internal atomic states as an intermediate step  
is of much interest. For trapped ions in particular, this amounts to a species-independent 
approach based entirely on the charge and electric forces. In contrast, $\pi$-pulse sequences require specific lasers for each system and a suitable level structure. 
In general, i.e., both for ions and neutral atoms, a method not using internal-state excitation   
suppresses decoherence and random kicks due to spontaneous decay. 
They may be important limiting factors 
to preserve  quantum dynamics with optical transitions \cite{Leibfried2003}. 
%
%
%
%
%
%
%
%

Among the possibilities to avoid decay from an intermediate state in a transition among motional states, 
one might think of using Stimulated Raman adiabatic passage (STIRAP) \cite{Bergmann2015}, which in principle 
does not populate the upper, intermediate state. 
This technique, however, is best suited for transitions involving a change in internal state, and its application to purely vibrational transitions (within the same internal state) is not straightforward.  Numerical simulations show that several motional states are  populated \cite{vogelius2005cooling}, and 
in fact the experimental applications of STIRAP for
trapped ions have been only used for inducing carrier or sideband
transitions that involve changing the internal state \cite{Gebert2015}.
 
The objective of this chapter 
is to design fast controlled bias inversions of a double well so that the lowest states 
in each well remain as  lowest states  
without residual excitation. Unlike multiplexing, 
however, there is no truly adiabatic slow process that achieves this state transformation. In the bias
inversion depicted within the dashed-line frame of  Fig. \ref{general_scheme},  for example, a slow bias change would preserve the 
order of the states according to their energy,
so that the states represented in the third potential configuration would be interchanged (i.e., the gray state in the right well, and the white one in the left well).     
Nevertheless, in the limit in which the two wells are effectively independent, which in practice means, for times 
shorter than the tunneling time among the wells, the intended state transition might indeed be done slowly enough to be considered adiabatic. If we approximate  each ``isolated'' well by a harmonic oscillator, the intended transformation amounts to a ``horizontal'' displacement along the interwell axis together with a rising or lowering of the energy of the wells.  The latter effects, however, do not affect the intrawell dynamics, so we may focus on the displacement. In other words, within the stated approximations the bias inversion amounts to  transporting  
a particle in a harmonic potential. Thus, to achieve a fast transition 
without residual excitation we may use STA designed to perform fast transport \cite{Torrontegui2013d}.
Specifically, we shall use a compensating-force approach \cite{Torrontegui2011, Palmero2013}, equivalent to the FF scaling technique \cite{Masuda2010}, based on adding  to the potential a linear ramp with time-dependent
slope  to compensate for the effect of the trap motion in the noninertial frame of the trap. 
We shall compare this approach with a sudden bias switch,  a FAQUAD 
approach \cite{Martinez-Garaot2015}, or a smooth polynomial connection without compensation. 
In Sec. \ref{compensating} we introduce the compensating-force approach and Sec. \ref{alme} describes the  alternative methods.
In Sec. \ref{exam} numerical examples are presented with parameters appropriate for trapped ions in 
multisegmented, linear Paul traps, and for neutral atoms in optical traps.  
Finally, in Sec. \ref{discussion} we discuss the results and open questions.
\section{Compensating-force approach}
\label{compensating}

If the double-well potential with nearly independent wells is subjected to a bias inversion such that the trap frequencies of each well are essentially equal and constant
throughout, 
and the trajectories of the well minima move in parallel, the process may be described by a parallel transport 
of two rigid harmonic oscillators, one for each well. 
The Hamiltonian potential near the  minima may be approximated as   
\beq
\label{harmonic_V}
V_{0}(x-x_0)=\frac{1}{2}m\Omega_{0}^2(x-x_{0})^2,
\eeq
where $\Omega_{0}$ is the angular frequency and $x_0=x_0(t)$ is the common notation for either of the two minima.\footnote{We disregard purely time-dependent functions in each well. Differential
phases among the wells depending on these functions can be ignored since the traps are assumed to be independent.} When needed, we may distinguish the minima as  $x_{0,\pm}$, with $x_{0,+}>x_{0,-}$. 
The Hamiltonian $H_{0}=p^2/2m+V_{0}(x-x_0)$ has eigenenergies 
\beq
E_n=\left(n+\frac{1}{2}\right)\hbar \Omega_{0},
\eeq
and well-known normalized eigenstates $\phi_n(x-x_0)$, proportional to Hermite polynomials \cite{schiffquantum}.  

Adding to the Hamiltonian a linear term with an appropriate time-dependent slope, 
the noninertial effect of the motion of the well will be compensated in the trap frame \cite{Torrontegui2011, Palmero2013}. 
To define the trap frame consider the following  position and momentum displacement unitary operator
\beq
\label{u_t}
\mathcal{U}=e^{ipx_0(t)/\hbar}e^{-im\dot x_0(t)x/\hbar},
\eeq
where the overdot represents a time derivative. 
Starting from the Schr\"odinger equation
\beq
i\hbar\partial_t|\psi\rangle=H_{0}|\psi\rangle,
\eeq
the transformed wave function $|\Phi\rangle=\mathcal{U}|\psi\rangle$ obeys
\beq
\label{int_H}
i\hbar\partial_t|\Phi\rangle=\mathcal{U}H_{0}\mathcal{U}^\dag|\Phi\rangle+i\hbar(\partial_t\mathcal{U})\mathcal{U}^\dag|\Phi\rangle=H'_{0}|\Phi\rangle, 
\eeq
where 
the IP (trap frame) 
Hamiltonian is 
\beq
\label{H_trap}
H'_{0}=\frac{p^2}{2m}+V_{0}(x)+mx \ddot x_0+\frac{1}{2}m{\dot x_0}^2,
\eeq
and $V_0(x)=\frac{1}{2}m\Omega_{0}^2x^2$.
The term $m{\dot x_0}^2/2$ only depends on time; it does not affect the dynamics and can be ignored. 
To compensate the motion of the trap, we add $-mx\ddot x_0$ to $H_{0}$. This produces $-m(x+x_0)\ddot{x}_0$ in the trap frame and the resulting potential in that frame is reduced to $V_0(x)$, 
again neglecting purely time-dependent functions.  $V_0(x)$ does not depend  on time, so any stationary state in this trap frame will remain stationary, and excitations are avoided. 
\section{Alternative methods\label{alme}}

In this section we discuss three simple alternative approaches to perform the bias inversion. They are all quite natural and simple approaches
whose performance can be compared to that of the compensating force approach.  

\subsection{Sudden approach}
In the sudden approach the potential is changed abruptly from the initial  to the final configuration, but  
the state of the system remains unchanged immediately after the potential change (in general it will evolve afterwards). 
If the target state is $\psi_{tar}$ the resulting fidelity is 
\beq
\label{F_sudden}
F_s=|\langle \psi(0)|\psi_{tar}\rangle|.
\eeq

\subsection{Fast quasi-adiabatic approach}

A quasi-adiabatic method to speed up adiabatic processes when there is one control parameter $\lambda(t)$ 
is based on distributing the adiabaticity parameter homogeneously in time (see Chapter \ref{Chapter3}). 
For instantaneous levels 0 and 1 this means 
\beq
\label{adia}
\hbar \left | \frac{\la \phi_0|\partial_t\phi_1\ra}{E_0-E_1}\right | =c,
\eeq
where the instantaneous eigenstates $\phi_0$, $\phi_1$ and eigenenergies $E_0$, $E_1$ 
depend on time through their dependence on $\lambda$,
and $c$ is constant. By the chain rule this becomes a first order differential equation for $\lambda(t)$, and  
$c$ is set so that the boundary conditions for $\lambda(t)$ at initial time, $t=0$,  and final time $t_f$ are satisfied. 
In the transport of a particle with a harmonic oscillator of angular frequency $\Omega_0$ 
centered at $x_0(t)$ we  set $\lambda(t)=x_0(t)$.  
Using the energies and eigenstates of the first two levels of the harmonic oscillator in Eq. (\ref{adia}), the FAQUAD condition 
becomes simply
\beq
\label{adia2}
\frac{m \dot x_0(t)}{\sqrt{2\hbar m \Omega_0}}=c.
\eeq
The solution is the linear connection 
\beq
x_0(t)=x_0(0)+[x_0(t_f)-x_0(0)]\frac{t}{t_f}.
\eeq
The minimal $t_f$ for which a zero of excitation energy 
appears is $2\pi/\Omega_0$ \cite{Bowler2012,Martinez-Garaot2015}. 
 
\subsection{Polynomial connection without compensation}
The final and initial values of the control parameter may as well be smoothly connected 
without applying any compensation, for example, using a fifth order polynomial that assures the vanishing 
of first and second derivatives of the parameter at the boundary times.      
\section{Examples\label{exam}}

In the following examples, the potentials and parameters are adapted for a trapped ion in a multisegmented Paul trap, and for 
a neutral atom in a dipole trap.  

\subsection{Trapped ions}

For a trapped ion we consider a simple double-well potential of the form
\beq
\label{V_ions}
V(x,t)=\beta x^4+ \alpha x^2+\gamma x,
\eeq
with $\alpha(t)<0$ and $\beta(t)>0$ \cite{Home2006,Nizamani2012,Kaufmann2014}.  $\alpha$ and $\beta$ are assumed to be constant 
and $\gamma\equiv\gamma(t)$ time dependent. The bias inversion implies  a change of sign of $\gamma(t)$ from $\gamma_0>0$ to $-\gamma_0$. 

From 
$
\frac{\partial V}{\partial x} =0
$
the condition for the extrema becomes 
\beq
4\beta x^3+2\alpha x+\gamma=0.
\eeq
It is useful to define
\beqa
\label{const_1}
A&=&0,\nonumber \\ 
B&=&\frac{2\alpha}{4\beta},\nonumber \\ 
C&=&\frac{\gamma}{4\beta},
\eeqa
and
\beqa
\label{const_2}
Q&=&\frac{A^2-3B}{9},\nonumber \\
R&=&\frac{2A^3-9AB+27C}{54}.
\eeqa
When $R^2<Q^3$ there are two minima and one maximum. With  $\alpha<0$ and $\beta>0$, 
this is satisfied for  
\beq
\label{limits_gamma}
|\gamma|<\left(\frac{2}{3}\right)^{3/2}\sqrt{-\frac{\alpha^3}{\beta}}.
\eeq
The trajectories of the minima are
\label{sol_extrema}
\beqa
\label{ext_1}
x_{0,\pm}=-2\sqrt{Q}\cos\left({\frac{\theta+(1\pm1) \pi}{3}}\right)-\frac{A}{3},
\eeqa
where $\theta=\arccos\left ({\frac{R}{\sqrt{Q^3}}}\right )$, $0\leqslant \theta \leqslant \pi$ 
and the roots are taken to be positive.
Up to second order in $\gamma$ they are 
\beqa
\label{trajec_approx_1}
x_{0,-}&\approx&-\frac{1}{\sqrt{2}}\sqrt{-\frac{\alpha}{\beta}}+\frac{\gamma}{4\alpha}-\frac{3\gamma^2\sqrt{-\alpha\beta}}{16\sqrt{2}\alpha^3}, \\
\label{trajec_approx_2}
x_{0,+}&\approx&\frac{1}{\sqrt{2}}\sqrt{-\frac{\alpha}{\beta}}+\frac{\gamma}{4\alpha}+\frac{3\gamma^2\sqrt{-\alpha\beta}}{16\sqrt{2}\alpha^3}.
\eeqa
The quadratic term in $\gamma$ is negligible with respect to the linear term when  
\beq
\label{wef_bias}
|\gamma| \ll \frac{4\sqrt{2}}{3}\sqrt{-\frac{\alpha^3}{\beta}},
\eeq
which implies that the trajectories for the minima move in parallel.
Note that this inequality automatically implies the one in Eq. (\ref{limits_gamma}). 
Neglecting the quadratic term, the two minima are given by 
\beq
x_{0,\pm}=\pm\frac{1}{\sqrt{2}}\sqrt{-\frac{\alpha}{\beta}}+\frac{\gamma}{4\alpha}.
\eeq
The distance between the minima is 
\beqa
\label{D_minima}
D&=&2\sqrt{Q}\left\{ \cos{ \left( \frac{\theta}{3} \right) }+ \sin{\left [ \frac{1}{6} ( \pi + 2 \theta)\right ]} \right \}
\nonumber\\
&\approx& \sqrt{2} \sqrt{-\frac{\alpha}{\beta}}+ \frac{3\sqrt{-\alpha \beta}}{8 \sqrt{2} \alpha^3} \gamma^2. 
\eeqa
We may also compute the energy bias between the two wells as
\beq
\label{bias}
\delta=\gamma D.
\eeq
The distance travelled by each well is, when (\ref{wef_bias}) is fulfilled, 
$d=\gamma_0/(2\alpha)$ 
[see Eqs. (\ref{trajec_approx_1}) and (\ref{trajec_approx_2})], and
the effective frequency at each minimum  
\beq
\label{w_ef}
\omega_{0}=\sqrt{\frac{1}{m}\left( \frac{d^2V}{dx^2}\right)_{x=x_{0}}}.
\eeq
For Eq. (\ref{V_ions}) the effective frequencies are
\beqa
\label{w_ef_exp}
\omega_{0,\pm}&=&\sqrt{\!\frac{2}{m}} \sqrt{\!\alpha\!+\!\frac{2}{3}\beta \left \{\!A\!+\! 6\sqrt{Q} \cos\!{\left[\frac{1}{3}\! \left (\! \frac{\theta+(1\pm 1)\pi}{3}\! \right)\!\right] }\! \right\}^{\!2}\!}
\nonumber\\
\label{w_ef2}
&\approx& 2 \sqrt{-\frac{\alpha}{m}} \mp \frac{3}{2\sqrt{2}}\sqrt{\frac{\beta}{\alpha^2 m}} \gamma.
\eeqa
Hence, comparing the two terms, the condition for the frequencies to be essentially constant 
\beq
\omega_{0,\mp}\approx\Omega_0\equiv2 \sqrt{-\frac{\alpha}{m}}
\eeq
is again the inequality in  Eq. (\ref{wef_bias}). 

\begin{figure}[t]
\begin{center}
\includegraphics[width=0.5 \linewidth]{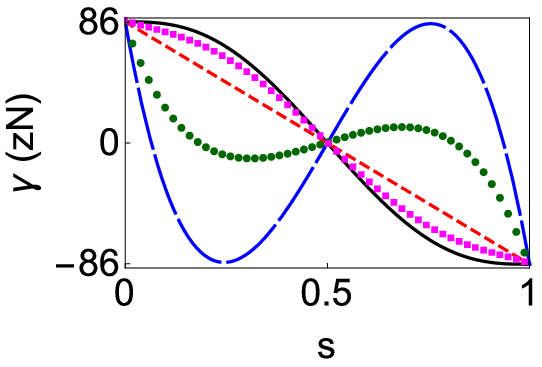}
\end{center}
\caption{\label{displacement_i}
$\gamma$ versus $s=t/t_f$ for the polynomial in Eq. (\ref{connection_i}) (solid black line) and FAQUAD (short-dashed red line). 
$\gamma_0=86.4$ zN, $\gamma(t_f)=-\gamma_0$, $\alpha=-4.7$ pN/m, and $\beta=5.2$ mN/m$^3$.
Also shown are the different effective slopes adding a compensation to the polynomial, 
$\gamma_{eff}(t)=\gamma(t)-m\ddot{\gamma}(t)/(4\alpha)$, 
for the mass of $^{9}$Be$^{+}$ and times $t_f=0.07$ $\mu$s (long-dashed blue line);  $t_f=0.1$ $\mu$s (green dots);  
and $t_f=0.3$ $\mu$s (magenta squares). 
}
\end{figure}

In the regime where the inequality (\ref{wef_bias}) holds, we can apply the compensating force approach to implement a fast bias inversion.
Since the compensating term $-mx\ddot x_0$ is equal for both harmonic traps, we add it to  
$V$ in Eq. (\ref{V_ions}), and the resulting Hamiltonian $H$ is
\beq
\label{H_comp_ions}
H=\frac{p^2}{2m}+\beta x^4+ \alpha x^2+ (\gamma-m\ddot x_0) x.
\eeq
Note that the compensation amounts to changing the time dependence of the slope 
of the linear term from 
the reference process defined by $\gamma(t)$ to 
\beq
\gamma_{eff}(t)\equiv\gamma(t)-m\ddot x_0(t)=\frac{\gamma(t)-m\ddot{\gamma}(t)}{4\alpha}.  
\eeq
To set $\gamma(t)$ we design a connection between the initial and final configurations.
First note the boundary conditions  
\beqa
\label{b_c1}
&&\gamma (0)=\gamma_0>0, \nonumber \\
&&\gamma (t_f)=-\gamma_0,
\eeqa
which we complement by 
\beqa
\label{b_c2}
&&\dot \gamma (t_b)=0, \nonumber \\
&&\ddot \gamma (t_b)=0, \nonumber \\ 
&&t_b=0,t_f, 
\eeqa
so that $\dot{x}_0(t_b)=\ddot{x}_0(t_b)=0$. This implies that ${\cal{U}}(t_b)=e^{ip x_0(t_b)/\hbar}$ and  $\dot{\cal{U}}(t_b)=0$. Therefore, 
the Hamiltonians and the wave functions in interaction and Schr\"odinger pictures transform into each other by a simple coordinate displacement. 
At intermediate times, we interpolate the function as $\gamma (t)=\sum_{n=0}^{5}c_n t^n$, where the coefficients are found by solving Eqs. (\ref{b_c1}) and (\ref{b_c2}). Finally,
\beqa
\label{connection_i}
\gamma (t)&=&\gamma (0)+ 10 [\gamma (t_f)-\gamma (0)]s^3
\nonumber 
\\ 
&-&15 [\gamma (t_f)-\gamma (0)]s^4+ 6 [\gamma (t_f)-\gamma (0)]s^5,
\eeqa
where $s=t/t_f$. This function and examples of $\gamma_{eff}$ are shown in Fig. \ref{displacement_i}. 
%
%
%
%
%
%
%
%

In order to compare the robustness of the compensating force method against the alternative ones we consider a  $^{9}$Be$^{+}$ ion in the double well with the realistic parameters $\alpha=-4.7$ pN/m and $\beta=5.2$ mN/m$^3$, similar to those in \cite{Wilson2014}. 
For a moderate initial bias compared to the vibrational quanta, such as 
\beq
\label{moderate_bias}
\gamma_0 \sim \frac{\hbar \Omega_0}{D},
\eeq
the fidelity provided by the sudden approach is one for all practical purposes 
so we can change the bias abruptly and reach the target state.
The  displacement of the trap $d$ may be compared with the oscillator characteristic length 
$a_0=\sqrt{\hbar/m\Omega_0}$. Their ratio is 
\beq
\label{des_rel}
R=\frac{d}{a_0}=\frac{\gamma_0}{2\alpha}\sqrt{\frac{m\Omega_0}{\hbar}}.
\eeq
For the Paul trap $R\approx 0.00065$,  
which explains the high fidelity of the sudden approach for a moderate bias inversion of the ion.   
At these bias values there is really no need to apply a more sophisticated protocol than the sudden switch.    

Henceforth,  we assume a much larger $\gamma_0$, but still satisfying the condition (\ref{wef_bias}).  
In particular, for an initial bias of 1000 $\Omega_0\hbar$ (corresponding to $\gamma_0=86.4$ zN), the ratio becomes $R\approx0.65$.
The maximum variation of the difference between the trajectories of the minima is $3$ pm, about three orders of magnitude less than the displacement of each minimum ($9.2$ nm), so the minima follow parallel trajectories. Furthermore, the maximum variation of the frequencies in Eq. (\ref{w_ef_exp}) with respect to $\Omega_0=2\pi\times 5.6$ MHz
is  $2\pi\times 3.7$ kHz, so the effective frequency is nearly constant.  
%
%
%
%
\begin{figure}[t]
\begin{center}
\includegraphics[width=0.82 \linewidth]{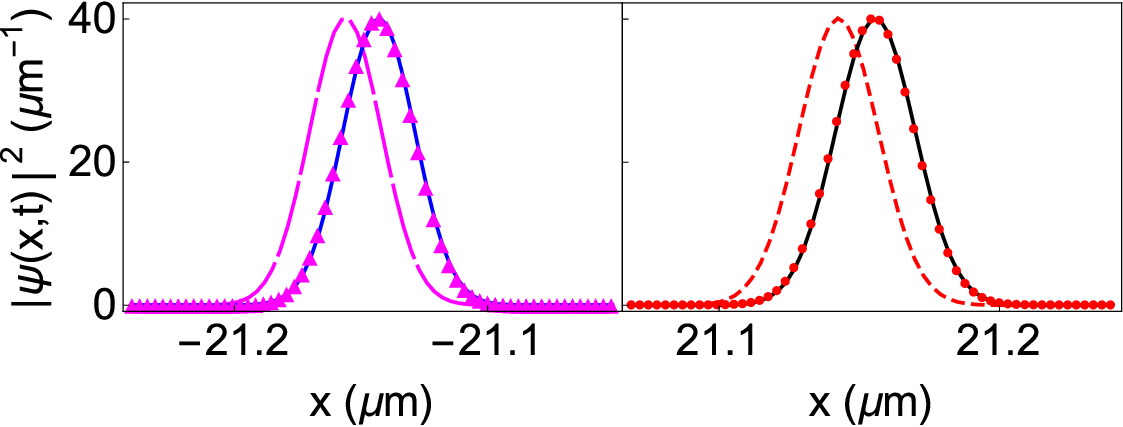}
\end{center}
\caption{\label{cf_evolution_i}
Left: Ground state of the left well  at $t=0$ (long-dashed magenta line) and at $t=t_f$ (magenta triangles), and final state with the compensating force applied on the double well (solid blue line). 
Right: Ground state of the right well: at $t=0$ (short-dashed red line) and at $t=t_f$ (red dots) and final state with the 
compensating force applied (solid black line). 
$t_f=4$ ns and other parameters as in Fig. \ref{displacement_i} for $^{9}$Be$^{+}$.}
\end{figure}
%
%
%
%

Figure \ref{cf_evolution_i} demonstrates  the effect of the 
compensating-force approach. Starting from the ground state of the lower (left) well,  the final evolved state following the shortcut with compensation stays as the ``ground state'' of the left well. This is actually defined as the lowest  state of the double-well system predominantly located on the left. 
There is a similar process for the right well. 
The final states represented are obtained by solving the Schr\"odinger equation with the full Hamiltonian (\ref{H_comp_ions}). 
%
%
%
%
\begin{figure}[t]
\begin{center}
\includegraphics[width=0.5 \linewidth]{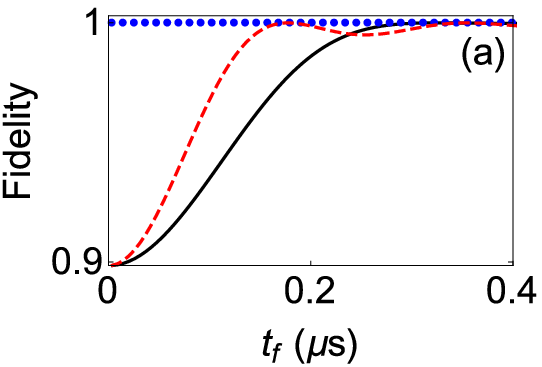}
\includegraphics[width=0.5 \linewidth]{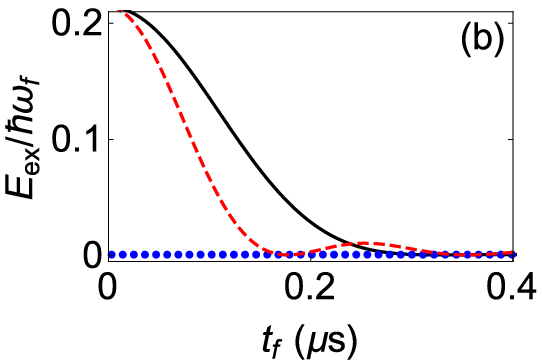}
\end{center}
\caption{\label{dynamics_tf_i}
(a) Fidelity $|\langle\phi_L(t_f)|\psi(t_f) \rangle|$, where $|\phi_L(t_f)\rangle$ is the lowest state located in the 
left well 
in the final time configuration, and $|\psi(t_f) \rangle$ is the evolved state following the shortcut at final time. (b) Final excitation energy for the process on the left well using  
compensating-force (blue dots), fifth degree polynomial in Eq. (\ref{connection_i}) (solid black line), and FAQUAD (short-dashed red line).
The parameters  are for $^{9}$Be$^{+}$ as in Fig. \ref{displacement_i}.}
\end{figure}
%
%
%
%
%

Figure \ref{dynamics_tf_i} is for the process in the left well. 
It compares the fidelity at final time and the excitation energy, defined as $E_{ex}=E(t_f)-E_0(t_f)$ where $E(t_f)$ is the final energy after the quantum evolution following the shortcut and $E_0(t_f)$ is the ground state final energy of the upper harmonic well,
using the polynomial (\ref{connection_i}) for $\gamma$ 
with and without compensation, as well as the results of the  FAQUAD approach.  
The fidelity without compensation tends to the fidelity of the sudden approach ($0.89$) for very short final times. 
The method with compensation clearly outperforms the others. 
In principle, a fundamental limitation of the approach is due to the fact that the inequality (\ref{wef_bias}), which guarantees 
parallel motion and stable frequencies of the wells, should as well be satisfied by $\gamma_{eff}$, but,  at very short times, the dominant 
term of $\gamma_{eff}\sim-m\ddot{\gamma}/{4\alpha}$ may be too large. To estimate this short time limit we 
combine the mean-value theorem inequality for the maximum \cite{Torrontegui2011}, $|\ddot{\gamma}|_{max}\geqslant 4\gamma_0/t_f^2$, with Eq. (\ref{wef_bias}) for $\gamma_{eff}$ 
to find  the condition
\beq
 t_f\gg \left( \frac{3m\gamma_0}{4\sqrt{2}} \sqrt{-\frac{\beta}{\alpha^5}}  \right)^{1/2}.
\eeq
The factor on the right-hand side is 
$10^{-9}$ s for this example (see Fig. \ref{dynamics_tf_i}), which is  several orders of magnitude smaller than $2\pi/\Omega_0$ and does not affect the fidelity in the scale of times shown.  
\subsection{Neutral atoms\label{na}}

The potential introduced in Chapter \ref{Chapter4} (Sec. \ref{mapp}),
\beq
\label{V_Oberthaler}
V_{na}(x,t)=\frac{1}{2}m\omega^2x^2+V_0\cos^2 \left [ \frac{\pi(x-\Delta x)}{d_l} \right ],
\eeq
forms also a double well. It was implemented in \cite{Gati2006} with optical dipole potentials, combining a harmonic confinement due to a crossed beam dipole trap with a periodic light shift potential provided by the interference pattern of two mutually coherent laser beams. $\omega$ is the angular frequency of the dipole trap
about the waist position, $V_0$ the amplitude, 
$\Delta x$ the displacement of the optical lattice relative to the center of the harmonic well,  and $d_l$ is the lattice constant. (Double wells with a controllable bias may be also realized by  
two optical lattices of different periodicity with controllable
intensities and relative phase \cite{Folling2007}). Here, the bias inversion implies a change of sign of $\Delta x(t)$ from $\Delta x_0>0$ to $-\Delta x_0$. 

To check if the conditions to apply the compensating force approach hold here 
as well, an analysis similar to the one in the previous example is now performed.
We approximate the potential around each minimum, $V^{(\pm)}$, up to fourth order.  
From $\frac{\partial V^{(\pm)}}{\partial x}=0$ we get a cubic equation for each minimum. 
The positions of the minima are thus given by 
\beq
\label{sol_an}
x_{0,\pm}=-2\sqrt{Q}\cos{\left ( \frac{\theta^{(\pm)}-2\pi}{3} \right )}-\frac{A^{(\pm)}}{3},
\eeq
where
\beqa
Q&=&\frac{2d_l^2\pi^2V_0+d_l^4m\omega^2}{4\pi^4V_0}, 
\nonumber \\
A^{(\pm)}&=&-\frac{3}{2}(2\Delta x \pm d_l), 
\nonumber \\
\theta^{(\pm)}&=&\cos{ \left [ \frac{-3d_l(2\Delta x \pm d_l) m \pi^2\sqrt{V_0}\omega^2}{2(2\pi^2V_0+d_l^2m\omega^2)^{3/2}} \right ]}^{-1}. 
\eeqa
Up to a second order in $\Delta x$,
\beq
\label{sol_an_2}
x_{0,\pm}\approx\pm a+b \Delta x \pm c\Delta x^2, 
\eeq
where the coefficients are known explicitly but too lengthy to be displayed here. 
Whenever the quadratic term is negligible with respect to the linear term ($c\Delta x^2\ll b\Delta x$), we can approximate $x_{0,\pm}=\pm a+b\Delta x$ (parallel motion). 
The distance between the minima is given by
\beqa
\label{D_an}
D&=&\frac{1}{3}\left \{ A^{(-)}-A^{(+)}+6\sqrt{Q} \left [ -\cos{\left ( \frac{\pi+\theta^{(-)}}{3} \right )} \right. \right. \nonumber \\ 
&+&\left. \left. \cos{\left (\frac{\pi+\theta^{(+)}}{3} \right )} \right ] \right \}
\approx 2a + 2c\Delta x^2.
\eeqa
Moreover, 
$\omega_{0,\pm}\approx f \mp g \Delta x$, again with known but lengthy 
coefficients $g$ and  $f$. 
As long as $g\Delta x \ll f$,
we may set $\omega_{0,\pm}\approx\Omega_0\equiv f$.

For realistic parameters the conditions for parallel transport of the minima and constant frequency are indeed satisfied.  
We consider a ${}^{87}$Rb atom in the trap and
set the parameters in Chapter \ref{Chapter4} after the demultiplexing process, namely, $d_l=5.18$ $\mu$m,  $\omega=2\pi\times 59.4$ Hz, and
$V_0/h=1.4$ kHz; the time-dependent displacement  $\Delta x=\Delta x(t)$ is the control parameter
to perform the bias inversion, such that 
\beqa
\label{b_c_d0}
&&\Delta x (0) =\Delta x_0, \nonumber \\
&&\Delta x (t_f)=-\Delta x_0,
\eeqa
with 
$\Delta x_0=200$ nm. 
We also impose 
\beqa
\label{b_c_d}
&&\dot \Delta x (0)=0, \nonumber \\
&&\ddot \Delta x (0)=0,  \nonumber \\
&&\dot \Delta x (t_f)=0, \nonumber \\  
&&\ddot \Delta x (t_f)=0 
\eeqa
to achieve similar conditions in the derivatives of the minima $x_0$.
At intermediate times, we interpolate the function as $\Delta x (t)=\sum_{n=0}^{5}d_n t^n$, where the coefficients are found by solving Eqs. (\ref{b_c_d0}) and (\ref{b_c_d}). Consequently, the connection between the initial and final Hamiltonians is given by the same polynomial 
in Eq. (\ref{connection_i}) changing $\gamma(t)\to\Delta x(t)$. 
The double wells are much deeper and tight for trapped ions than for neutral atoms;
compare an intrawell angular frequency $\Omega_0$ of  $2\pi\times5.6$ MHz for the ions versus $2\pi\times0.35$ kHz
for the optical trap. Therefore, in this case, for a moderate initial bias compared to the vibrational quanta, the ratio between the displacement of the trap $d$ and the oscillator characteristic length $a_0$ is $R\approx0.67$.

With the parameters given at time $t=0$, the separation of the minima is $D=5$ $\mu$m, the bias between minima $\delta=2.02\times10^{-32}$ J, and the effective angular frequency $\Omega_0=2\pi\times 0.35$ kHz, whereas the maximum variation of the frequencies along the process is approximately $2\pi\times 0.2$ Hz.
Furthermore, the maximum deviation from $D$ of the minima separation is $1.8$ nm, whereas the displacement of each minimum is about $0.4$ $\mu$m. In summary, the minima do move in parallel with constant, equal frequencies for practical purposes.  
%
%
%
%
%
\begin{figure}[t]
\begin{center}
\includegraphics[width=0.5 \linewidth]{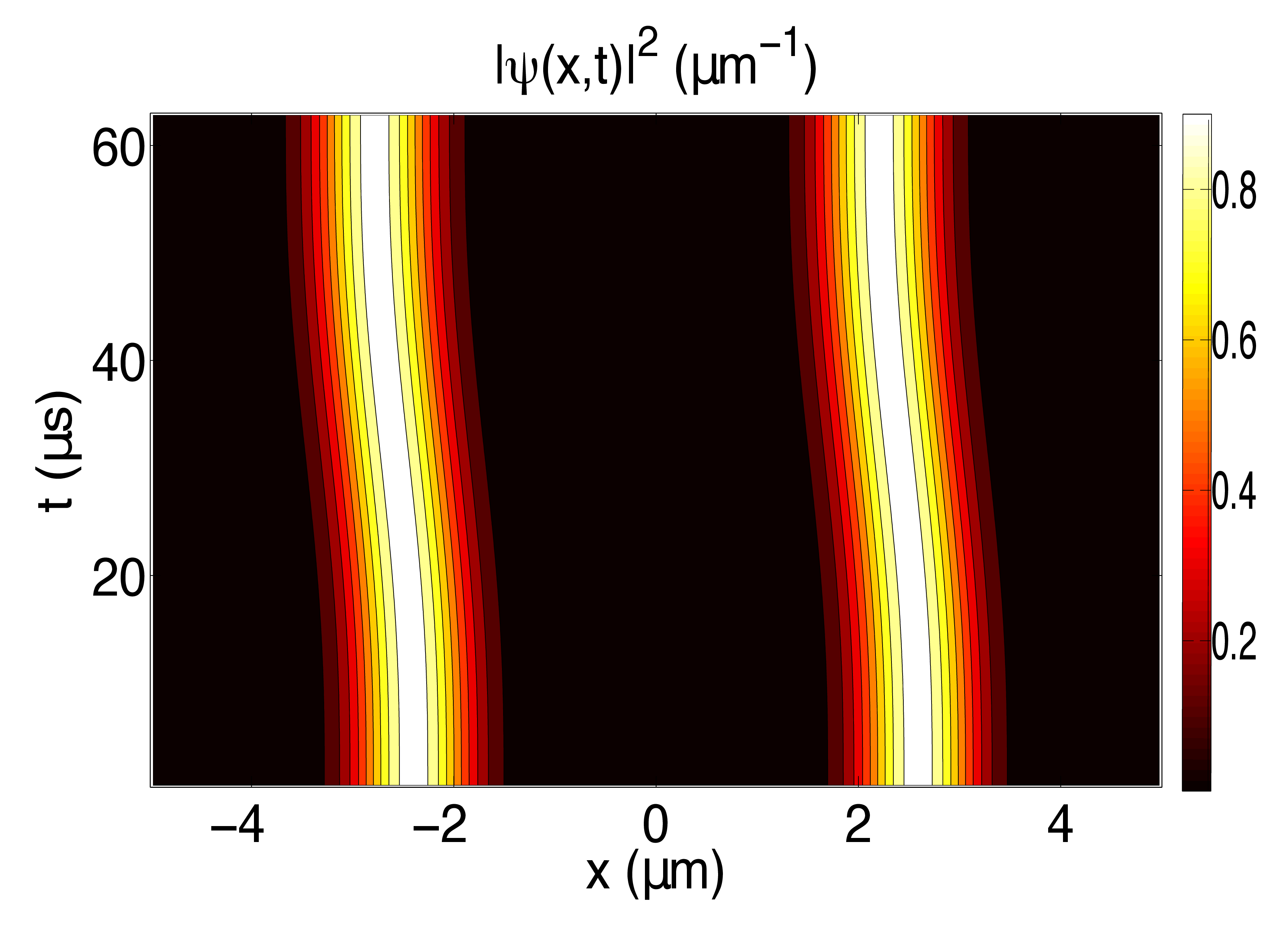}
\end{center}
\caption{\label{contour}
Evolution of the wave function densities  following the shortcut in Eq. (\ref{H_comp}) for 
states in left and right wells.
The parameters are for ${}^{87}$Rb: $d_l=5.18$ $\mu$m, $\omega=59.4\times 2\pi$ Hz, $V_0/h=1.4$ kHz,  $\Delta x_0=200$ nm, and $t_f=63$ $\mu$s.}
\end{figure}
%
%
%
%

To accelerate the bias inversion we add the compensating term to $V$ in Eq. (\ref{V_Oberthaler}), 
\beq
\label{H_comp}
H=\frac{p^2}{2m}+\frac{1}{2}m\omega^2x^2+V_0\cos^2 \left [ \frac{\pi(x-\Delta x)}{d_l} \right ]-mx\ddot x_0.
\eeq

Figure \ref{contour} shows the evolution of the densities. 
%
%
%
%
\begin{figure}[t]
\begin{center}
\includegraphics[width=0.5 \linewidth]{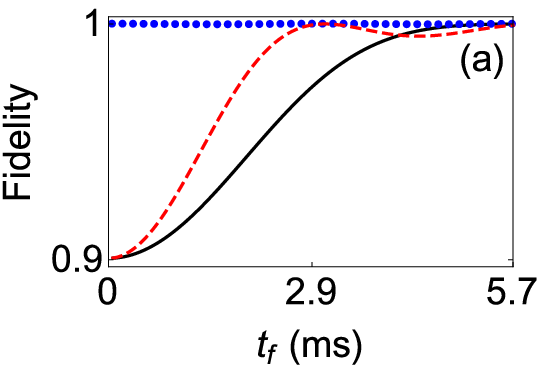}
\includegraphics[width=0.5 \linewidth]{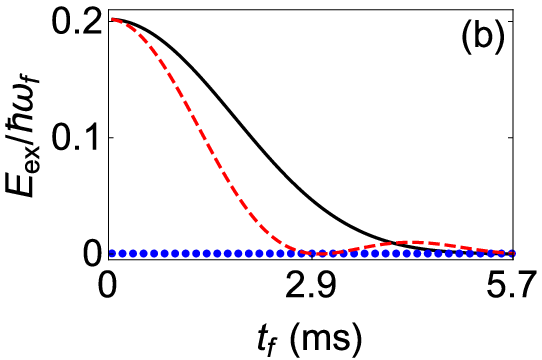}
\end{center}
\caption{\label{dynamics_tf}
(a) Fidelity $|\langle\varphi_1(t_f)|\psi(t_f) \rangle|$, where $|\varphi_1(t_f)\rangle$ is the lowest state 
predominantly of the left well at final time (the first excited state of the double well) and $|\psi(t_f) \rangle$ is the evolved state following the shortcut at final time.
(b) Final excitation energy.
Compensating-force approach (blue dots), fith degree polynomial in Eq. (\ref{connection_i}) with the change $\gamma(t)\to\Delta x(t)$ without compensation (solid black line), and FAQUAD approach (short-dashed red line).
The parameters are chosen for ${}^{87}$Rb: $d_l=5.18$ $\mu$m, $\omega=59.4\times 2\pi$ Hz, $V_0/h=1.4$ kHz, and $\Delta x_0=200$ nm.
}
\end{figure}
%
%
%
%
Focusing on the left well, 
Fig. \ref{dynamics_tf}(a) demonstrates that  the fidelity is exactly one (blue dots) with the compensating force. However, using for the inversion  
the fifth degree polynomial in Eq. (\ref{connection_i}) [with the change $\gamma(t)\to\Delta x(t)$] without compensation, the fidelity at short final times decreases until the value of the sudden approach, 0.9.
Furthermore, the excitation (residual) energy $E_{ex}$
is approximately zero for the shortcut protocol, compared to the excitation without compensation 
in Fig. \ref{dynamics_tf}(b). 
\section{Discussion
\label{discussion}}

We have proposed a method to invert the bias of a double-well potential, in the regime of independent wells,  to keep the final states motionally unexcited within the same original well. The method  treats the bias inversion as a rigid transport of the wells, which is justified for realistic parameters, and applies  
a ``compensating-force'' to cancel the excitations. 
Examples have been worked out 
for ions or neutral atoms, and comparisons have been provided with a sudden approach, a FAQUAD approach, or a smooth polynomial connection of initial and final bias 
without compensation. The compensating-force method clearly outperforms the others and gives  perfect fidelities  under ideal conditions, 
up to very small times compared, e.g., with the time 
$2\pi/\Omega_0$ (one oscillation period) where FAQUAD provides a first zero of the excitation energy.    
The feasibility of the approach may be analyzed in the light of current technology
in the two systems examined: 

\begin{enumerate}
\item{
For trapped ions we have considered initial and final values differing by $\Delta\gamma\approx 200$ zN, 
whereas resolutions of $15$ zN of have been  reported \cite{Ruster2014}. 
As for the timing, much effort is being put into achieving suboscillation-period resolutions for the potential update \cite{Alonso2013,bowler2015coherent,Alonso2015}
in ion transport experiments. The  
possibility to switch on and off potentials in a few nanoseconds, much faster than the ion oscillation period, has been demonstrated \cite{Alonso2015}. The designed bias inversion 
is thus in reach with current technology.}         

\item{
For neutral atoms, the minimal process times $t_f$ are not limited by the method per se but by the technical capabilities to implement the maximal compensating force. This force depends on the maximal acceleration of the well, whose lower bound is known to be $a_{\rm max}=2d/t^2_f$ \cite{Torrontegui2011}. 
To implement the compensation with a magnetic field gradient $G$, the gradient should be of the order of $G\simeq m a_{\rm max}/\mu_B$ in an amount of time $t_f$ ($\mu_B$ is Bohr's magneton). For rubidium atoms polarized in the magnetic level $F=m_F=2$ and the transport parameters considered above, this requires a magnetic field gradient on the order of $40$ T/m shaped on a time interval $t_f=63$ $\mu$s. This is definitely challenging from an experimental point of view. Alternatively, one can use the dipole force of an out of axis Gaussian laser beam. If the double well is placed on the edge at a distance of $w/2$ where $w$ is the waist and if $\alpha_p$ denotes the polarizability, the local potential slope experienced by the atoms is on the order of $\alpha_p P/w^3$, where $P$ is the power of the beam. The compensation requires that $P/w^3=m/\alpha_p$. For instance, with an out-of-resonance beam at a wavelength of 1 $\mu$m, the polarizability of rubidium-87 atoms is $\alpha_p \simeq 1.3\times 10^{-36}$ m$^2$s, and the compensation can be performed using a 1W laser with a waist of 20 $\mu$m. As for the timing, a submicrosecond  time scale for shaping the offset potential is  perfectly achievable using a control of the intensity based on acousto-optics modulators.}  
\end{enumerate}

A relevant feature of the proposed approach is that the reference process used to design the corresponding compensation
(we have used a polynomial for simplicity) 
may be chosen among a broad family of functions satisfying Eqs. (\ref{b_c1}) and (\ref{b_c2}). As in other STA approaches, this flexibility may be used to enhance robustness versus noise and perturbations \cite{Ruschhaupt2012a,Lu2014,Guery-Odelin2014}.

The bias inversion put forward here and the multiplexing and demultiplexing protocols developed in \cite{Martinez-Garaot2013}, 
see Chapter \ref{Chapter4}, provide  the necessary toolkit to perform vibrational state inversions \cite{Bucker2013,Bucker2011}, or Fock state preparations \cite{Martinez-Garaot2013}. Applications in optical waveguide design are also feasible \cite{Tseng2012}.  As well, the fast bias inversion 
is directly applicable to Bose-Einstein condensates \cite{Masuda2012,Torrontegui2012b}.  
Generalizations for conditions in which rigid transport does not hold 
are also possible using invariant theory \cite{Torrontegui2011}, which allows for finding processes without final excitation where both the frequency and position of the well depend on time \cite{Palmero2015b}.


\chapter*{Conclusions} 
\label{Conclusions}
\lhead{\emph{Conclusions}} 

In this Thesis a set of Shortcuts-to-Adiabaticity (STA) techniques have been developed and applied to speed up adiabatic processes in  systems confines by double-well potentials. The main results can be summarized as follows:

\begin{itemize}

\item {\bf Engineering fast and stable splitting of matter waves}
\begin{itemize}
\item{Wave-packet splitting is very unstable with respect to slight trap asymmetries. The adiabatic following produces the collapse of the wave into the lower well. We used the fast-forward (FF) approach to accelerate and stabilize the separation.}

\item{We also introduced a simple moving two-mode model, which combined with sudden and adiabatic approximations provides a stability criterion. This model has also played an important role in the rest of the Thesis.}

\item{Furthermore, we applied the shortcut to speed up the splitting of a condensate in the mean-field regime.}
\end{itemize}

\item {\bf Shortcuts to adiabaticity in three-level systems using Lie transforms}
\begin{itemize}
\item{The shortcuts based on the counterdiabatic approach are, in most cases, difficult to implement in the laboratory, so we developed alternative, realizable protocols making use of the dynamical symmetry of the Hamiltonian.}

\item{The new approaches have been designed by means of a Lie transform. Although the transformations are formally equivalent to 
interaction-picture (IP) transformations, the resulting IP Hamiltonian and state represent a different physical process from the original one.}

\item{Mott-insulator transitions and beam splitter implementations have been stabilized and accelerated thanks to the new, Lie-based STA.}
\end{itemize}

\item {\bf Fast quasi-adiabatic dynamics}
\begin{itemize}
\item{General properties of a ``fast-quasi-adiabatic'' (FAQUAD) method based on using the time dependence of a control parameter to delocalize in time the transition probability among adiabatic levels have been found.}

\item{The approach has been applied to different systems where other approaches are not available. In particular, in a two-site boson system and in a many-particle system.}

\item{Another important result is the discovery that FAQUAD is optimal  within the sequence of iterative superadiabatic frames.}
\end{itemize}

\item {\bf Vibrational mode multiplexing of ultracold atoms}
\begin{itemize}
\item{Processes to achieve fast vibrational-state multiplexing or demultiplexing have been designed. The invariant-based inverse engineering protocol has been applied in a two-mode model and then mapped onto a realistic potential.}

\item{While protocols calculated with an Optimal Control Theory (OCT) algorithm are quite difficult to implement experimentally, our proposal relies on a smooth potential deformation.}
\end{itemize}

\item {\bf Compact and high conversion efficiency mode-sorting asymmetric Y junction using shortcuts to adiabaticity}
\begin{itemize}
\item{The power of the approach used to reproduce fast multiplexing and demultiplexing processes is demonstrated in the context of optical waveguides. Specifically, a short mode-sorting asymmetric Y junction has been worked out.}
\end{itemize}

\item {\bf Fast bias inversion of a double well without residual particle excitation}
\begin{itemize}
\item{The compensating-force method has been applied to realize a fast bias inversion, both for trapped ions and for neutral atoms.}

\item{The combination of the bias inversion and multiplexing/demultiplexing processes can be used to induce vibrational state inversions based on trap deformations only. The possibility of achieving a population inversion without using internal-state excitations is of much interest. In particular, for trapped ions this amounts to a species-independent approach based entirely on the charge and electric forces.}

\item{The implementation of a fast and stable bias inversion could be useful also to produce vibrationally excited Fock states from an initial ground state.}

\end{itemize}

\end{itemize}

\addcontentsline{toc}{chapter}{Conclusions}


\addtocontents{toc}{\vspace{2em}} 

\appendix 
\part*{Appendix}
\addtocontents{toc}{\vspace{0.6em}} 


\chapter{Interaction versus asymmetry for adiabatic following}
\label{Interaction versus asymmetry for adiabatic following}
\lhead{Appendix A. \emph{Interaction versus asymmetry for adiabatic following}}

Making some simplifying assumptions, we find the conditions under which  
the interacting condensate ground state splits adiabatically, instead of collapsing into the deepest well. 
We consider complete splitting of the trap into separated wells  and also $\delta(t_f)\ll \lambda$, 
so that the noninteracting wave would collapse (see Sec. \ref{2mode}).
In atomic interferometry, the two split branches of the condensate have
to be individually addressed 
and manipulated during the differential phase accumulation stage, so that 
tunnelling must be negligible \cite{Pezze2005,Shin2004,Grond2010}.    
We also assume that the two ground states 
of the final wells can be approximated by ground states
of harmonic oscillators at $\pm x_f$, with the right one lifted by $\lambda$:     
\beqa
V_L&=&\frac{1}{2}m\omega^2(x+x_f)^2,
\\
V_R&=&\frac{1}{2}m\omega^2(x-x_f)^2+\lambda.
\eeqa
The total energy is approximated as 
$E_{tot}=E_L+E_R$. For $j=L,R$, 
\beq
E_{j}=N_j\int dx \phi_j\left[-\frac{\hbar^2\partial_x^2}{2m}+V_j\right]\phi_j
+\frac{1}{2}g_1 N_j^2\int dx |\phi_j|^4, 
\eeq
where 
\beq
\phi_j(x)=\frac{1}{[\sqrt{\pi}a_0]^{1/2}}e^{[-(x\pm x_f)^2/2a_0^2]},
\eeq
and the total number of particles is $N=N_R+N_L$. 
The result is 
\beqa
E_L&=&N_L\frac{\hbar\omega}{2}+\frac{\widehat{g_1}}{2\sqrt{2\pi}}\hbar\omega N_L^2,
\\
E_R&=&N_R\left(\frac{\hbar\omega}{2}+\lambda\right)+\frac{\widehat{g_1}}{2\sqrt{2\pi}}\hbar\omega N_R^2,
\eeqa
where 
\beq
\widehat{g_1}=g_1/(\hbar \omega a_0).
\eeq
From the minimum-energy condition, 
$\partial {E_{tot}}/\partial{N_R}=0$, it follows that  
\beq
\frac{\Delta N}{N}=\sqrt{2\pi} \frac{\lambda/\hbar\omega}{\widehat{g_1}N},
\eeq
%
\begin{figure}[t]
 \begin{center}
   \includegraphics[width=0.54 \linewidth]{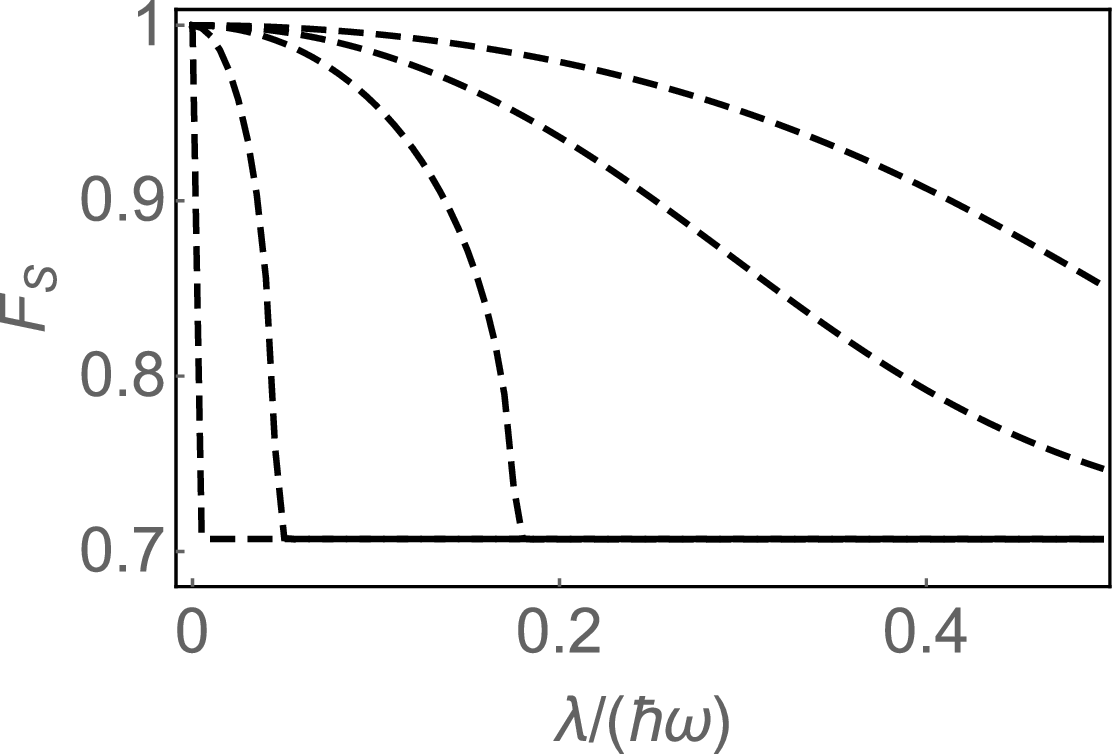}
    \end{center}
\caption{
Structural fidelities for the Bose-Einstein condensate. 
From left to right, $\widehat{g_1}N= 0, 0.138, 0.55, 0.69$, and $1.38$.   
In all curves
$x_{f}=4$ $\mu$m and $\omega=780$ rad/s. Equation (\ref{eq11}) was used to design the potential $V_{FF}$. 
}
\label{f5}
\end{figure} 
with $\Delta N=N_L-N_R$. (See \cite{Hall2007} for a similar treatment in the Thomas-Fermi regime.) Thus, collapse into one well is avoided when $\lambda/(\hbar\omega \widehat{g_1}N)\ll1$. This relation sets the scale for the uncontrolled and, possibly, unknown asymmetry that may be tolerated to achieve balanced splitting.
Adiabatic  control of  population imbalance  requires control of the energy splitting of the order 
$\lambda\lesssim (\hbar\omega \widehat{g_1}N)$.

Figure \ref{f5} shows the structural fidelity $F_S(\lambda)$ for several    
values of $\widehat{g_1}N$. The sharp drop at $\widehat{g_1}N=0$ is substituted by more and more stable curves as $\widehat{g_1}N$ increases. 
For  the splitting described in \cite{Albiez2005} and \cite{Gati2006} using $^{87}$Rb, 
we get $\widehat{g_1}N\approx 9.5$, quite large compared to the values in Fig. \ref{f5}. Under these conditions, 
adiabatic splitting is very stable with respect to minor asymmetries.
Moreover, $F_S$ decays slowly with respect to $\lambda$,
so that 
the relative population imbalance may be prepared at will  by controlling the asymmetry. 
In \cite{Albiez2005} and \cite{Gati2006} the asymmetry is due to a potential shift 
that can be controlled with a standard deviation of 100 nm, whereas  a displacement of $\sim1\mu$m is required for the 
total collapse into one of the wells.


\chapter{Lie algebra}
\label{algebra}
\lhead{Appendix B. \emph{Lie algebra}}

The algebra of this three-level system is a four-dimensional Lie Algebra U3S3 according to the classification of four-dimensional Lie algebras in \cite{maccallum1999classification}. 
(For comparison with that work it is useful to rewrite the generators in the skew-Hermitian base  $\tilde G_k=-i G_k$, $k=1,2,3,4$.) 
$U3S3$ 
is a direct sum of the one-dimensional algebra spanned by the invariant $G_4-G_3$, that commutes with all members of the algebra, 
and a three-dimensional SU(2) algebra spanned by $\{G_1, G_2, G_3\}$.  
Notice that this realization of the three-dimensional (3D) algebra is not spanned by  the matrices 
\beq
\label{generators_spin1}
J_x=\frac{1}{\sqrt{2}}\left ( \begin{array}{ccc}
0 & 1 & 0\\
1 & 0 & 1 \\
0 & 1 & 0
\end{array} \right),
J_y=\frac{1}{\sqrt{2}}\left ( \begin{array}{ccc}
0 & -i & 0\\
i & 0 & -i \\
0 & i & 0
\end{array} \right), 
J_z=\left ( \begin{array}{ccc}
1 & 0 & 0\\
0 & 0 & 0 \\
0 & 0 & -1
\end{array} \right), 
\eeq
which correspond, in the subspace ${|2,0\rangle,|1,1\rangle,|0,2\rangle}$, to the operators
\beqa
\label{spin1}
J_x&=&\frac{1}{2}\left ( a_1^\dag a_2+a_2^\dag a_1 \right ), 
\\
J_y&=&\frac{1}{2i}\left ( a_1^\dag a_2-a_2^\dag a_1 \right ), 
\\
J_z&=&\frac{1}{2} \left ( a_1^\dag a_1-a_2^\dag a_2 \right ).
\eeqa
In particular we cannot get the matrices for $J_y$ or $J_z$ by any linear combination of our $G_k$
matrices [see Eqs. (\ref{G1_G4}-\ref{generators})].
A second-quantized form for the $G_k$ consistent with the matrices includes quartic terms in annihilation and creation operators: 
\beqa
G_1&=&\frac{1}{4}\left ( a_1^\dag a_2+a_2^\dag a_1 \right ), \nonumber
\\
G_2&=&\frac{1}{4i}\left [ a_1^\dag a_2^\dag  \left ( a_1 a_1 + a_2 a_2 \right )
-\left (a_1^\dag a_1^\dag+ a_2^\dag a_2^\dag \right ) a_1 a_2  \right ], \nonumber
\\
G_3&=&\frac{1}{8} \left [ \left (a_1^\dag a_1^\dag+ a_2^\dag a_2^\dag \right) a_1 a_1 - 4 a_1^\dag a_2^\dag a_1 a_2
+\left (a_1^\dag a_1^\dag+ a_2^\dag a_2^\dag \right ) a_2 a_2 \right ] , \nonumber
\\
G_4&=&\frac{1}{4} {\left( a_1^\dag a_1 - a_2^\dag a_2 \right )} ^2.
\eeqa
These second-quantized operators do not form a closed algebra under commutation, but their matrix elements
for two particles do. 

An invariant (defined in a Lie-algebraic sense) commutes with any member  of the algebra.
There are generically two independent invariants for $U3S3$ \cite{Patera1976}.   
For the matrix representation in Eqs. (\ref{G1_G4}) and (\ref{generators}) they are 
\beqa
&&I_1=G_1^2=G_2^2=G_3^2=\frac{1}{8}\left ( \begin{array}{ccc}
1 & 0 & 1\\
0 & 2 & 0 \\
1 & 0 & 1
\end{array} \right),
\nonumber
\\
&&I_2=G_4-G_3=\frac{1}{4}\left ( \begin{array}{ccc}
3 & 0 & -1\\
0 & 2 & 0\\
-1 & 0 & 3
\end{array} \right).
\eeqa
$I_1$, which is not in the algebra,  has eigenvalues
\beqa
\lambda_1^{(2)}&=&1,\nonumber
\\
\lambda_1^{(1,3)}&=&\frac{1}{2}, 
\eeqa
and $I_2$, a member of the algebra,  has eigenvalues
\beqa
\lambda_2^{(2)}&=&0, \nonumber
\\
\lambda_2^{(1,3)}&=&\frac{1}{4}. 
\eeqa
The two invariants have the same eigenvectors,
\beqa
|u^{(1)}\rangle&=&\frac{1}{\sqrt{2}}(|2,0\rangle + |0,2\rangle),
\nonumber
\\
|u^{(2)}\rangle&=&\frac{1}{\sqrt{2}}(|2,0\rangle- |0,2\rangle),
\nonumber
\\
|u^{(3)}\rangle&=&|1,1\rangle,
\eeqa
with $|u^{(1)}\rangle$ and $|u^{(3)}\rangle$ spanning a degenerate subspace. 

Lie-algebraic invariants constructed with time-independent coefficients 
satisfy  as well the equation 
\beq
\label{di}
i\hbar \frac{\partial I_{1,2}}{\partial t} +  [H(t),I_{1,2}]=0
\eeq
so they are also dynamical invariants \cite{Lewis1969} [i.e., operators that satisfy  Eq. (\ref{di}) whose expectation values remain constant]. 
The  degenerate subspace of eigenvectors allows the existence of time-dependent eigenstates 
of time-independent invariants. 
In particular, in all the examples in the main text, the dynamics takes place within the degenerate subspace:  
the initial state is $|u^{(3)}\ra$ at $t=0$ and ends up in some combination of $|u^{(1)}\ra$ and $|u^{(3)}\ra$ at $t_f$.
The specific state as a function of time is known explicitly, 
$|\psi_I(t)\ra=e^{i\alpha(t)G_4}e^{-i\int_0^{t} E_1 dt'} |\phi_1(t)\ra$; see Eq. (\ref{I_state}).   
Note that $|\phi_1\ra$ and $|\phi_3\ra$ in Eqs. (\ref{eigenstates_1}) and (\ref{eigenstates_3}) 
are two orthogonal combinations of $|u^{(1)}\ra$ and $|u^{(3)}\ra$. Also 
$|u^{(2)}\ra=|\phi_2\ra$; see Eq. (\ref{eigenstates_2}). In the nondegenerate subspace spanned by 
$|u^{(2)}\ra$ ``nothing evolves'', other than
a phase factor, but the initial states in the examples do not overlap with it.


\addtocontents{toc}{\vspace{0.6em}}  
\backmatter
\pagestyle{empty}  

\label{Bibliography}
\lhead{\emph{Bibliography}}  
\bibliographystyle{sofia} 
\bibliography{Bibliography}  

\end{document}